\documentclass[prl,nofootinbib,showpacs]{revtex4}

\usepackage{graphicx}

\begin{document}

{\sl To appear in Physics Uspekhi}

\title{High -- Temperature Superconductivity in Iron Based Layered Compounds}

\author{M.\,V.~Sadovskii}

\affiliation{Institute for Electrophysics, Russian Academy of Sciences, 
Ural Branch, 620016 Ekaterinburg, Russia\\
}

\begin{abstract}

We present a review of basic experimental facts on the new class of high --
temperature superconductors --- iron based layered compounds like
REOFeAs (RE=La,Ce,Nd,Pr,Sm...), AFe$_2$As$_2$ (A=Ba,Sr...), 
AFeAs (A=Li,...) and FeSe(Te). We discuss electronic structure,
including the role of correlations,  spectrum and role of collective
excitations (phonons, spin waves), as well as the main models, describing
possible types of magnetic ordering and Cooper pairing in these compounds
\footnote{{\sl Extended version of the talk given on the 90th anniversary 
celebration of Physics Uspekhi, P.N. Lebedev Physical Institute, Moscow, 
November 19, 2008.}}.

\end{abstract}

\pacs{74.25.Jb, 74.20.Rp, 74.25.Dw, 74.25.Ha,
74.20.-z, 74.20.Fg, 74.20.Mn, 74.62.-c, 74.70.-b}

\maketitle

\tableofcontents


\section{Introduction}

The discovery more than 20 years ago of high -- temperature superconductivity 
(HTSC) in copper oxides \cite{BM86} attracted much interest and has
lead to publication of thousands of experimental and theoretical papers.
During these years a number of reviews were published in Physics Uspekhi, which
were devoted to different aspects of superconductivity in cuprates and 
perspectives of further progress in the growth of the critical temperature of
superconducting transition $T_c$, which appeared both immediately after this
discovery \cite{GK88,I89,I91} and much later 
\cite{I99,M2000,MS01,K2006,M2008,K2008}. Physics of HTSC and of a number of new
and unusual superconductors was discussed in a number of individual and
collective monographs \cite{Plak,And,Schr,Kett}.

Unfortunately, despite these unprecedented efforts of researchers all over the
world, the physical nature of high -- temperature superconductivity in cuprates
is still not completely understood. Basic difficulties here are attributed to
the significant role of electronic correlations -- in the opinion of the majority
of authors cuprates are strongly correlated systems, which lead to anomalies
of the normal state (inapplicability of Fermi -- liquid theory) and the wide
spectrum of possible explanations of microscopic mechanism of superconductivity
-- from relatively traditional \cite{M2000,M2008} to more or less exotic
\cite{And}.

In this respect, the discovery in the early 2008 of a new class of high --
temperature superconductors, i.e. layered compounds based on iron \cite{Hos08},
has attracted a tremendous interest. This discovery has broken cuprate
``monopoly'' in the physics of HTSC compounds and revived hopes both on
further progress in this field related to the synthesis of new perspective
high -- temperature superconductors, as well as on more deep theoretical
understanding of mechanisms of HTSC. Naturally, the direct comparison of
superconductivity in cuprates and new iron based superconductors promises
the identification of common anomalies of these systems, which are relevant for
high values of $T_c$, as well as important differences not directly related 
to the phenomenon of high -- temperature superconductivity and, in fact,
complicating its theoretical understanding.

The aim of the present review is a short introduction into the physics of new
iron based layered superconductors and comparison of their properties with
well established facts concerning HTSC in copper oxides. Progress in this field
is rather spectacular and rapid, so that this review is in no sense exhaustive 
\footnote{It is sufficient to say that during the first six months of the
studies of new superconductors about 600 original papers (preprints) were
published. Obviously, in this review it is even not possible just to cite all 
these works and our choice of citations is rather subjective. The author pays
his excuses in advance to those authors of important papers, which remained
outside our reference list, as this is attributed to obvious limitations of size
and author's own ignorance.}. However, the author hopes that 
it may be a kind of elementary introduction into this new field of research
and may help a more deep studies of original papers.

\newpage

\subsection{Anomalies of high--temperature superconductivity in copper
oxides}

At present dozens of HTSC compounds based on copper oxides are known to have
temperature of superconducting transition $T_c$ exceeding 24K \cite{Cry}.
In Table I we present critical temperatures for number of most ``popular''
cuprates.

\begin{center}
Table I. Temperatures of superconducting transition in copper oxides\footnote{
Under pressure $T_c$ of $HgBa_2Ca_2Cu_3O_{8+\delta}$ reaches $\sim$ 150K.}.
\vskip 0.5cm
\begin{tabular}{| c | c |}
\hline
Compound    &  $T_c$(K)   \\ 
\hline 
$HgBa_2Ca_2Cu_3O_{8+\delta}$  & 134 \\  
$Tl_2Ca_2Ba_2Cu_3O_{10}$  & 127  \\
$YBa_2Cu_3O_7$ & 92  \\
$Bi_2Sr_2CaCu_2O_8$ & 89  \\
$La_{1.83}Sr_{0.17}CuO_4$ & 37 \\
$Nd_{1.85}Ce_{0.15}CuO_4$ & 24 \\
\hline
\end{tabular}
\end{center}

More than 20 years of experience has lead to rather deep understanding of the 
nature of superconductivity in these systems. It is clear for example, that
HTSC in cuprates is not connected with some ``essentially new'' physics,
which is different in comparison with other superconductors, superfluid
Fermi -- liquids like He$^3$, nucleons in atomic nuclei or nuclear matter
in neutron stars, dilute Fermi -- gases with Cooper pairing actively studied
at present, or even hypothetical ``color'' superconductivity of quarks.

In this respect let us list 

\subsubsection{What do we definitely know about cuprates:}

\begin{itemize}

\item{\sl The nature of superconductivity in cuprates $\rightleftharpoons$ 
Cooper pairing.}

\begin{enumerate}

\item{This pairing is anisotropic and of $d$ -- type, so that the energy gap 
acquires zeroes at the Fermi surface: $\Delta\cos 2\phi$, where $\phi$ is
the polar angle, determining momentum direction in the two -- dimensional
Brillouin zone.}

\item{The size of Cooper pairs (coherence length at zero temperature $T=0$) 
is relatively small:  $\xi_0\sim 5-10 a$, where $a$ is the lattice constant in 
$CuO_2$ plane.}

\end{enumerate}

\item{\sl There exists a relatively well defined (at least in the
part of the Brillouin zone) Fermi surface, in this sense these systems are
metals.}

\item{\sl However, appropriate stoichiometric compounds are antiferromagnetic
insulators  --- superconductivity is realized close to the Mott metal -- 
insulator phase transition (controlled by composition) induced by strong
electronic correlations.}

\item{\sl The strong anisotropy of all electronic properties is observed ---
conductivity (and superconductivity) is realized mainly within $CuO_2$ layers
(quasi two--dimensionality!).}

\end{itemize}

At the same time, many things are still not understood. Accordingly, we
may list 

\subsubsection{What do we not really know about cuprates:}

\begin{itemize}

\item{{\sl Mechanism of Cooper pairing (a ``glue'' leading to formation of
Cooper pairs).} 

Possible variants:

\begin{enumerate}

\item{Electron -- phonon mechanism.}

\item{Spin fluctuations.}

\item{Exchange -- $RVB$, $SO(5)$, or something more ``exotic''.}

\end{enumerate}

}

The difficulties of choice here are mainly due to the unclear

\item{{\sl Nature of the normal state.}

\begin{enumerate}

\item{Is Fermi -- liquid theory (Landau) valid here?}

\item{Or some more complicated scenario is realized, like
``marginal'' or ``bad'' Fermi -- liquid?}

\item{Possible also is some variant of Luttinger liquid, essentially different 
from Fermi -- liquid.}

\item{Still mysterious is the nature of the pseudogap state.}

\item{Not completely clear is the role of internal disorder and local 
inhomogeneities.}

\end{enumerate}

}

\end{itemize}

Of course, we observe continuous, though slow, progress. For example, most
researchers are now leaning towards spin -- fluctuation (non phonon) mechanism
of pairing. Pseudogap state is most likely connected with fluctuations of
some competing (with superconductivity) order parameter (antiferromagnetic or
charged) \cite{lpisem, 2gap}. Fermi -- liquid description is apparently
applicable in the most part of the Brillouin zone, where Fermi surface remains
not ``destroyed'' by pseudogap fluctuations etc. However, the consensus in
HTSC community is still absent, which is obviously related to the complicated
nature of these systems, controlled beforehand by strong electronic correlations,
which control this nature and complicate theoretical understanding.
Just because of that, the discovery of a new class of HTSC compounds promises
definite hopes as new possibilities of HTSC studies in completely different
systems appear, where some of these difficulties may be just absent.

\subsection{Other superconductors with unusual properties}

Obviously, during all the years since the discovery of HTSC in cuprates 
active efforts continued in the search of new compounds with potentially high
temperatures of superconducting transition. A number of systems obtained 
during this search are listed in Table II.

\begin{center}
Table II. Temperatures of superconducting transition  
in some ``unusual'' superconductors.
\vskip 0.5cm
\begin{tabular}{| c | c |}
\hline
Compound    &  $T_c$(K)   \\ 
\hline 
$MgB_2$  & 39 \\  
$RbCs_2C_{60}$ &  33  \\
$K_3C_{60}$  & 19  \\
$Sr_2RuO_4$ & 1.5 \\
\hline
\end{tabular}
\end{center}

Despite the obvious interest from the point of view of physics and unusual
properties of some of these systems there were no  significant progress on this
way. This was mainly due to the fact that all the systems listed in Table II
are, in some sense, ``exceptional'' -- none is a representative of a wide class
of compounds with a possibility of a change of system properties (parameters) 
in a wide range, as in cuprates. All these systems were studied in detail, 
and summaries of these studies can be found in the relevant reviews
\cite{Penn,Maeno,Ovch,Vin,Row}. In fact, all these studies have added very
little to our understanding of the cuprates. Up to now there was kind of
``cuprate monopoly'' in the physics of ``real'' HTSC materials with wide
perspectives of further studies and search of compounds with even higher
values of $T_c$ and practical applications\footnote{Surely, when we speak 
about practical applications we should not underestimate the perspectives
of $MgB_2$ compound.}.


\section{Basic experimental facts on new superconductors}

\subsection{Electrical properties and superconductivity}

\subsubsection{$REOFeAs$ ($RE=La,Ce,Pr,Nd,Sm,...$) system}

Discovery of superconductivity with $T_c=26$K in $LaO_{1-x}F_xFeAs$ 
($x=0.05-0.12$) \cite{Hos08} was preceded by the studies of electrical
properties of a number of oxypnictides like $LaOMPn$ ($M=Mn,Fe,Co,Ni$ 
and $Pn=P,As$) highlighted by discovery of superconductivity in
$LaOFeP$ with $T_c\sim 5$K \cite{Hos06} ¨ $LaONiP$ á $T_c\sim 3$K \cite{Hos07},
which has not attracted much attention from HTSC community. 
This situation has changed sharply after Ref. \cite{Hos08} has appeared and
shortly afterwards a lot of papers followed (see e.g. 
\cite{Chen0128,Man1,Zhu1288,Chen3603,Chen3790,Chen4384,Ren4234,Ren2582,Lu3725, Ren3727}), 
where this discovery was confirmed and substitution of lanthanum by a number of
other rare -- earths, according to a simple chemical formula
(RE)$^{+3}$O$^{-2}$Fe$^{+2}$As$^{-3}$, has lead to more than doubling of
$T_c$ up to the values of order of 55K in systems based upon
$NdOFeAs$ and $SmOFeAs$, with electron doping via addition of fluorine or
creating oxygen deficit, or hole doping achieved by partial substitution of 
the rare -- earth (e.g. La by Sr) \cite{Wen3021}. Note also Ref. \cite{Wang4290},
where the record values of $T_c\sim 55$K were achieved by partial substitution
of Gd in $GdOFeAs$ by Th, which, according to the authors, also corresponds
to electron doping. In these early works different measurements of electrical
and thermodynamic properties were performed on polycrystalline samples.

\begin{figure}[!h]
\includegraphics[clip=true,width=0.45\columnwidth]{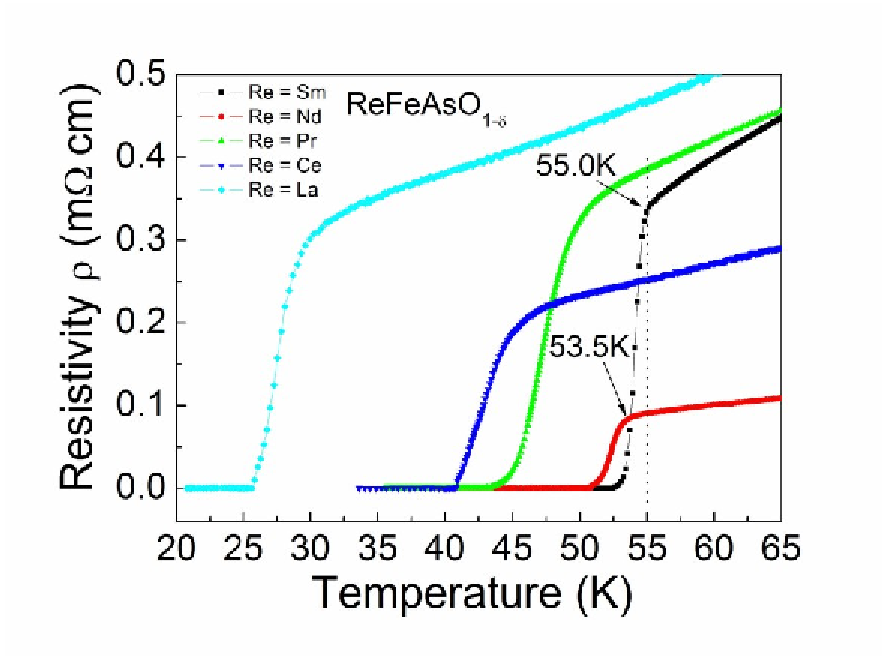}
\includegraphics[clip=true,width=0.5\columnwidth]{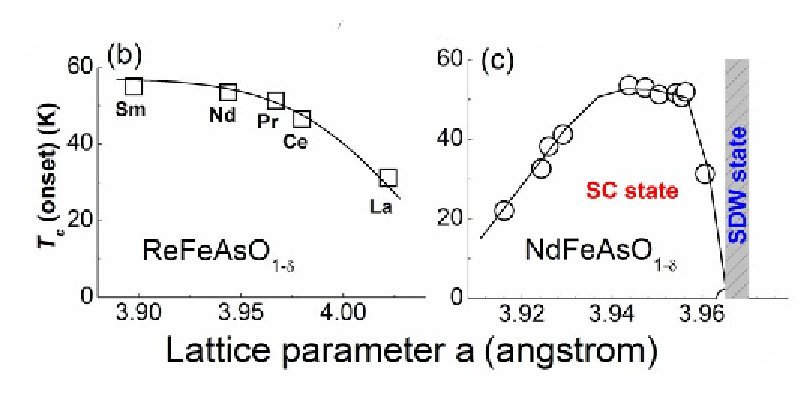}
\caption{(a) resistivity behavior in REOFeAs (RE=La,Ce,Pr,Nd, Sm) close to
superconducting transition, (b,c) $T_c$ dependence on the value of the
lattice constant \cite{Ren2582}.}  
\label{fig_1} 
\end{figure}

In Fig. \ref{fig_1} (a), taken from Ref. \cite{Ren2582}, we show typical
temperature dependences of electric resistivity in different $REOFeAs$ compounds. 
It can be seen that for most of these compounds $T_c$ is within the interval 
40-50K, while $LaOFeAs$ system drops out with its significantly lower
transition temperature $\sim$ 25K. In this respect we can mention Ref.
\cite{Lu3725} in which a synthesis of this system under high pressure was
reported, producing samples with $T_c$ (onset of superconducting transition) 
of the order of 41K. In Ref. \cite{NatPrs} analogous increase of $T_c$ 
in this system was achieved under external pressure $\sim$ 4GPa. 
Probably this last result is characteristic only for La system, as further
increase of pressure leads to the drop of $T_c$, while in other systems
(e.g. Ce based) $T_c$ lowers with the increase of external pressure from the
very beginning (see. e.g. Ref. \cite{Zoc4372,Gar1131}).

In many papers the growth of $T_c$, induced by La substitution by other
rare -- earths (with smaller ion radius) is often attributed to
``chemical'' pressure, which is illustrated e.g. by qualitative dependence
of $T_c$ on lattice spacing, shown in Fig. \ref{fig_1} (b) \cite{Ren2582}.
At the same time, Fig. \ref{fig_1} (c) taken from the same work shows that
lattice compression leads to the growth of $T_c$ only up to a certain limit,
and after that $T_c$ drops (compare it with results of high pressure 
experiments mentioned above).

\begin{figure}[!h]
\includegraphics[clip=true,width=0.45\columnwidth]{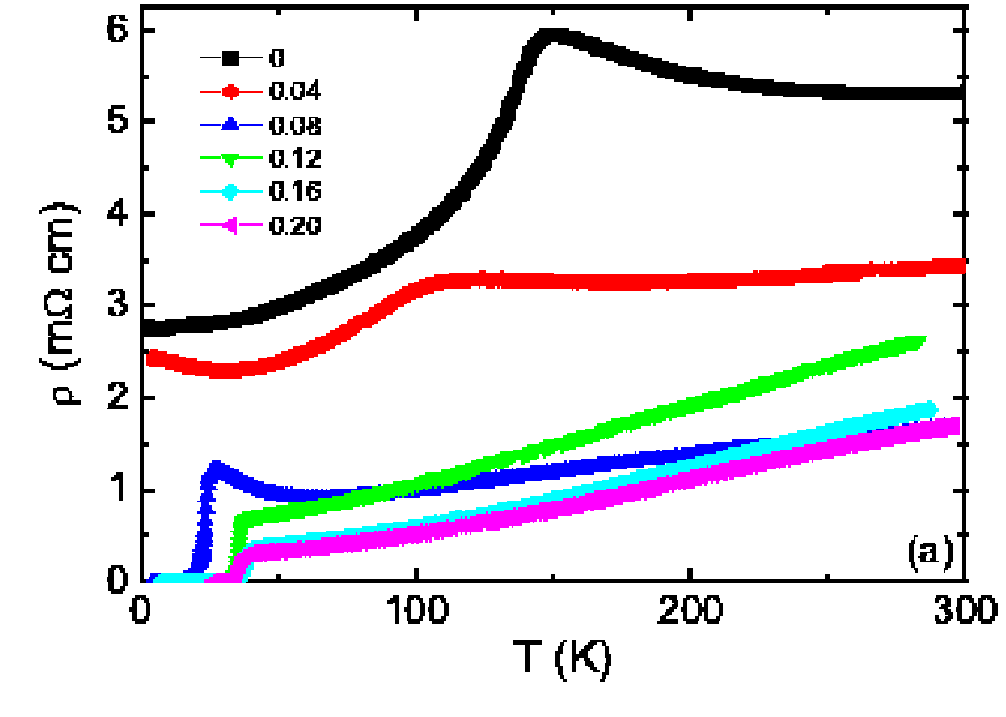}
\includegraphics[clip=true,width=0.45\columnwidth]{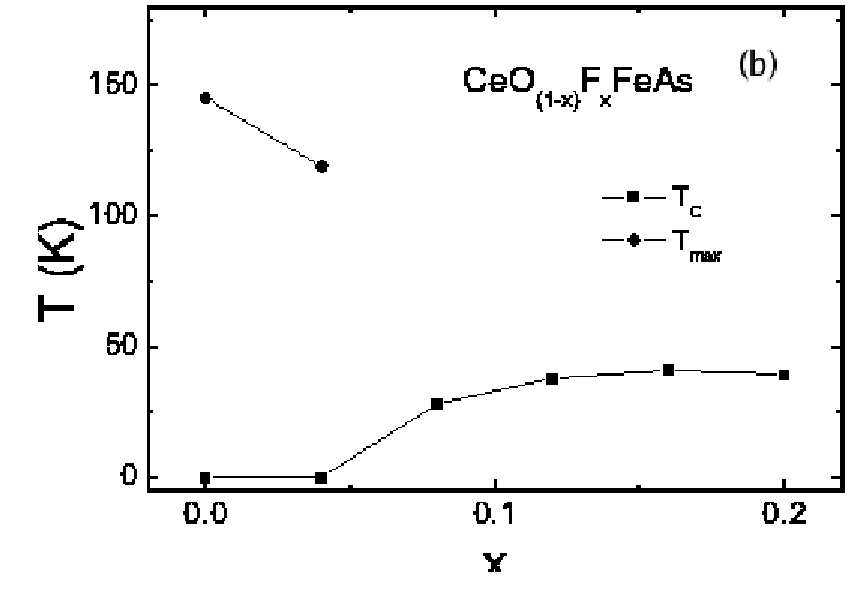}
\caption{(a) temperature dependence of resistivity in $CeO_{1-x}F_{x}FeAs$ for
different compositions $x$, shown on the graph, (b) phase diagram showing
superconducting region and concentration dependence of high -- temperature
anomaly $T_{max}$ of resistivity, associated with SDW transition \cite{Chen3790}.}  
\label{fig_2} 
\end{figure}

In more wide temperature interval typical temperature behavior of resistivity
is illustrated by the data for $CeO_{1-x}F_xFeAs$ shown in Fig. \ref{fig_2}, 
taken from Ref. \cite{Chen3790}. It can be seen that the prototype system
$CeOFeAs$ is characterized by {\em metallic} behavior of resistivity up to
the lowest temperatures achieved, with characteristic anomaly in the vicinity of 
$T\sim 145$K and sharp drop of resistivity at lower temperatures. Metallic
nature of prototype $REOFeAs$ systems contrasts with insulating nature of
stoichiometric cuprates. After doping, e.g. by fluorine, the value of
resistivity drops, its anomaly becomes less visible and disappears at higher
dopings, where superconductivity appears. The highest value of $T_c=41$K 
is achieved for $x=0.16$, and the resulting F concentration phase diagram is
shown in Fig. \ref{fig_2}. High temperature anomaly of resistivity in
prototype and slightly doped system in most of the works is attributed to
structural phase transition and (or) accompanying spin density wave (SDW)
transition, while its degradation under doping is usually connected with the
breaking of ``nesting'' of the Fermi surfaces (cf. below). This behavior of
resistivity is rather typical and is observed in all $REOFeAs$ systems
(cf. e.g. Ref. \cite{Chen4384}), which are for shortness called now 
1111 -- systems.

In fact, to the same class belongs the recently synthesized compound
Sr(Ca)FFeAs~\cite{Tegel,Han}. Here the typical SDW anomaly of resistivity is
observed in the vicinity of 175 K. Further doping of this system with Co has
lead to the appearance of superconductivity with T$_c \sim$5~K~\cite{Matsuishi}, 
while in Sr$_{1-x}$La$_x$FFeAs superconducting transition with 
T$_c$=36~K~\cite{Zhu} was obtained. 
   
\subsubsection{$AFe_2As_2$ ($A=Ba,Sr,...$) system}

\begin{figure}
\includegraphics[clip=true,width=0.45\columnwidth]{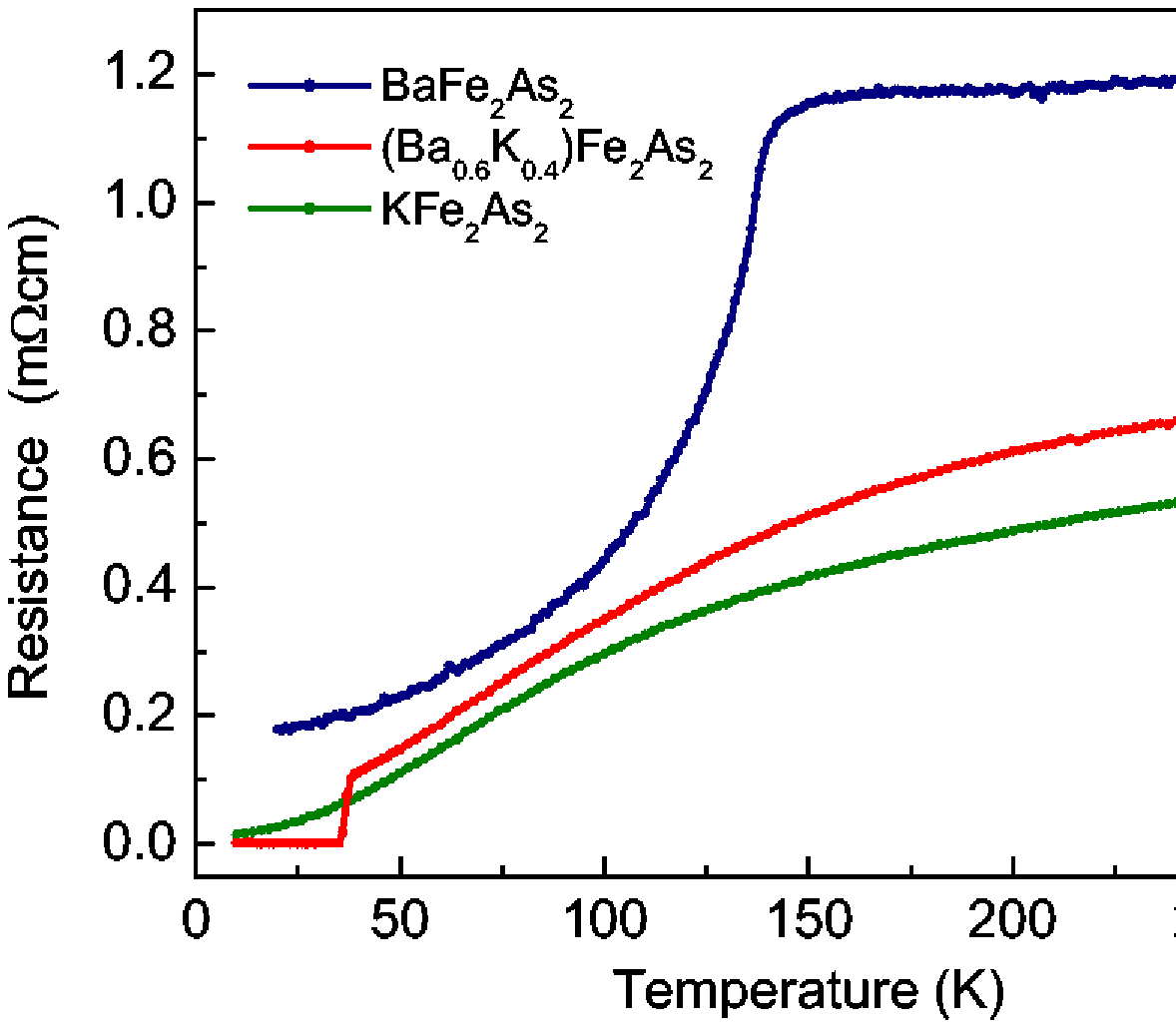}
\caption{Temperature dependence of resistivity in  
$Ba_{1-x}K_{x}Fe_2As_2$ for different compositions $x$, shown on the graph 
\cite{rott}.}  
\label{bafeas} 
\end{figure}

More simple (structurally and chemically) class of iron based superconductors
was discovered in Ref.  \cite{rott}, by the synthesis of
$Ba_{1-x}K_{x}Fe_2As_2$ compound, superconductivity with $T_c=38$K was observed 
for $x=0.4$. The relevant data on the temperature dependence of resistivity are
shown in Fig. \ref{bafeas} \cite{rott}.

In the prototype compound $BaFe_2As_2$ temperature dependence of resistivity
demonstrates typically metallic behavior with characteristic anomaly in the
vicinity of $T\sim 140$K, which is connected to a spin density wave (SDW) and 
structural phase transition (cf. below). According to a simple chemical formula
$Ba^{+2}(Fe^{+2})_2(As^{-3})_2$, the partial substitution of $Ba^{+2}$ by 
$K^{+1}$ leads to hole doping, which suppresses SDW transition and leads to
superconductivity in some concentration interval. Obviously, these results
are quite similar to quoted above for $REOFeAs$ systems (1111).

In Ref. \cite{Chen1209} a similar behavior was obtained in
$Sr_{1-x}K_{x}Fe_2As_2$ with maximal $T_c\sim 38$K, while in Ref. \cite{Chu1301}
the systems like $AFe_2As_2$ with $A=K,Ca,K/Sr,Ca/Sr$ were studied, and electron
doping by $Sr$ of compounds with $A=K$ ¨ $A=Cs$ produced values of
$T_c\sim 37$K followed by SDW transition. This new class of Fe based HTSC
is sometimes denoted for shortness as 122 systems.

Note also an interesting paper \cite{Alireza1896}, where superconductivity
with $T_c$ up to 29K was obtained in prototype (undoped) compounds
$BaFe_2As_2$ and $SrFe_2As_2$ under external pressure, but only in a limited
interval of pressures. Superconductivity also appears under electron doping
of these systems by Co \cite{Sefat2237}.

\subsubsection{$AFeAs$ ($A=Li,...$) system}

One more type of Fe based superconductors is represented by
$Li_{1-x}FeAs$ (111) system, where superconductivity appears at 
$T_c\sim 18$K \cite{Wang4688,Tapp2274}. Up to now there are few works on
this system, however, it is clear that it is quite similar to
1111 and 122 systems and is rather promising in a sense of comparison of its
properties with those of other systems, discussed above.

\subsubsection{$FeSe(Te)$ system}

Finally, rather unexpectedly, superconductivity was discovered in 
very ``simple'' $\alpha-FeSe_x$ ($x<1$) system \cite{Hsu2369} with $T_c=8$K,
reaching 27K under pressure of 1.48GPa \cite{Miz4315}. Next the system
$Fe(Se_{1-x}Te_{x})_{0.82}$ was also studied, and the maximum value of
$T_c=14$K was achieved for $0.3<x<1$ \cite{Fang4775}. 
Here there are still very few detailed
studies, though it is quite clear that electronic structure these systems are
again similar to those discussed above, so that in the nearest future we can
expect a lot of papers devoted to the comprehensive studies of their electronic
properties.

\subsubsection{Search for new systems}

Naturally, at present an active search of other similar systems is underway with
the hope to obtain even higher values of $T_c$. Up to now, outside the domain
of FeAs based systems, successes are rather modest. We shall mention only few
works, were new superconductors were synthesized.

We have already mentioned compounds like $LaOFeP$ with $T_c\sim 5$K 
\cite{Hos06} and $LaONiP$ with $T_c\sim 3$K \cite{Hos07}, which are 
representatives of the same class of superconducting oxypnictides (1111), 
as the whole series of $REOFeAs$. Superconductivity with $T_c\sim 4$K 
was discovered in $LaO_{1-\delta}NiBi$ compound \cite{Kozh}, while in
$GdONiBi$ and $Gd_{0.9}Sr_{0.1}ONiBi$ the value of $T_c\sim 0.7$K 
was obtained in Ref. \cite{Ge5045}. Transition temperature $T_c\sim 2.75$K 
was obtained in $LaO_{1-x}F_xNiAs$ \cite{Li2572}.

In the class of 122 systems the value of $T_c\sim 0.7$K was obtained in
$BaNi_2As_2$ \cite{Ronn3788}.

An interesting new system $La_3Ni_4P_4O_2$ with different crystal structure 
demonstrated superconductivity at $T_c\sim 2.2$K \cite{Klim1557}. In this
reference a number of promising systems of the same type were also discussed.

Surely, all these results are not very exciting. However, there is rather large 
number of systems somehow similar to already known FeAs compounds and where we 
can also expect. A wide set of promising compounds was discussed in Ref.
\cite{Oz1158}.  

\subsection{Crystal structure and anisotropy}

\begin{figure}[!h]
\includegraphics[clip=true,width=0.35\columnwidth]{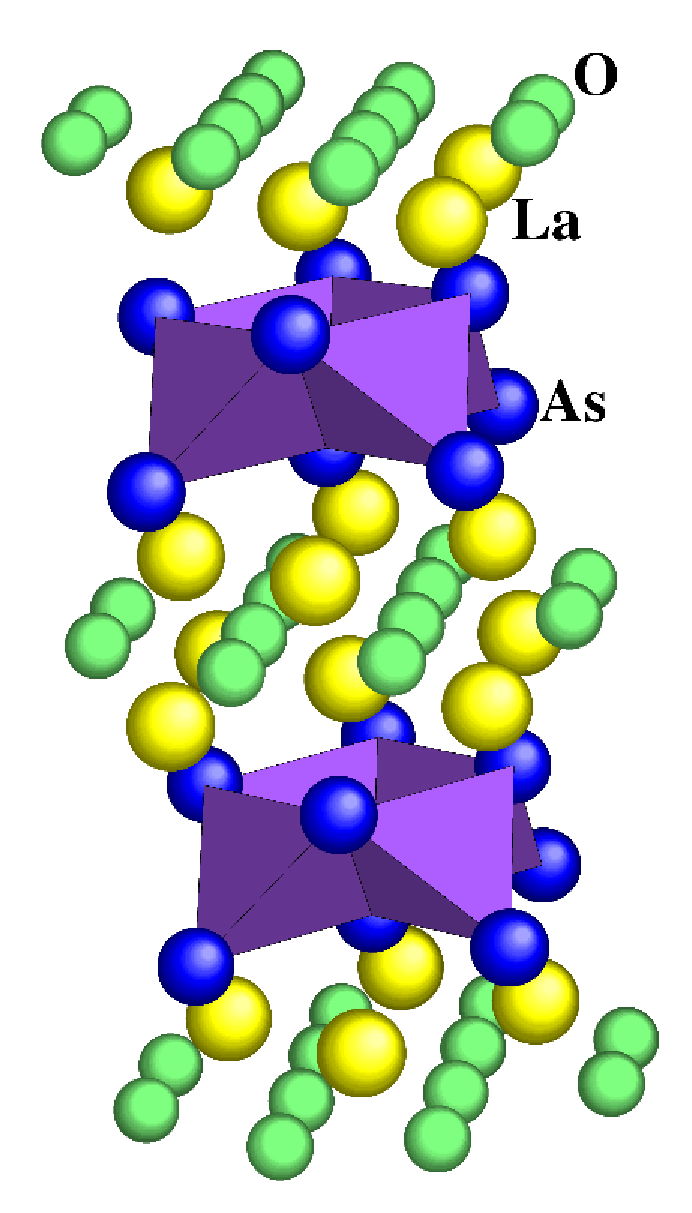}
\includegraphics[clip=true,width=0.25\columnwidth]{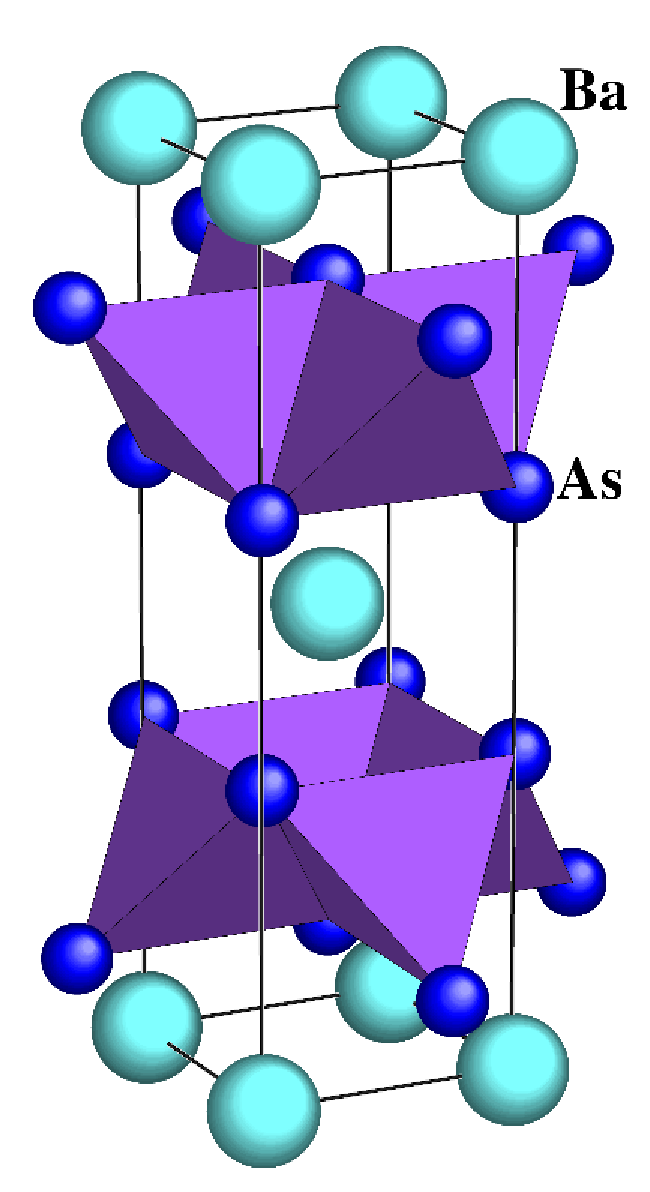}
\caption{Crystal structure of LaOFeAs (left) and BaFe$_2$As$_2$ 
(right).  FeAs tetrahedra form two - dimensional layers, surrounded by the 
layers of LaO or Ba. Fe ions inside tetrahedra form square lattice.} 
\label{ba_la_struc}
\end{figure}

Crystal structure of BaFe$_2$As$_2$ and LaOFeAs compounds and their analogs
corresponds to tetragonal symmetry and space groups $I$4/$mmm$ and $P$4/$nmm$. 
Both compounds are formed by the layers of (FeAs)$^{-}$ with covalent bonding,
interlaced by the layers of Ba$_{0.5}^{2+}$ or (LaO)$^{+}$, while interlayer
bonding is ionic. Ions of Fe$^{2+}$ are surrounded by four ions of As,
which form tetrahedra. The general view of crystal structures of
LaOFeAs and BaFe$_2$As$_2$ is shown in Fig. \ref{ba_la_struc}. Layered
(quasi two -- dimensional) nature of these compounds is similar to that of HTSC
cuprates. At 140K BaFe$_2$As$_2$ undergoes the structural phase transition from
tetragonal ($I$4/$mmm$) to orthorhombic ($Fmmm$) structure~\cite{rotter_4021}. 
Analogous transition takes place also in LaOFeAs at 150 K:  $P$4/$nmm$ 
(tetragonal) $\rightarrow$ $Cmma$ (orthorhombic)~\cite{Nomura_3569}.  
Experimental atomic positions in
BaFe$_2$As$_2$ are:  Ba (0, 0, 0), Fe (0.5, 0, 0.25), As (0, 0, $z$). For
LaOFeAs these are: La(0.25, 0.25, $z$), Fe (0.75, 0.25, 0.5), As (0.25, 0.25, $z$), 
O (0.75, 0.25, 0). The rest of crystallographic data for both compounds are
given in Table III. It is seen that in BaFe$_2$As$_2$ the Fe-As distance is
smaller than in LaOFeAs. Thus in BaFe$_2$As$_2$ we can expect stronger 
Fe-$d$-As-$p$ hybridization in comparison with LaOFeAs and, correspondingly,
the larger $d$-band width for Fe. Similarly, the distance between adjacent
atoms of Fe within the layers of FeAs in BaFe$_2$As$_2$ is also significantly
smaller than in LaOFeAs (and related compounds). After the phase transition of
BaFe$_2$As$_2$ into orthorhombic structure the four (initially equal)
Fe-Fe distances are separated into two pairs of bonds with the widths of
2.808 \AA~and 2.877 \AA. Moreover, two As-Fe-As angles are significantly
different in LaOFeAs system (113.6$^\circ$ and 107.5$^\circ$) and very
close ($\sim 109^\circ$) in BaFe$_2$As$_2$. Such differences in the nearest
neighborhood of Fe ions should lead to appropriate changes in their electronic
structure.


\begin{center}
Table III.   
Structural data for BaFe$_2$As$_2$ and LaOFeAs. 
\vskip 0.5cm
\begin{tabular}{|l|c|c|}                                             
\hline 
Parameters &BaFe$_2$As$_2$ & LaOFeAs \\ \hline Group     &I4/mmm          & 
P4/nmm  \\ $a$, \AA  & 3.9090(1)    &4.03533(4) \\ $c$, \AA  &13.2122(4)    
&8.74090(9) \\ $z_{La}$  &  -           &0.14154(5) \\ $z_{As}$  &0.3538(1)       
&0.6512(2)  \\ Reference    & \cite{rott}& \cite{Hos08}\\ \hline 
Ba-As, \AA&3.372(1)$\times$8&  -        \\ La-As, \AA&-         
&3.380$\times$4 \\ Fe-As, \AA&2.388(1)$\times$4&2.412$\times$4     \\ Fe-Fe, 
\AA&2.764(1)$\times$4&2.853$\times$4     \\ As-Fe-As  &109.9(1)$^\circ$ 
&113.6$^\circ$ \\ &109.3(1)$^\circ$ &107.5$^\circ$ \\ 
\hline 
\end{tabular} 
\end{center}

Doping of prototype REOFeAs compounds by fluorine (or by creation of oxygen 
deficit) or BaFe$_2$As$_2$ by substitution of Ba(Sr) by K etc., leads to
the suppression of transition from tetragonal to orthorhombic
structure and appearance of superconductivity in tetragonal phase. 

\begin{figure}[!h]
\includegraphics[clip=true,width=0.30\columnwidth]{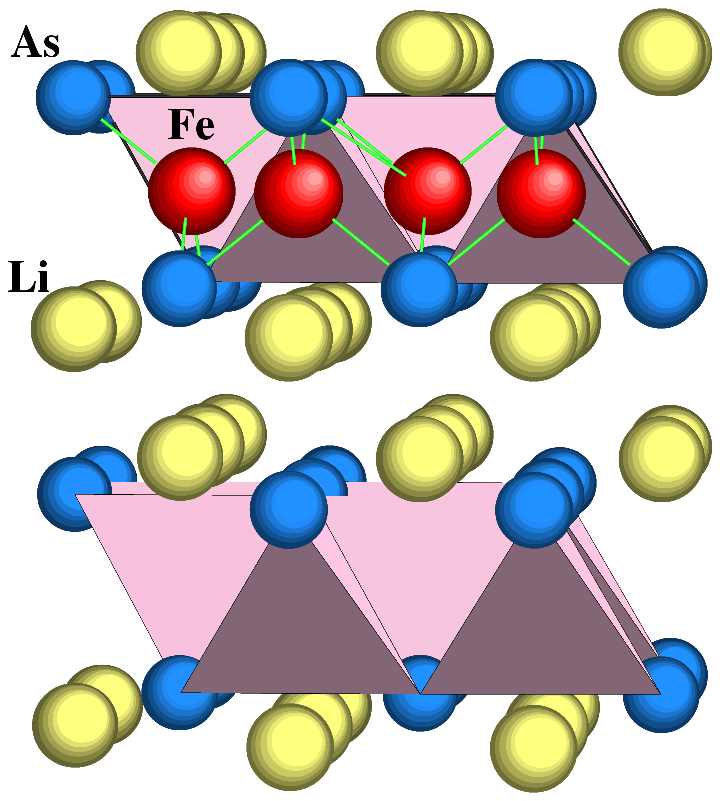}
\includegraphics[clip=true,width=0.45\columnwidth]{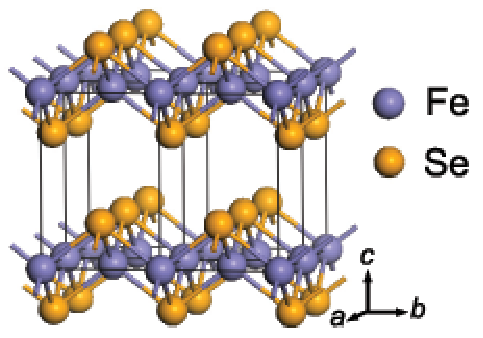}
\caption{Crystal structure of LiFeAs (left). FeAs tetrahedra form two --
dimensional layers separated by layers of Li ions. At right -- crystal
structure of $\alpha$-FeSe.} 
\label{lifeas_struc} 
\end{figure}

Recently, the crystal structure of LiFeAs compound was also refined \cite{Tapp2274}. 
LiFeAs forms tetragonal structure with space group $P$4/$nmm$ and lattice
parameters $a=3.7914(7)$~\AA, $c=6.364(2)$~\AA. Experimentally determined
atomic positions are: Fe(2b) (0.75, 0.25, 0.5), Li(2c)  (0.25, 0.25, z$_{Li}$),
As(2c) (0.25, 0.25, z$_{As}$), z$_{As}$=0.26351,
z$_{Li}$=0.845915~\cite{Tapp2274}. Crystal structure of LiFeAs is shown in
Fig. \ref{lifeas_struc} (a) and is again characterized by its layered nature,
which suggests quasi two -- dimensional electronic properties and is clearly
analogous to the structure of LaOFeAs~\cite{Hos08} and
BaFe$_2$As$_2$~\cite{rotter_4021}. Most important Fe-Fe and Fe-As distances are
2.68 and 2.42 \AA~correspondingly. At present this structure is most spatially
compact among similar compounds. As-Fe-As angles in LiFeAs has the values of
$\sim 103.1^\circ$ and $\sim 112.7^\circ$, which also can lead to some fine
differences of electronic structure in comparison with LaOFeAs and 
BaFe$_2$As$_2$.

Crystal structure of $\alpha$-FeSe is especially simple -- it forms the layered
tetragonal phase (like PbO) with square sublattice of Fe and space group
$P$4/$nmm$. The crystal of $\alpha$-FeSe consists of layers of FeSe$_4$ 
edge-sharing tetrahedra as shown in Fig. \ref{lifeas_struc} (b). Experimentally
determined \cite{Hsu2369} lattice constants are:
 $a=3.7693(1)$~\AA, $c=5.4861(2)$~\AA\ for FeSe$_{0.82}$ and 
$a=3.7676(2)$~\AA, $c=5.4847(1)$~\AA\  for FeSe$_{0.88}$. 
Analogy with oxypnictides like REOFeAs and systems like BaFe${_2}$As${_2}$ 
is obvious. At $T\sim 105$K this system undergoes the phase transition
from tetragonal to triclinic structure (group P-1) \cite{Hsu2369}.

\begin{figure}[!h]
\includegraphics[clip=true,width=0.45\columnwidth]{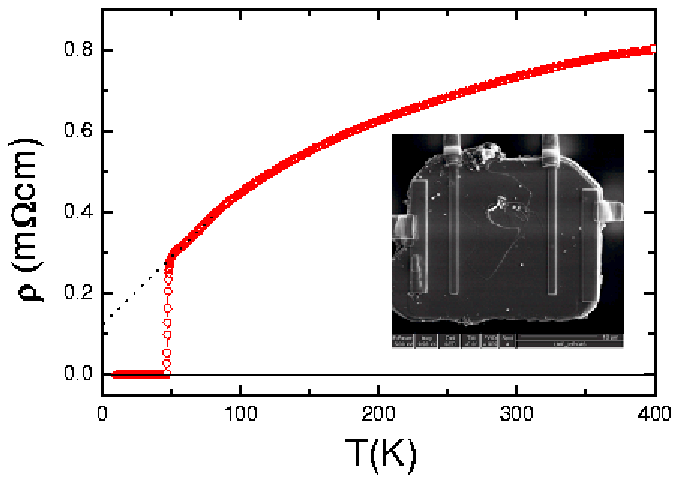}
\includegraphics[clip=true,width=0.45\columnwidth]{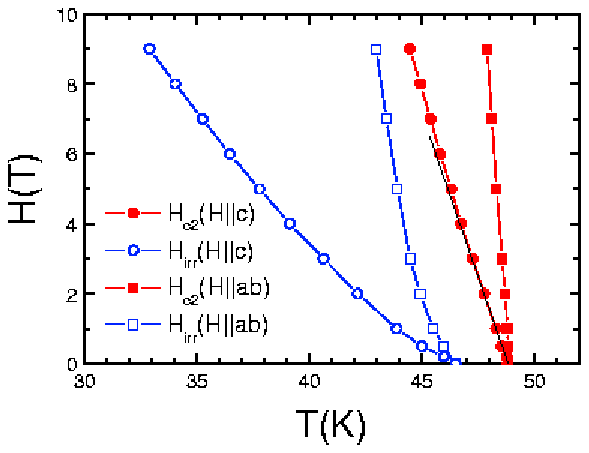}
\includegraphics[clip=true,width=0.35\columnwidth]{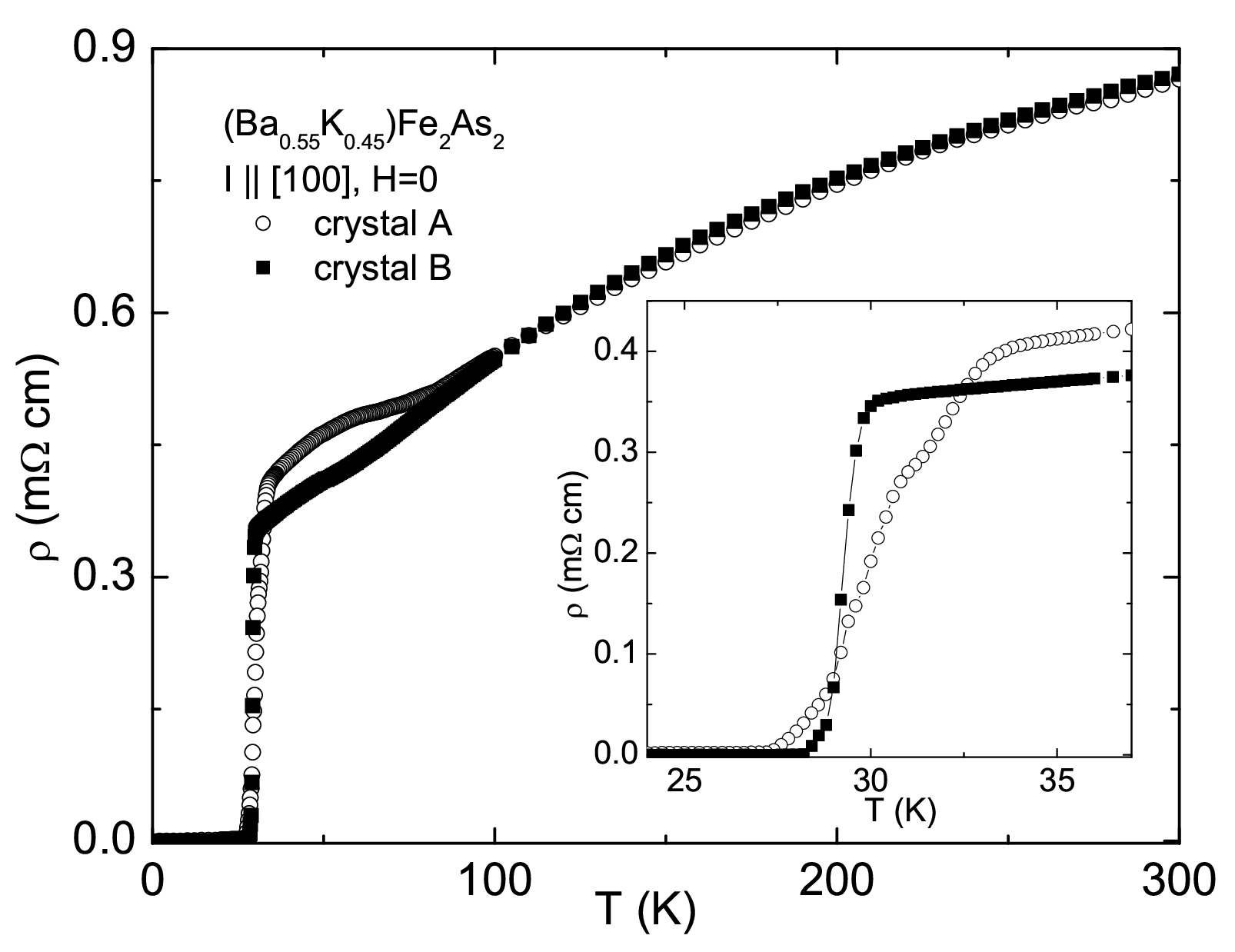}
\includegraphics[clip=true,width=0.15\columnwidth]{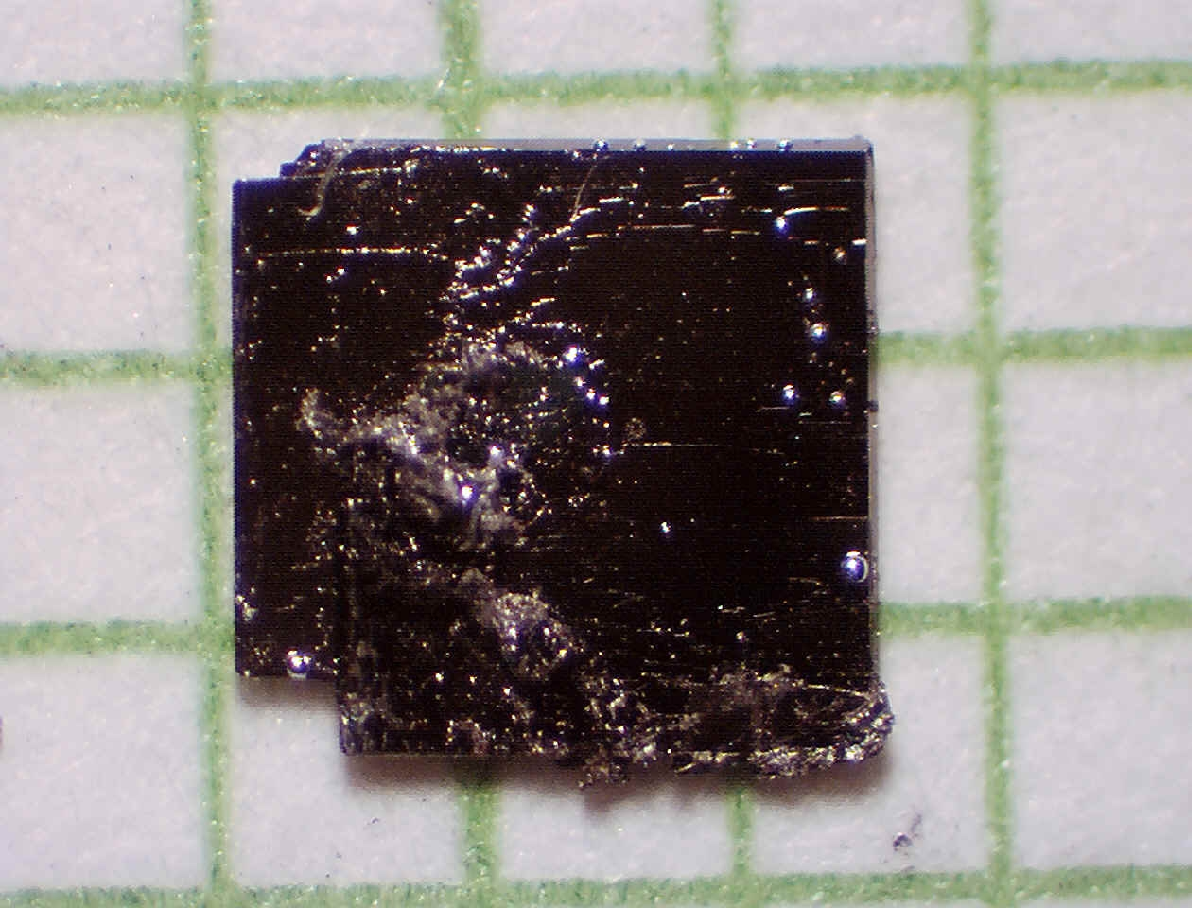}
\includegraphics[clip=true,width=0.35\columnwidth]{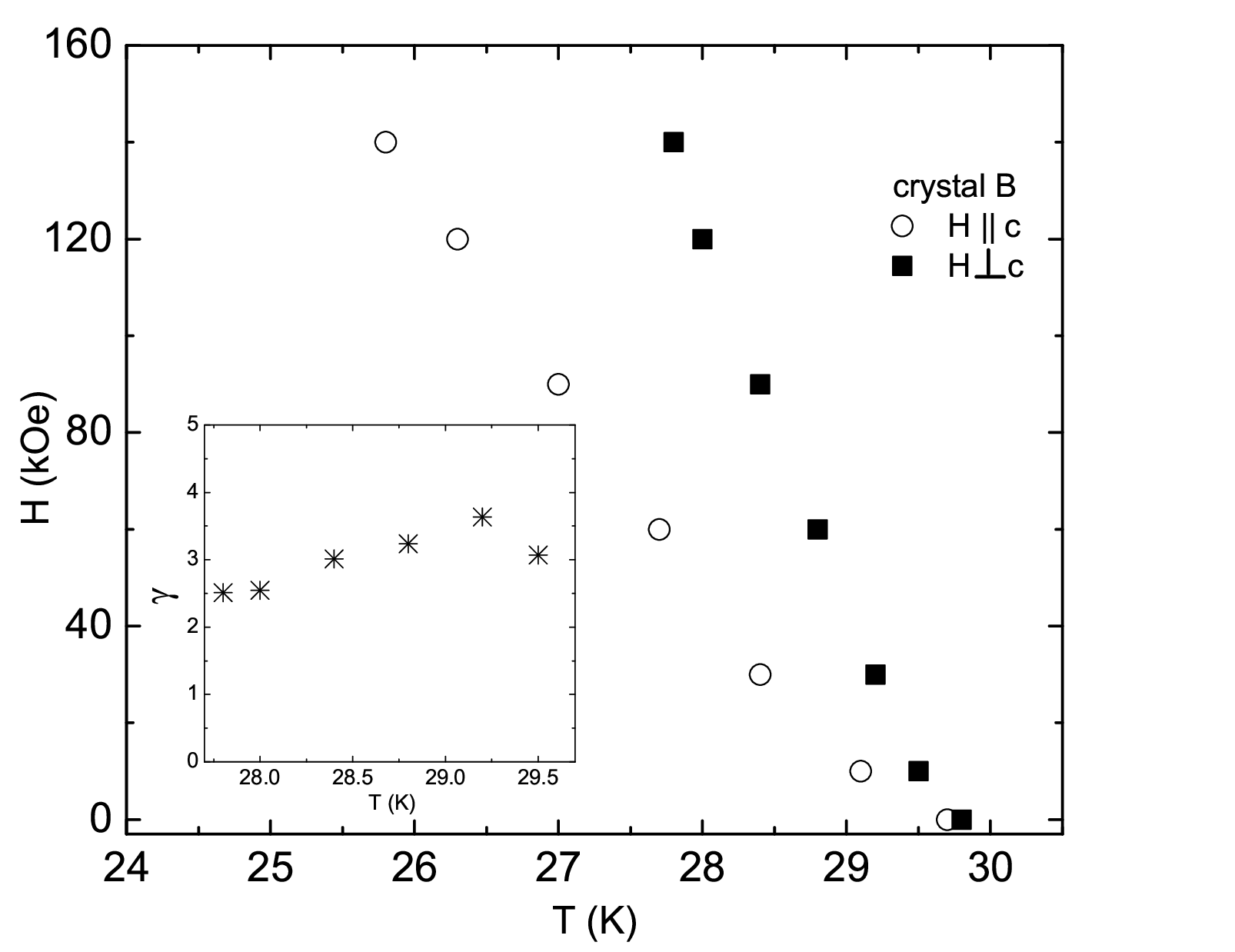}
\caption{(a) superconducting transition and anisotropic behavior of the upper
critical field $H_{c2}$ (and irreversibility field $H_{irr}$ ) (b) in the
single crystal of $NdO_{0.82}F_{0.18}FeAs$ \cite{Jia0532}, (c) superconducting
transition and anisotropic behavior of the upper critical field
$H_{c2}$ (d) in the single crystals of $Ba_{1-x}K_{x}FeAs$ \cite{Ni1874}, 
where at the insert we also show small temperature dependence of anisotropy of
$H_{c2}$. In the photo -- typical single crystal of $Ba_{1-x}K_{x}Fe_2As_2$.} 
\label{cryst} 
\end{figure}

Up to now only very small single crystal of 1111 compounds were successfully
synthesized ($\sim 100\times 100\ \mu m^2$) (cf. e.g. Refs.
\cite{Zhig0337,Jia0532}). Situation with 122 systems is much better, here
almost immediately the crystals of millimeter sizes were obtained
(see Fig. \ref{cryst}) \cite{Ni1874}. Thus, most of the measurements in the
following were made on single crystals of this system (though for 1111 system
a number of interesting studies on single crystals were also performed).
Recently rather small single crystals ($\sim 200\times 200\ \mu m^2$) of
$\alpha-FeSe$ with $T_c\sim 10$K \cite{Zhang1905} were also obtained, but
detailed physical measurements on this system are still to be done.

Typical results of measurements on single crystals of 1111 \cite{Jia0532} and
122 \cite{Ni1874} systems are shown in Fig. \ref{cryst}. In particular, the
anisotropy of the upper critical field $H_{c2}$ corresponds quasi
two -- dimensional nature of electronic subsystem of these superconductors,
which is already evident from their crystal structure. At the same time, we
can see that this anisotropy of critical fields is not too large.

\begin{figure}[!h]
\includegraphics[clip=true,width=0.5\columnwidth]{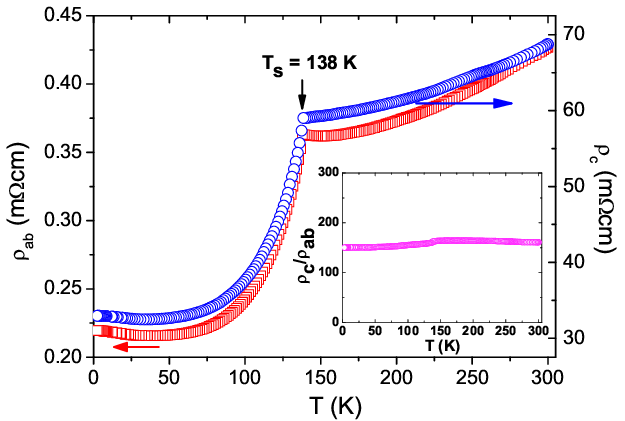}
\caption{Temperature dependence of resistivity $\rho_{ab}$ in the $ab$ plane
and transverse resistivity $\rho_{c}$ in orthogonal direction in the single
crystal of $BaFe_2As_2$ \cite{Wang2452}. At the insert -- temperature
dependence of resistivity anisotropy.} 
\label{anis_res} 
\end{figure}

In Fig. \ref{anis_res} from Ref. \cite{Wang2452} we show temperature 
dependence of resistivity $\rho_{ab}$ in $ab$ plane and of transverse
resistivity $\rho_{c}$ in orthogonal direction for the single crystal of
prototype (undoped) $BaFe_2As_2$ system \cite{Wang2452}.  It can be seen that
anisotropy of resistivity exceeds $10^2$, which confirms the quasi
two -- dimensional nature of electronic properties of this system. This
anisotropy is significantly larger than the value which can be expected from
the simple estimates\footnote{Anisotropy of $H_{c2}$ is usually of the order
of square root of the anisotropy of resistivity.}, based on the anisotropy of 
$H_{c2}$. However, we must stress that data on the anisotropy of resistivity
of superconducting samples in Refs. \cite{Ni1874,Wang2452} is just absent.
In the data shown in Fig. \ref{anis_res} we can also clearly see an anomaly
in temperature dependence of resistivity at $T_s=138$K, connected with
antiferromagnetic (SDW) transition.

The question of anisotropy of electronic properties has sharpened after in
Ref. \cite{Yuan3137} the measurements of $H_{c2}$ in single crystals of
$Ba_{1-x}K_{x}Fe_2As_2$ were performed in much wider temperature interval
than in \cite{Ni1874}, up to the field values of the order of $\sim$ 60T. 
According to Ref. \cite{Yuan3137} anisotropy of $H_{c2}$ is observed only in
relatively narrow temperature interval close to $T_c$, changing to almost
isotropic behavior as temperature lowers.

\subsection{Magnetic structure and phase diagram}

As we have already mentioned, as temperature lowers prototype (undoped) 1111 and 
122 compounds undergo structural transition accompanied by simultaneous or
later antiferromagnetic transition (probably of SDW type). Direct confirmation
of this picture was obtained in neutron scattering experiments. First results
for $LaOFeAs$ system were given in Ref. \cite{Cruz0795}. It was discovered that
the structural transition in this compound (according to Ref. \cite{Cruz0795} 
from orthorhombic $P4/nmm$ into monoclinic $P112/n$ structure, which differs
from the results of other authors \cite{Nomura_3569}) takes place at $T\sim 150$K 
(where anomaly in temperature dependence of resistivity is observed), and
afterwards at $T\sim 134$K antiferromagnetic ordering appears.

\begin{figure}[!h]
\includegraphics[clip=true,width=0.5\columnwidth]{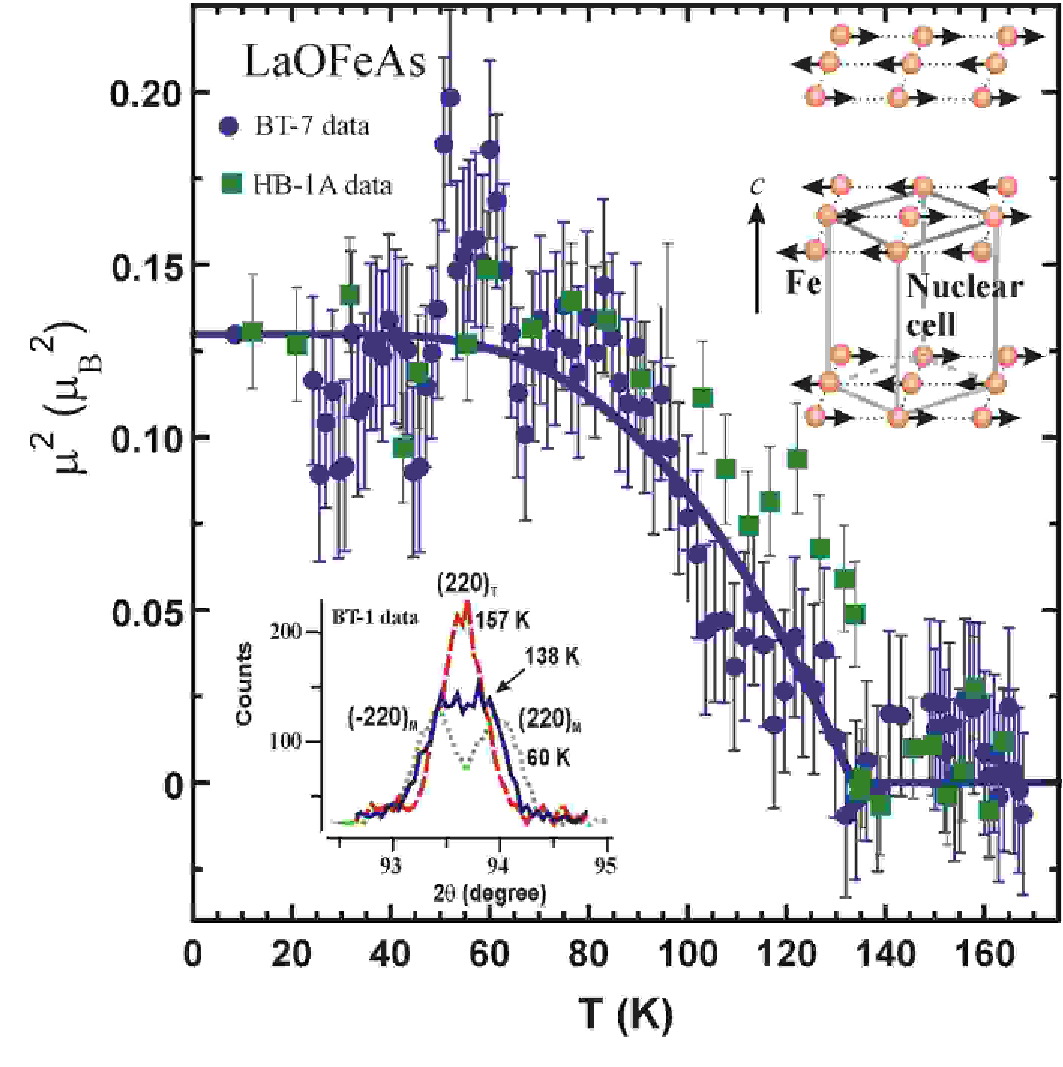}
\includegraphics[clip=true,width=0.45\columnwidth]{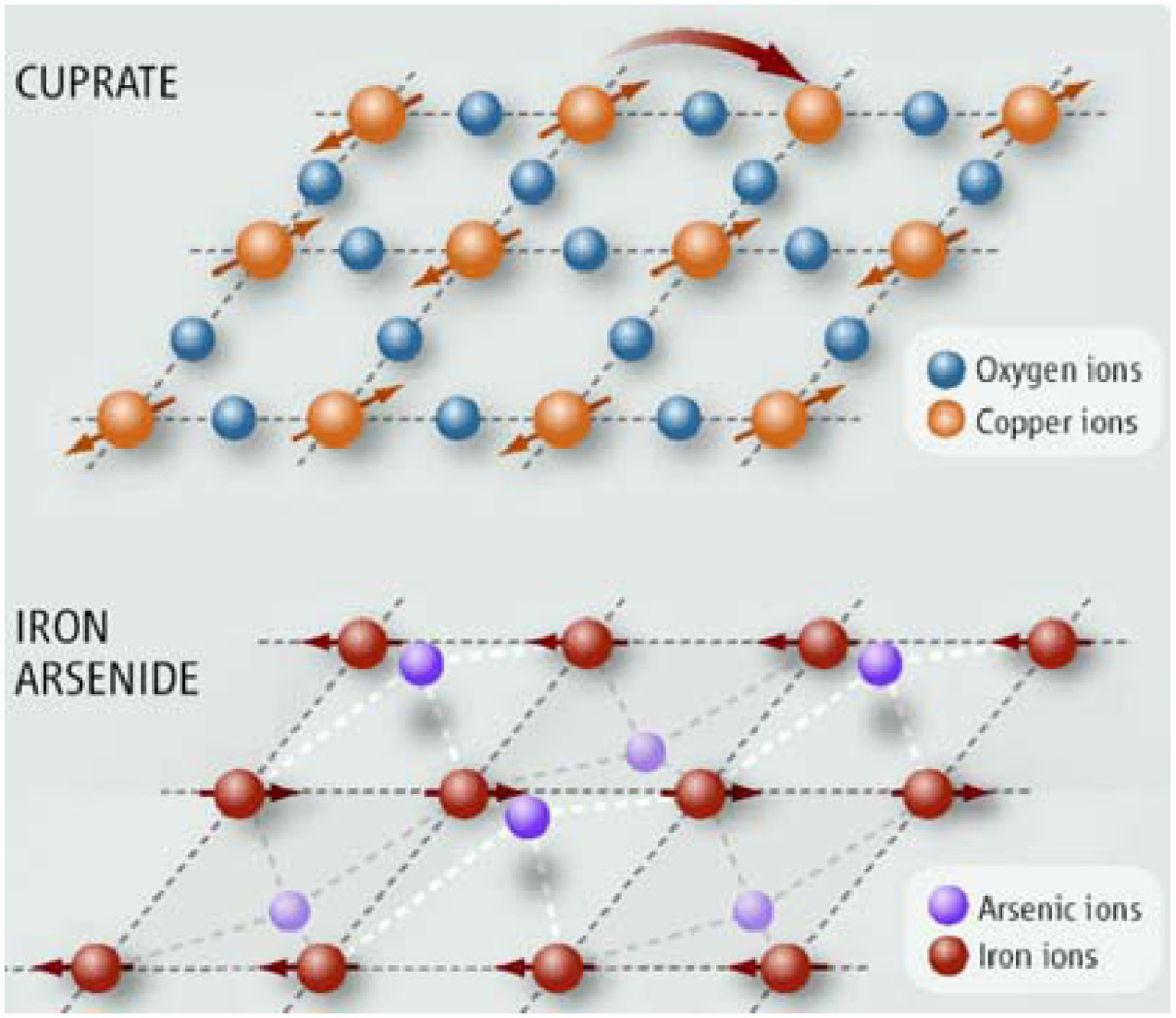}
\caption{(a) temperature dependence of the square of magnetic moment on
Fe in LaOFeAs obtained by neutron scattering (data from two spectrometers denoted
as BT-7 and HB-1A) \cite{Cruz0795}. At the insert in the upper right corner --
the experimentally determined antiferromagnetic structure in
$\sqrt{2}a\times\sqrt{2}b\times 2c$ lattice cell. Distortion of the nuclear
scattering peak shown at the insert in the lower left corner shows that
structural transition precedes magnetic transition. (b) comparison of
antiferromagnetic ordering in $CuO_2$ plane of cuprates and in
$FeAs$ plane of new superconductors.} 
\label{neutr_sctr} 
\end{figure}

In Fig. \ref{neutr_sctr} (a) we show antiferromagnetic structure obtained in
these experiments, as well as temperature dependence of the square of
magnetic moment at Fe site. The value of this moment is not larger than 
0.36(5) $\mu_B$. We see that spin ordering in the $ab$ plane takes the form
of characteristic chains of ferromagnetically oriented spins with opposite
spin orientations in neighboring chains (stripes). Along $c$ axis there is 
typical period doubling.

In Fig. \ref{neutr_sctr} (b) we compare the magnetic structures in $CuO_2$
plane of cuprates and in $FeAs$ plane of new FeAs based superconductors. 
Both analogy and significant difference can be clearly seen. Both structural
and antiferromagnetic transitions in $FeAs$ planes are suppressed by doping,
similarly to situation in cuprates. At the same time it should be stressed
that antiferromagnetic phase of cuprates is an insulator, while in
$FeAs$ superconductors it remains metallic, as is clearly seen from the
data on resistivity quoted above.

\begin{figure}[!h]
\includegraphics[clip=true,width=0.45\columnwidth]{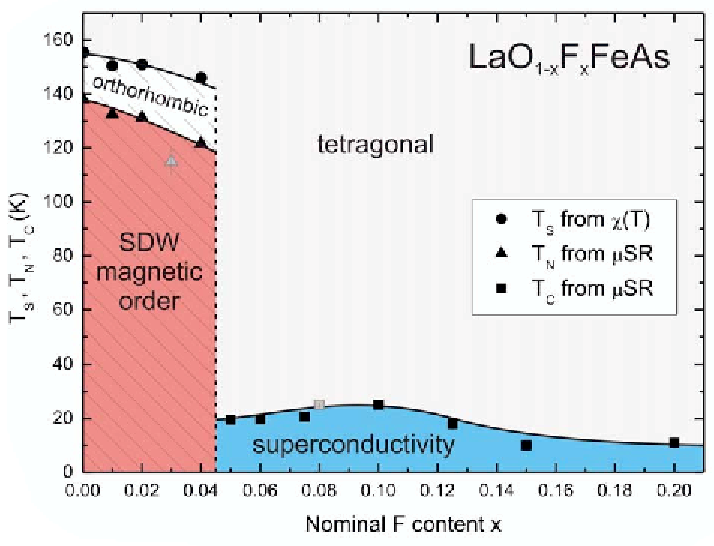}
\includegraphics[clip=true,width=0.5\columnwidth]{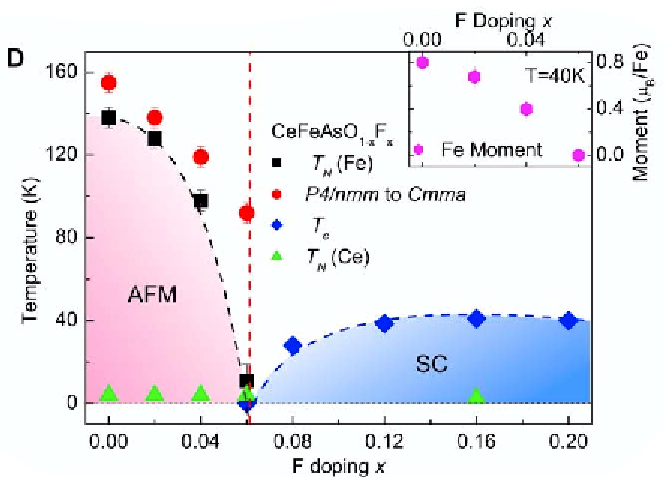}
\caption{(a) phase diagram of $LaO_{1-x}F_xFeAs$ obtained from $\mu SR$ 
experiments \cite{Luet3533}. It shows concentration dependence of critical
temperatures of superconducting ($T_c$), magnetic ($T_N$) and structural
transitions ($T_s$) (determined from resistivity measurements), (b) 
phase diagram of $CeO_{1-x}F_xFeAs$ system obtained from neutron scattering data
in Ref. \cite{Zhao2528}. At the insert -- concentration dependence of magnetic
moment of Fe.} 
\label{ph_d} 
\end{figure}

Phase diagram of new superconductors (with changing concentration of doping
element) is also significantly different from that of cuprates. In Fig.
\ref{ph_d} we show this phase diagram for $LaO_{1-x}F_xFeAs$ system, obtained
in Ref. \cite{Luet3533} from $\mu SR$ experiments. It can be seen that
temperatures of structural and magnetic transitions are clearly separated,
while superconductivity region does not overlap with antiferromagnetic region.

\begin{figure}[!h]
\includegraphics[clip=true,width=0.35\columnwidth]{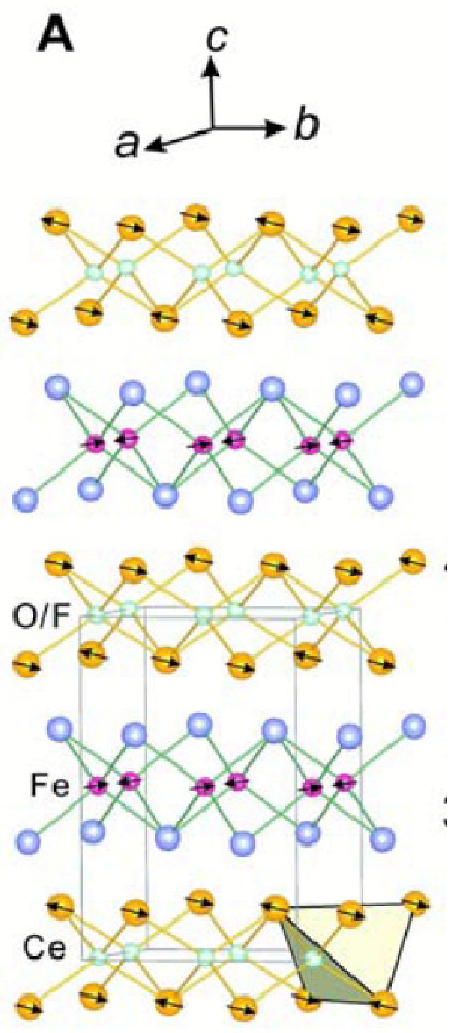}
\includegraphics[clip=true,width=0.6\columnwidth]{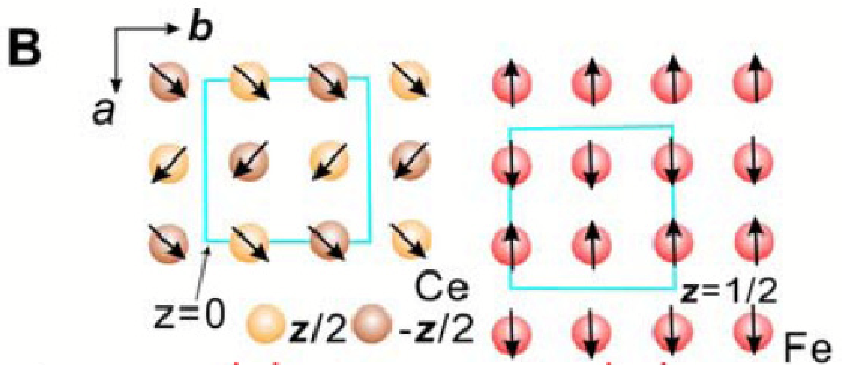}
\caption{Magnetic structure of CeOFeAs: (a) general picture of spin ordering at
low temperatures, (b) magnetic elementary cells of Fe and Ce.} 
\label{m_str} 
\end{figure}

Analogous phase diagram for $CeO_{1-x}F_xFeAs$, obtained in Ref. \cite{Zhao2528} 
from neutron scattering data is shown in Fig. \ref{ph_d} (b). According to this
work, structural tetra -- ortho transition ($P4/nmm\to Cmma$) is also well 
separated from antiferromagnetic transition, which takes place at lower 
temperatures, while superconductivity region does not overlap with 
the region of antiferromagnetic ordering on Fe. At the same time, magnetic
structure of $CeO_{1-x}F_xFeAs$, shown in Fig. \ref{m_str}, is different from
that of $LaO_{1-x}F_xFeAs$ (obtained by the same group \cite{Cruz0795}). 
In the $FeAs$ plane we again have a stripe -- structure, similar to that
in $LaOFeAs$, but spins on the adjacent planes are parallel and there is no
period doubling along $c$ axis. The value of magnetic moment on Fe is as
high as 0.8(1)$\mu_B$ at 40K, which is roughly twice as large as in $LaOFeAs$. 
Also in Ref. \cite{Zhao2528} a magnetic structure due to Ce spins ordering was 
determined at $T=1.7$K. According to this work, strong correlation between
spins of Fe and Ce appears already at temperatures below 20K. Note, that
spin ordering on rare -- earths in $REOFeAs$ series typically takes place at
temperatures of the order of few K, which is nearly an order of magnitude
higher than similar temperatures in cuprates like
$REBa_2Cu_3O_{7-\delta}$ \cite{MDM}, giving an evidence of significantly
stronger interaction of these spins. The general picture of spin ordering
in $CeO_{1-x}F_xFeAs$ is shown in Fig. \ref{m_str}.

Note that data obtained by neutron scattering are still sometimes contradictory.
For example, in Ref. \cite{Qiu2195}, where $NdO_{1-x}F_xFeAs$ system was
studied, it was claimed that spin ordering on Fe does not appear up to
temperatures as low as $\sim$ 2K, when ordering of Fe and Nd spins takes
place simultaneously, with the general picture of ordering similar to that
shown in Fig. \ref{m_str} for Fe and Ce. However, in Ref. \cite{Chen0662} 
it was shown that antiferromagnetic ordering on Fe, similar to discussed
above, in fact appears in $NdOFeAs$ at $T\sim 140$K, while difficulties with
its observability are due, apparently, to rather small value of magnetic
moment on Fe, which was found to be only 0.25 $\mu_B$.

\begin{figure}[!h]
\includegraphics[clip=true,width=0.45\columnwidth]{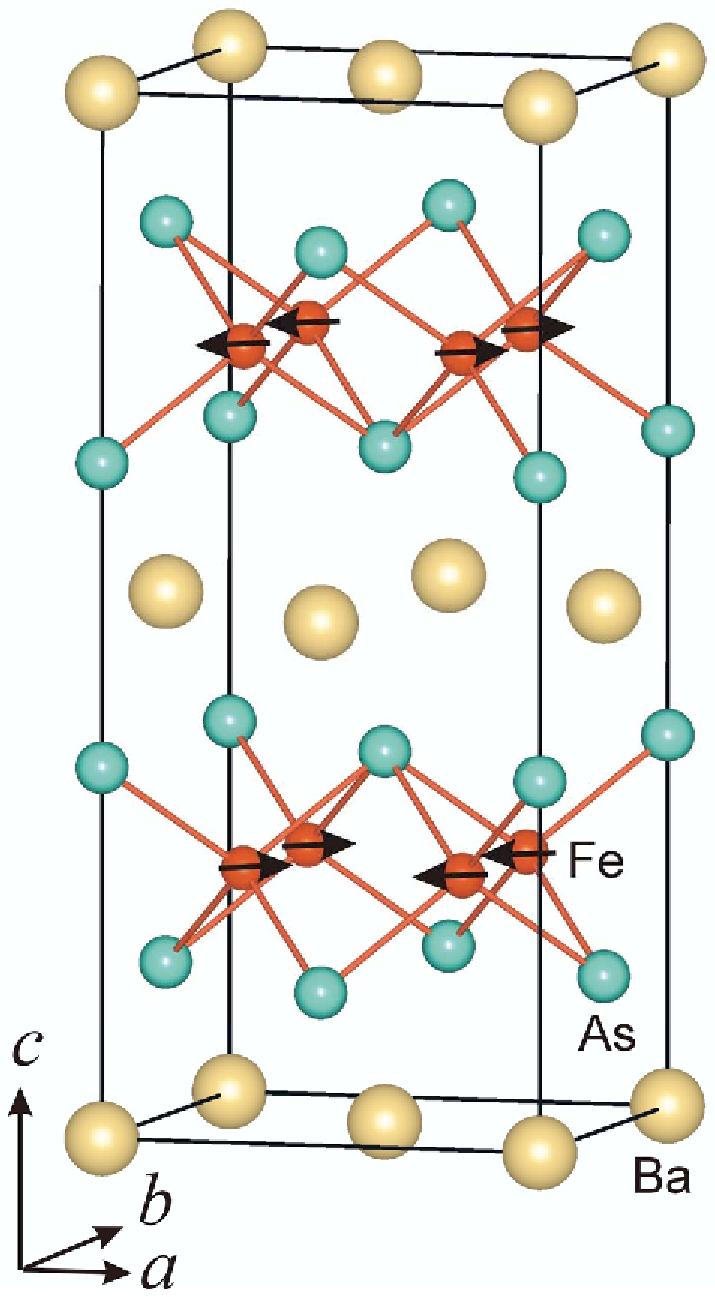}
\includegraphics[clip=true,width=0.5\columnwidth]{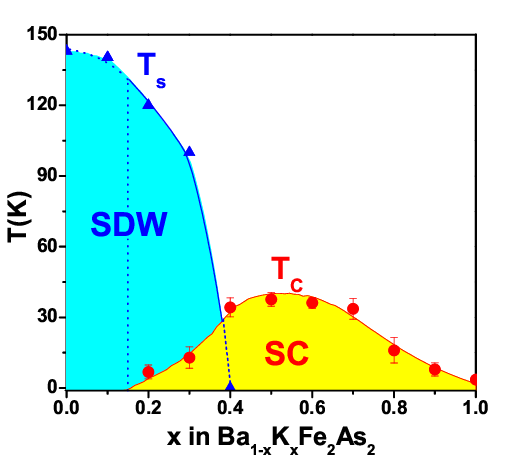}
\caption{(a) magnetic and crystal structure of $BaFe_2As_2$ in orthorhombic
($Fmmm$) cell \cite{Huang2776}, (b) phase diagram of
$Ba_{1-x}K_xFe_2As_2$ according to Ref. \cite{Chen3950}. $T_S$ is the
temperature of antiferromagnetic ordering (and structural transition),
$T_c$ is superconducting transition temperature.} 
\label{mm_str} 
\end{figure}

As to the phase diagram of 1111 systems, here also remain some questions.
In Ref. \cite{Drew4876} it was claimed that $\mu SR$ data on
$SmO_{1-x}F_xFeAs$ suggest an existence of some narrow region of coexistence
of superconductivity and antiferromagnetism at doping levels
0.1<$x$<0.15. Naturally, the complete understanding of this situation will 
appear only in the course of further studies.

Neutronographic studies were also performed on different 122 compounds. 
Polycrystalline samples of $BaFe_2As_2$ were studied in Ref. \cite{Huang2776}. 
It was shown that tetra -- ortho structural transition ($I4/mmm\to Fmmm$) 
takes place practically at the same temperature $T_S\approx 142$K as
antiferromagnetic transition, and spin ordering on Fe is the same as in
1111 systems, as shown in Fig. \ref{mm_str} (a). Magnetic moment on Fe at
$T=5$K is equal to 0.87(3) $\mu_B$. Similar data were obtained also on the
single -- crystal of $SrFe_2As_2$ \cite{Zhao1077}, where structural transition
at $T_S=220\pm 1$K is very sharp, indicating, in authors opinion, the 
first -- order transition. At the same temperature antiferromagnetic ordering
of spins on Fe appears, of the same type as in $BaFe_2As_2$ 
(Fig. \ref{mm_str} (a)), and this transition is continuous. Magnetic moment on
Fe at $T=10$K is equal to 0.94(4) $\mu_B$. These results unambiguously
demonstrate the same nature of antiferromagnetic ordering of Fe spins in
two -- dimensional FeAs planes in 1111 and 122 systems.

In Ref. \cite{Chen3950} a series of single crystals of $Ba_{1-x}K_xFe_2As_2$
with different $x$ content were studied by X-ray, neutron scattering and
electrical measurements. As a result, the authors have produced the phase
diagram shown in  Fig. \ref{mm_str} (b), where a region of coexistence of
superconductivity and antiferromagnetism is clearly seen for
0.2$<x<$0.4.

\begin{figure}[!h]
\includegraphics[clip=true,width=\columnwidth]{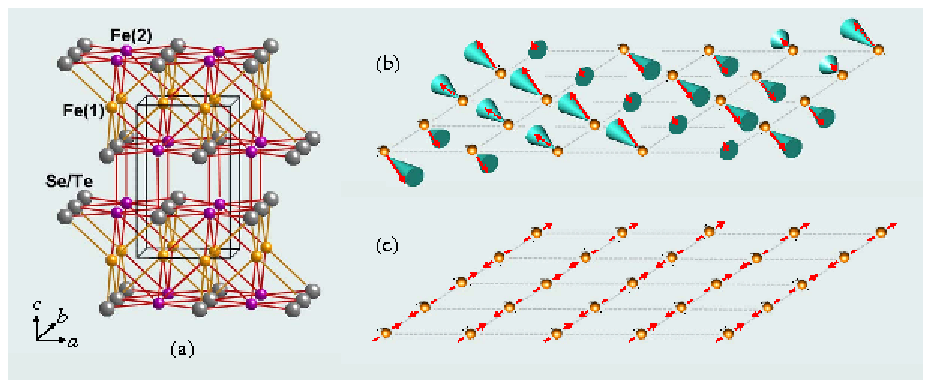}
\caption{(a) crystal structure of $\alpha-FeTe/Se$, excessive iron occupies
positions Fe(2), (b) spin ordering in $\alpha-FeTe$ according to the data of 
Ref. \cite{Bao2058} compared with antiferromagnetic ordering in $FeAs$ plane.} 
\label{fete_mstr} 
\end{figure}

Concluding this section, let us note the recent work \cite{Bao2058},  
where neutron scattering study was performed on $\alpha-Fe(Te_{1-x}Se_{x})$
system for the first time. It was found that in the prototype $\alpha-FeTe$
(with excess of Fe) ordering of Fe spins takes place at
temperatures lower than $T_S\approx 75$K (for $Fe_{1.076}Te$) and 
$T_S\approx 63$K (for $Fe_{1.141}Te$) and has a form of incommensurate
spin density wave accompanied by structural transition from tetragonal to
orthorhombic phase ($P4/nmm\to Pmmm$). Qualitative picture of spin ordering
compared with the case of FeAs plane is shown in Fig. \ref{fete_mstr}. 
At the same temperature $T_S$ rather sharp anomaly is observed also in the
temperature dependence of electrical resistivity. For superconducting
composition $Fe_{1.080(2)}Te_{0.67(2)}Se_{0.33(2)}$ with $T_c\approx 14$K 
spin ordering and structural transition are absent, though well developed
fluctuations of incommensurate SDW short -- range order are observed.

\subsection{Specific heat}

\begin{figure}[!h]
\includegraphics[clip=true,width=0.45\columnwidth]{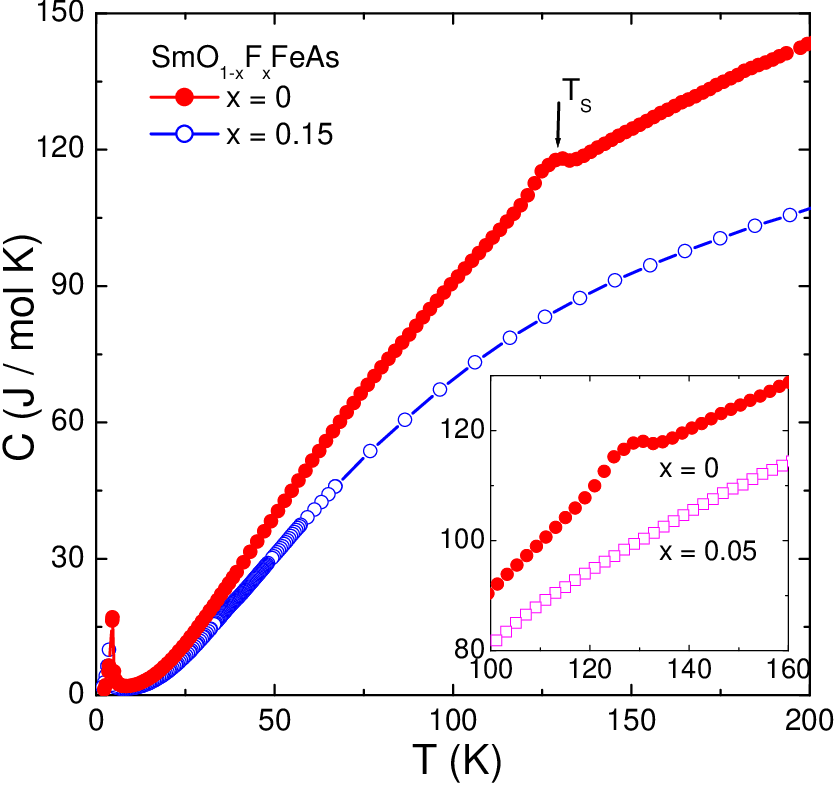}
\includegraphics[clip=true,width=0.45\columnwidth]{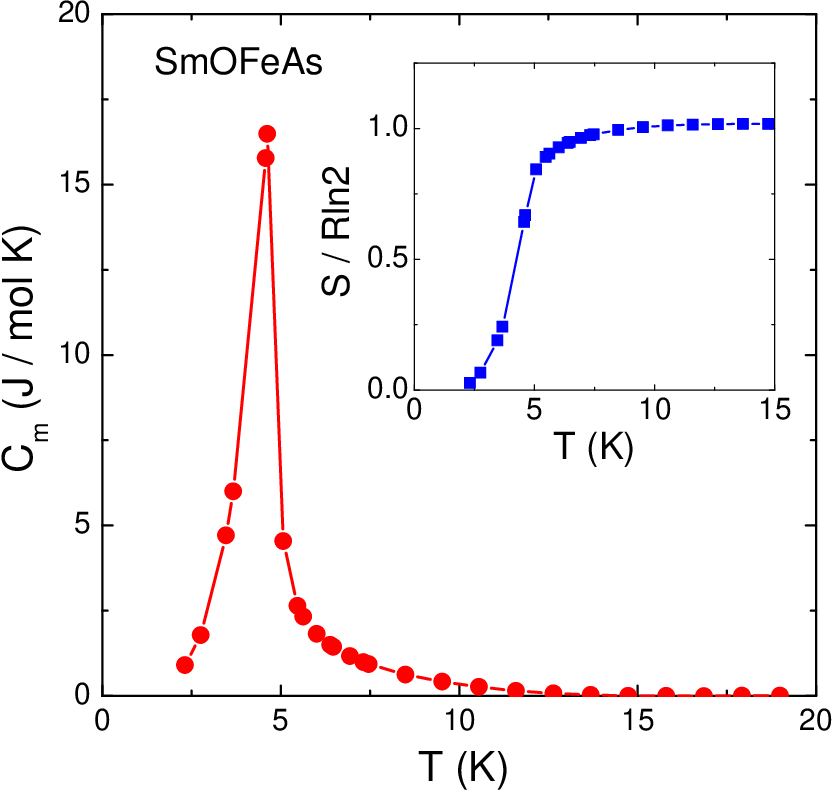}
\caption{(a) specific heat of $SmO_{1-x}F_xFeAs$ for $x=0$ and
$x=0.15$ (superconducting sample) \cite{Ding3642}. At the insert -- region
around $T_S=130$K, where (for $x=0$) an anomaly is observed, usually attributed 
to antiferromagnetic transition of spins on Fe, which is absent in 
superconducting sample, (b) magnetic contribution to specific heat of $SmOFeAs$ 
in the vicinity of Sm spins ordering temperature \cite{Ding3642}. At the insert 
-- entropy change due to this ordering.} 
\label{c_sm} 
\end{figure}

Specific heat measurements in new superconductors were performed starting from
the earliest works \cite{Chen3790,Ding3642}. As a typical example of specific
heat behavior in 1111 systems consider $SmO_{1-x}F_xFeAs$ data of Ref.
\cite{Ding3642}. In Fig. \ref{c_sm} (a) we show temperature behavior of
specific heat in this system for $x=0$ and $x=0.15$ (superconducting sample).
An anomaly of specific heat is observed at $T_S\approx 130$K, which is
obviously attributed to antiferromagnetic (SDW) (or structural tetra -- ortho)
transition. In superconducting sample ($x=0.15$) this anomaly is absent.
Besides this, in both samples there is a clear anomaly at $T\approx 5$K, 
which is connected with (antiferromagnetic) ordering of spins on Sm. As to
specific heat discontinuity at superconducting transition, it is rather hard to
separate, an is observed, according to Ref. \cite{Ding3642}, at temperatures
significantly lower than superconducting $T_c$, determined by resistivity
measurements. Apparently, this is due to rather poor quality of samples
(inhomogeneous content?).

\begin{figure}[!h]
\includegraphics[clip=true,width=0.6\columnwidth]{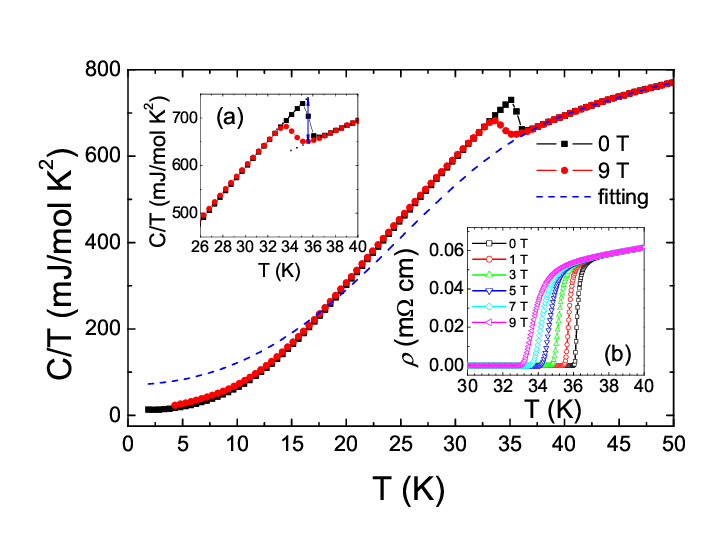}
\includegraphics[clip=true,width=0.6\columnwidth]{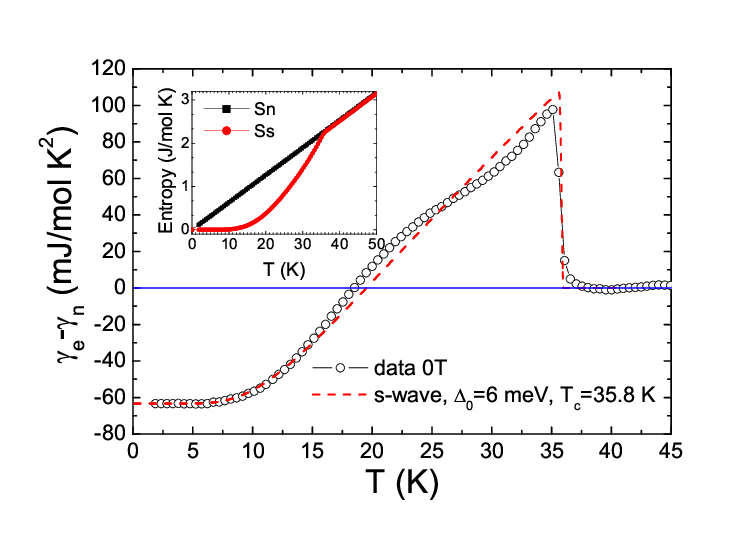}
\caption{(a) rough data on specific heat of $Ba_{0.6}K_{0.4}Fe_2As_2$ 
\cite{Mu2941}. At the inserts -- detailed behavior of specific heat in the
vicinity of $T_c$ and resistivity behavior close to superconducting transition
in different magnetic fields, (b) electronic specific heat discontinuity and
its comparison with predictions of BCS theory. At the insert -- temperature
behavior of entropy in normal and superconducting state \cite{Mu2941}.} 
\label{c_bk} 
\end{figure}

In 122 systems studies of specific heat were done on single - crystalline
samples \cite{Dong3573,Mu2941}. In Ref. \ref{c_bk} (a) we show temperature
behavior of specific heat in superconducting single - crystal of
$Ba_{0.6}K_{0.4}Fe_2As_2$ with $T_c=35.8$K \cite{Mu2941}. Detailed analysis
allowed authors to separate and study electronic specific heat coefficient
$\gamma_n$ in the normal phase induced by magnetic field and in superconducting
phase. Observed dependence of $\gamma_n$ on the value of magnetic field allowed
to come to a conclusion, that pairing in this system is of $s$ -- type, with
non-zero energy gap everywhere at the Fermi surface (absence of gap zeroes
characteristic e.g. for $d$-wave pairing in cuprates).

In Fig. \ref{c_bk} (b) a comparison is made of accurately separated electronic
specific heat discontinuity $\Delta C_e$ at superconducting transition with
predictions of BCS theory (weak coupling!). Agreement is quite impressing --
the use of thus determined value of $\gamma_n\approx 63.3$ mJ/mol K$^2$ gives  
the value of $\Delta C_e/\gamma_nT|_{T=T_c}\approx$ 1.55, to be compared with
BCS theory prediction of 1.43. The fit to BCS theory allowed the authors to 
determine also the value of energy gap at low temperatures, which was found to 
be $\Delta_0\approx 6$ meV. Below we shall see that this value is in good
agreement with other data (ARPES).

\subsection{NMR (NQR) and tunnelling spectroscopy}

There are already a couple of studies of NMR (NQR) in new superconductors,
leading to certain conclusions on the possible types of superconducting
pairing in these systems. In Ref. \cite{Nakai4765} $^{75}As$ and $^{139}La$ 
NMR was studied in $LaO_{1-x}F_xFeAs$ for $x=0.0, 0.04, 0.11$. 
In undoped $LaOFeAs$ a characteristic peak (divergence) of NMR relaxation
$1/T_{1}$ on $^{139}La$ is observed at the temperature of antiferromagnetic
transition $T_S\sim 142$K, while below the NMR spectrum is too wide, which is
attributed to antiferromagnetic ordering. In superconducting sample with
$x=0.4$ ($T_c=17.5$K) the value of $1/{TT_{1}}$ increases as temperature lowers
up to $\sim 30$K following Curie -- Weiss law: 
$1/{TT_{1}}\sim C/(T+\theta)$ with $\theta\sim 10$K, with no divergence at 
finite temperatures, so that appearance of superconductivity is accompanied by
suppression of magnetic ordering. The overall picture of nuclear spin
relaxation for the sample with $x=0.11$ ($T_c=22.7$)K is qualitatively different.
The value of $1/{TT_{1}}$ both on $^{139}La$ and $^{75}As$ decreases with
lowering temperature, which is similar to NMR picture of pseudogap behavior 
in underdoped cuprates, approaching a constant in the vicinity of $T_c$, 
and is well fitted by activation temperature dependence with pseudogap value 
$\Delta_{PG}=172\pm 17$K \cite{Nakai4765}. 

\begin{figure}[!h]
\includegraphics[clip=true,width=0.45\columnwidth]{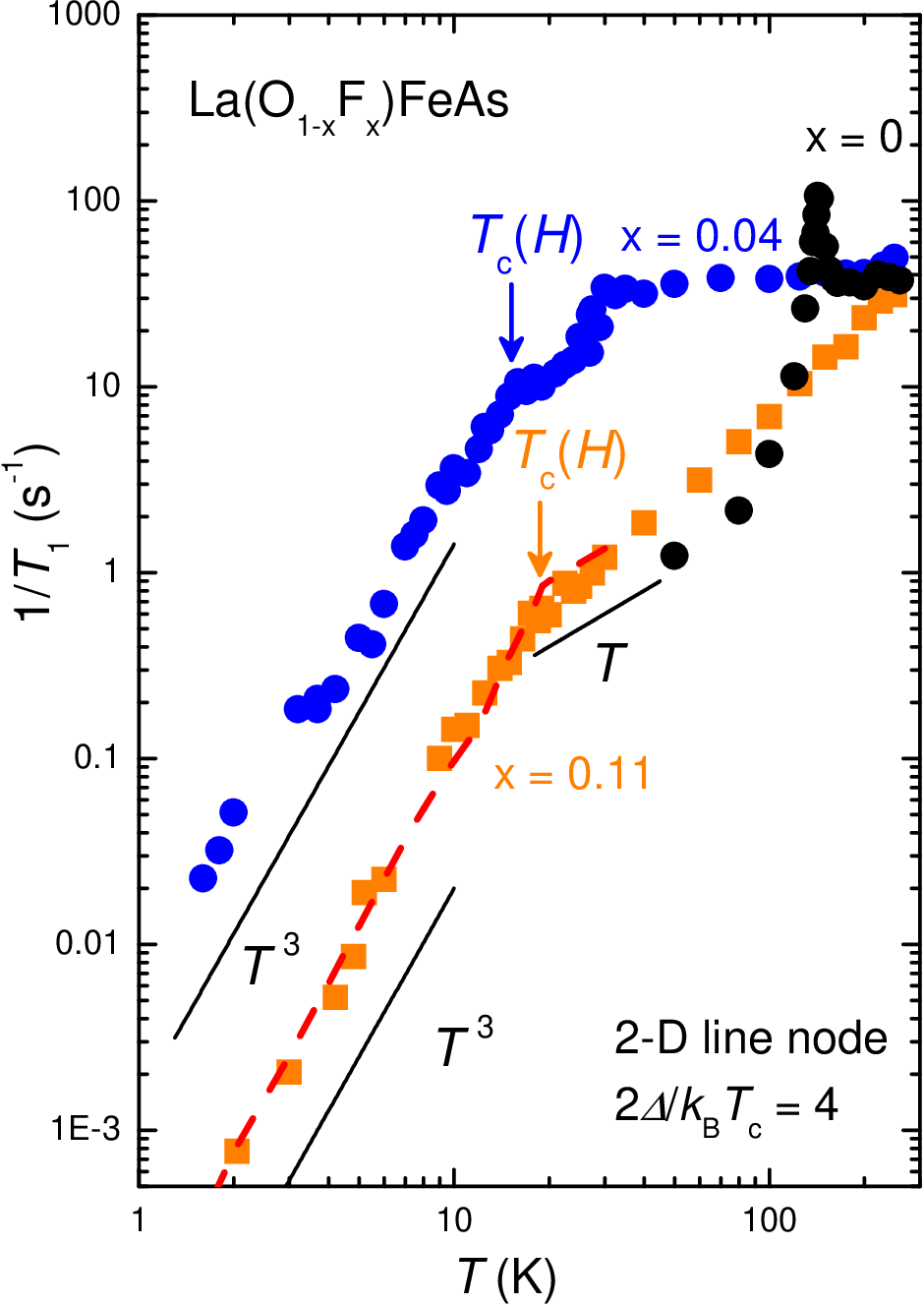}
\includegraphics[clip=true,width=0.45\columnwidth]{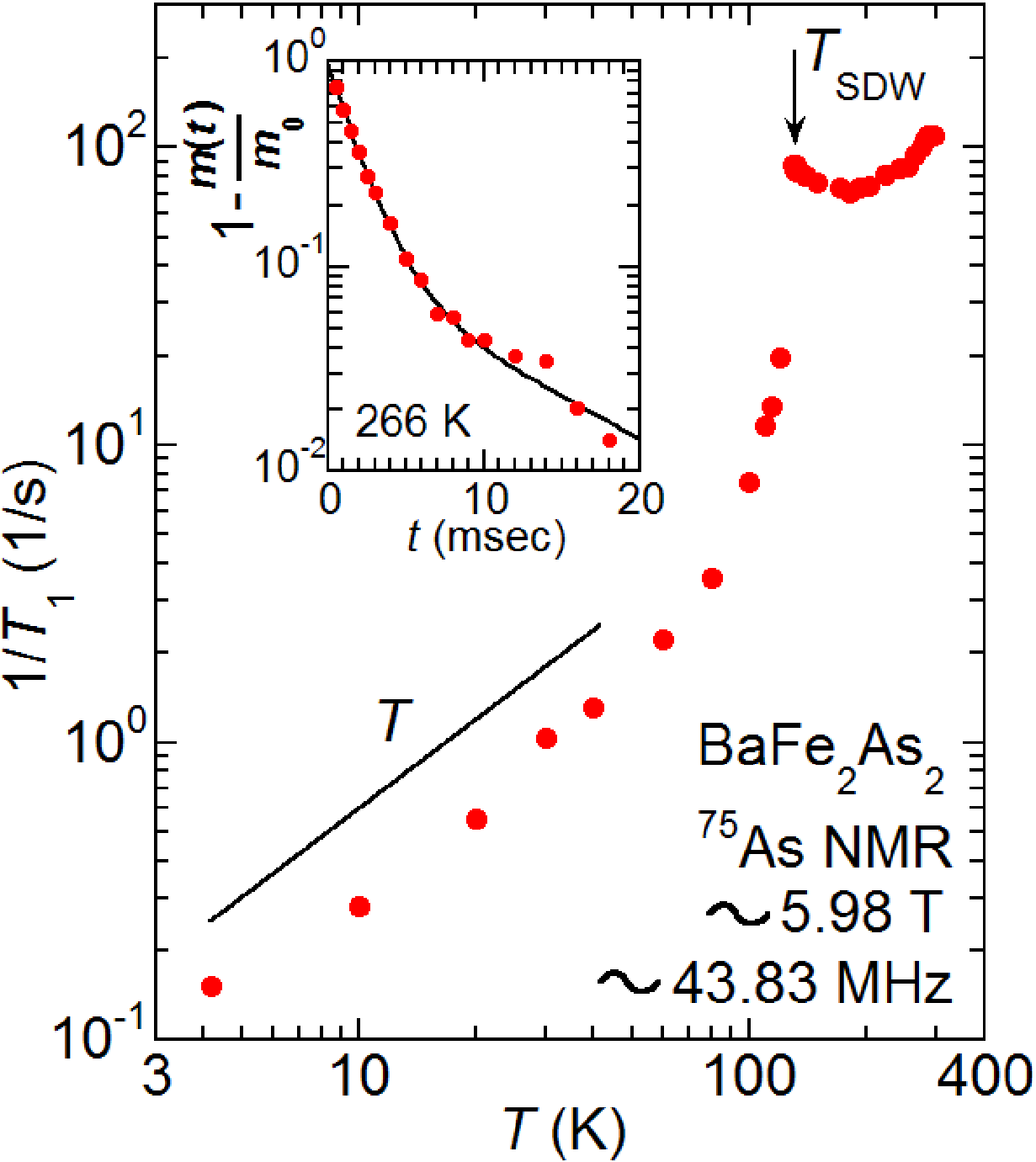}
\caption{(a) temperature dependence of NMR relaxation rate $1/T_1$ in
$LaO_{1-x}F_xFeAs$ for $x=0.0, 0.04, 0.11$ in low temperature region of the 
order of $T_c$ and below \cite{Nakai4765},
(b) temperature dependence of $1/T_1$ ­  $^{75}As$ in undoped $BaFe_2As_2$ 
\cite{Fuk4514}.} 
\label{nmr_1} 
\end{figure}

In Fig. \ref{nmr_1} (a) we show temperature dependence $1/T_1$ on
$^{75}As$ in $LaO_{1-x}F_xFeAs$ system in superconducting state ($x=0.04$ and 
$x=0.11$) and in undoped $LaOFeAs$ (on $^{139}La$ nuclei, normalized according 
to $^{139}(1/T_1)/^{75}(1/T_1)\sim$ 0.135) \cite{Nakai4765}. Note the absence
of Gebel -- Slichter peak of $1/T_1$ in the vicinity of $T_c$ and $T^3$ 
dependence of $1/T_1$ in superconducting region, characteristic for anomalous 
(non $s$ -- wave!) pairing with energy gap with zeroes on the Fermi surface,
e.g. like in the case of $d$ -- wave pairing. In fact, these dependences are
well fitted using $\Delta(\phi)=\Delta_0\sin (2\phi)$, where $\phi$ is polar,
angle determining direction of the momentum in two-dimensional inverse space,
corresponding to $FeAs$ -- plane, with the value of $2\Delta_0/T_c=4.0$ 
\cite{Nakai4765}.

Quite similar results were obtained in Ref. \cite{Muk3238} during the studies
of $^{75}As$ NMR and NQR in $LaO_{1-\delta}FeAs$ ($\delta=0, 0.25, 0.4$) 
with $T_c$ up to 28K and $NdO_{0.6}FeAs$ with $T_c=53$K. In particular, these
authors has come to a conclusion that their NQR relaxation data for
$LaO_{0.6}FeAs$ in superconducting phase, correspond to an energy gap of the
form $\Delta=\Delta_0\cos (2\phi)$ with $\frac{2\Delta_0}{T_c}\approx 5$, 
which corresponds to $d$ -- wave pairing with gap zeroes at the Fermi surface,
i.e. the same pairing symmetry as in cuprates. Temperature behavior of $1/T_1$
for $T>T_c$, found in this work, also gives an evidence of pseudogap behavior
with pseudogap value $\Delta_{PG}\approx 196$K.

In Ref. \cite{Mat0249} $^{75}As$ Knight shift measurements were done in 
$PrO_{0.89}F_{0.11}FeAs$. The sharp drop of Knight shift for $T<T_c$ 
is a definite evidence of singlet nature of pairing. Details of temperature
dependence of Knight shift were well fitted using the model with two $d$-wave
superconducting gaps:
$\Delta=\Delta_0\cos (2\phi)$, $\Delta_0=\alpha\Delta_1+(1-\alpha)\Delta_2$
with $2\Delta_1/T_c\approx 7$ ¨ $2\Delta_2/T_c\approx 2.2$, $\alpha=0.4$. 
In the same model the authors of Ref. \cite{Mat0249} has successfully described
temperature dependence of $1/T_1$ on $^{19}F$ for $T<T_c$.  

NMR study of undoped $BaFe_2As_2$ was performed in Ref. \cite{Fuk4514}. 
The temperature dependence of $1/T_1$ is shown in Fig. \ref{nmr_1} (b). 
A clear anomaly is observed at $T=131$K, connected to antiferromagnetic 
(SDW) transition. Sharp drop of $1/T_1$ is attributed to SDW -- gap opening
at the part of Fermi surface. Linear over $T$ behavior of relaxation rate
for $T<100$K is due to relaxation of nuclear spins on conduction electrons,
remaining on the ``open'' parts of Fermi surface. In general, situation here is
reminiscent of similar behavior in $LaOFeAs$, though there are some significant
quantitative differences \cite{Fuk4514}.

\begin{figure}[!h]
\includegraphics[clip=true,width=0.8\columnwidth]{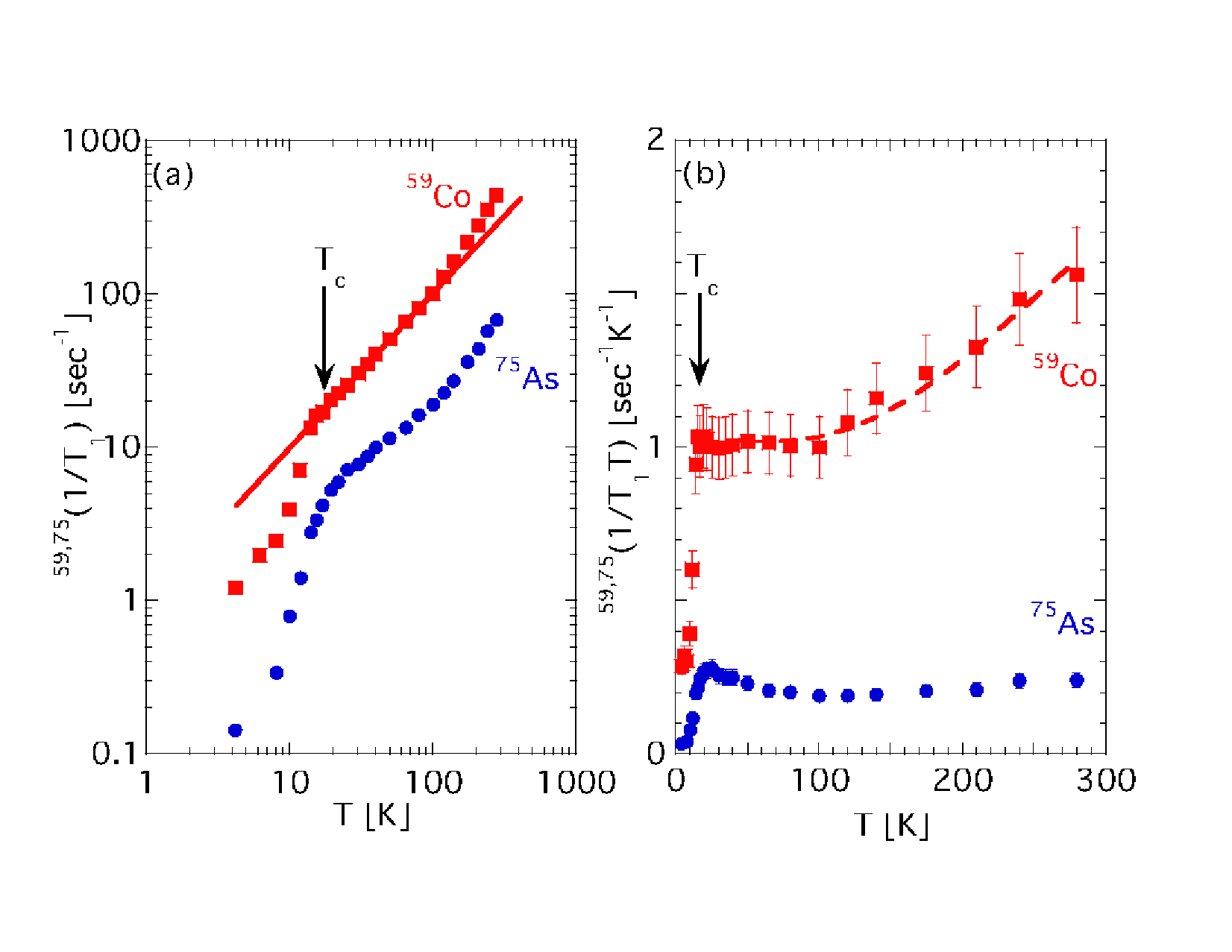}
\includegraphics[clip=true,width=0.6\columnwidth]{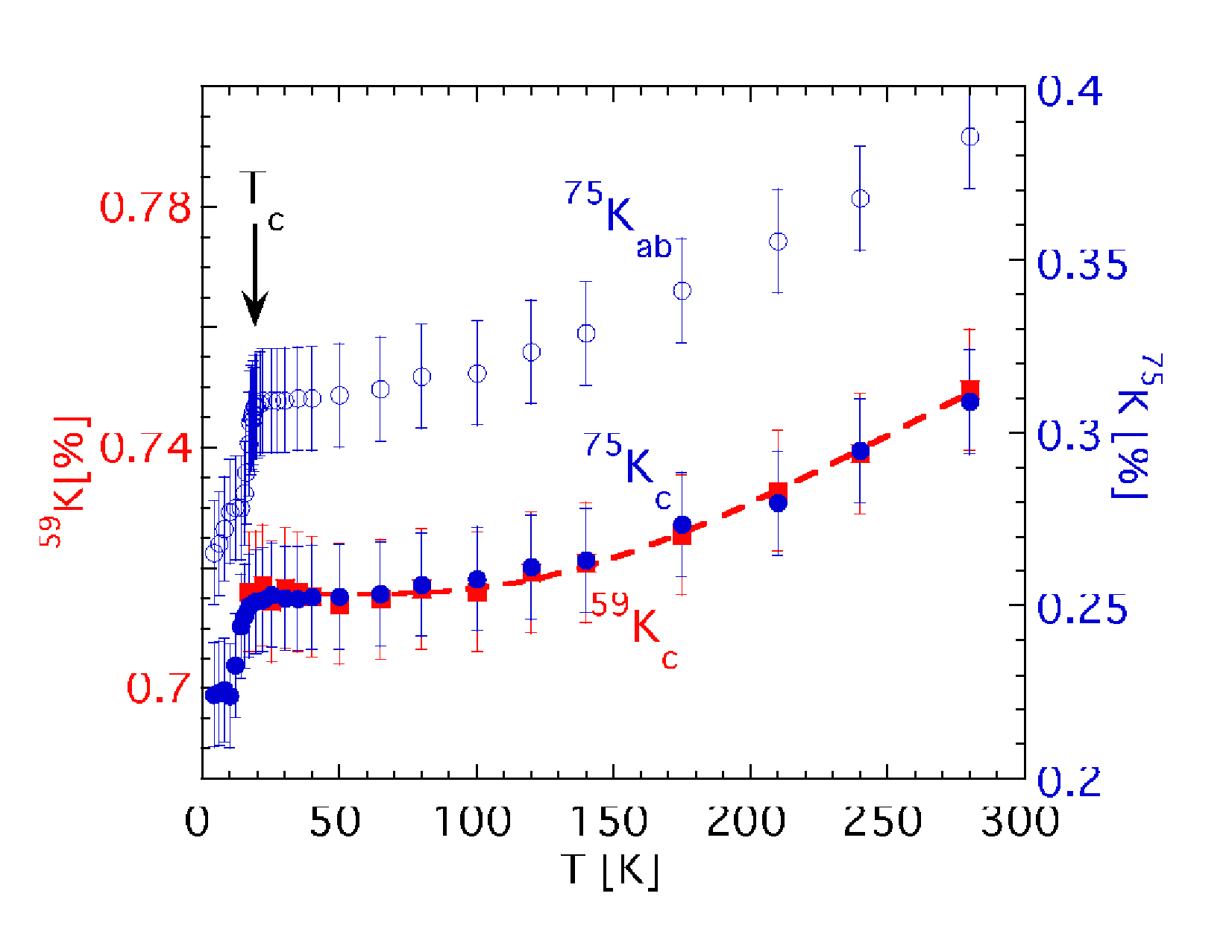}
\caption{(a) temperature dependence of $1/T_1$ on $^{59}Co$ and $^{75}As$ 
in $BaFe_{1.8}Co_{0.2}As_2$ and similar dependence of $1/TT_1$ (b), where
dashed curve corresponds to activation dependence with pseudogap
$\Delta_{PG}=560\pm 150$K \cite{Ning1420},
(c) temperature dependence of Knight shift on $^{59}Co$ and $^{75}As$ in the
same system, where dashed curve is described by activation dependence with the
same pseudogap width.} 
\label{nmr_2} 
\end{figure}

NMR study of superconducting phase in 122 system was done on
$BaFe_{1.8}Co_{0.2}As_2$ with $T_c=22$K in Ref. \cite{Ning1420}.
In Fig. \ref{nmr_2} results on nuclear spin relaxation rate and Knight shift
are shown. For temperatures below 280K a drop of $1/TT_1$ is observed, 
which is approximated by activation dependence with (pseudo)gap
$\Delta_{PG}\approx 560$K (Fig. \ref{nmr_2} (b)), which is significantly
larger than pseudogap width estimates for $LaO_{0.9}F_{0.1}FeAs$ 
obtained in Ref. \cite{Nakai4765} and quoted above. Data on Knight shift
(Fig. \ref{nmr_2} (c)), similarly to  1111 case, give an evidence of singlet
pairing. Temperature dependence of Knight shift above $T_c$ also show
pseudogap behavior with the same pseudogap width as obtained from fitting the
data on relaxation rate.

Thus NMR data on 1111 and 122 systems are in many respects similar.
Singlet pairing follows unambiguously, asa well as an evidence for the anomalous
nature of pairing with probable gap zeroes on the Fermi surface (and probably
the presence of two superconducting gaps). However, below we shall see that
the conclusion on gap zeroes contradicts some other experiments and 
interpretation of NMR data may be quite different.

As to $\alpha-FeSe$ system, up to now there is only one NMR study on
$FeSe_{0.92}$ with $T_c=8$K \cite{Koteg0040} (NMR on $^{77}Se$ nuclei), 
demonstrating the absence of Gebel -- Slichter peak in the temperature
dependence of $1/T_1$ in the vicinity of $T_c$ and also the absence of any
anomaly which can be attributed to any kind of magnetic ordering at higher
temperatures, as well as the absence there of any kind of ``pseudogap'' 
behavior (validity of Korringa relation). At the same time, in the $T<T_c$ 
region $\sim T^3$ -- behavior of $1/T_1$ is observed, probably giving an
evidence for zeroes of superconducting gap at the Fermi surface.

Among a number of studies using different kinds of tunnelling spectroscopy, 
let us mention Refs. \cite{Tes4616,Millo0359,Pan0895}. Up to now these
experiments were performed on polycrystalline samples and results are
somehow contradictory. 

In Ref. \cite{Tes4616} the method of Andreev spectroscopy was applied to
$SmO_{0.85}F_{0.15}FeAs$ with $T_c=42$K. Only one superconducting gap was
observed with $2\Delta=13.34 \pm 0.3$ meV (close to $T=0$), which corresponds 
to $2\Delta/T_c=3.68$, i.e. quite close to a standard BCS value of 3.52. 
Temperature dependence of the gap, determined in Ref. \cite{Tes4616}, also
closely followed BCS theory. In authors opinion, these results give an
evidence of the usual ($s$-wave) order parameter with no zeroes at the Fermi
surface, in obvious contradiction with NMR results quoted above.

In Ref. \cite{Millo0359} similar system -- $SmO_{0.85}FeAs$ with $T_c=52$K was
studied by scanning tunnelling spectroscopy at $T=4.2$K. Good quality
tunnelling characteristics were obtained only from some parts of the sample
surface, and these were fitted to tunnelling characteristics of $d$-wave
superconductor with $\Delta=8\div 8.5$ meV, which corresponds to 
$2\Delta/T_c\sim 3.55\div 3.8$.

The same method of scanning tunnelling spectroscopy (microscopy) was
applied also in Ref. \cite{Pan0895} to $NdO_{0.86}F_{0.14}FeAs$ with $T_c=48$K, 
and measurements were done at different temperatures. At temperatures
significantly lower than $T_c$ on different parts of sample surface two gaps
were observed -- a bigger one $\Delta\sim 18$ meV and smaller $\Delta\sim 9$ meV. 
Both gaps closed at the transition point $T=T_c$, with smaller gap more or less
following the BCS temperature dependence. At the same parts of the surface, 
where the smaller gap was observed below $T_c$, slightly above $T_c$ a jumplike
opening pseudogap appeared closing only at $T=120$K.  At present there is no
clear interpretation of this unexpected behavior.

Note also Ref. \cite{Wang1986} where $SmO_{0.9}F_{0.1}FeAs$ with $T_c=51.5$K 
was studied by point contact spectroscopy. The authors also observed two
superconducting gaps -- a larger one $\Delta=10.5\pm 0.5$ meV and smaller one
$\Delta=3.7\pm 0.4$ meV, with both gaps following BCS -- like temperature
dependence.

Up to now there are no systematic tunnelling data for 122 systems.

Contradictory nature of existing data of tunnelling spectroscopy is more or 
less obvious. It seems we have to wait the results of experiments on
single crystals.

\subsection{Optical properties}

The measurements of optical properties of new superconductors were done in a
number of works, both on polycrystalline samples of 1111 systems
1111 \cite{Dub2415,Boris1732} and on single crystals of 122 system 
\cite{Yang1040,Li1094}.

In Ref. \cite{Dub2415} ellipsometry measurements of dielectric permeability
of $REO_{0.82}F_{0.18}FeAs$ ($RE=Nd,Sm$) were made in the far infrared region.
It was shown that electronic properties of these systems are strongly
anisotropic (quasi two-dimensional) and, in these sense, analogous to those of
cuprates. A noticeable suppression of optical conductivity in superconducting
state was discovered, which was attributed by the authors to the opening of
superconducting gap $2\Delta\approx 300$cm$^{-1}$ (37 meV), corresponding to
$2\Delta/T_c\sim 8$, i.e. the strong coupling limit.

In Ref. \cite{Boris1732} the same method was applied to the studies of 
dielectric permeability of $LaO_{0.9}F_{0.1}FeAs$ with $T_c=27$K in a wide
frequency interval of 0.01$\div$6.5 eV at temperatures $10\leq T\leq 350$K.
Unusually narrow region of Drude behavior was observed, corresponding to the
density of free carriers as low as $0.040\pm 0.005$ per unit cell, as well as
signatures of pseudogap behavior at 0.65 eV. Besides that, the authors have
also observed a significant transfer of spectral weight to the frequency 
region above 4 eV. These results allowed to conclude that electronic 
correlations (and (or) electron -- phonon coupling) are very important in 
these systems.

The studies of reflectance and real part of optical conductivity of a single
crystal of $Ba_{0.55}K_{0.45}Fe_2As_2$ \cite{Yang1040} has shown that the
absorption spectrum of this system consists of a noticeable Drude peak at
low frequencies and a wide absorption band with maximum at 0.7 eV, which the
authors attributed to carrier scattering by collective (Boson) excitations
(e.g. spin fluctuations) with energies of the order of 25 meV and strongly 
temperature dependent coupling constant (with carriers).

Most convincing and interesting optical data were obtained in Ref. \cite{Li1094}, 
where detailed measurements of reflectance were performed on the single crystal
of $Ba_{0.6}K_{0.4}Fe_2As_2$ with $T_c=37$K in the infrared region and in the
wide temperature interval 10$\div$300K.

\begin{figure}[!h]
\includegraphics[clip=true,width=0.5\columnwidth]{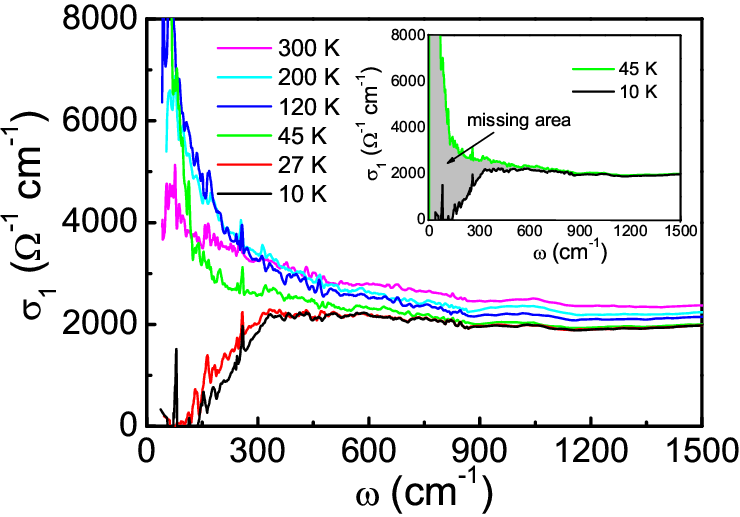}
\caption{Real part of optical conductivity of $Ba_{0.6}K_{0.4}Fe_2As_2$ 
at different temperatures \cite{Li1094}. At the insert -- comparison of the
data at 10K and 45K, where the ``missing area'', connected with superconducting
gap opening and formation of condensate of Cooper pairs is clearly seen.} 
\label{op_cond} 
\end{figure}

In Fig. \ref{op_cond} we show the data on real part of optical conductivity
$\sigma_1(\omega$) obtained in Ref. \cite{Li1094}. It is seen that close to and
below T$_c$ (curves, corresponding to 27 K and 10 K) demonstrate rapid drop of 
$\sigma_1(\omega$) at frequencies below 300 cm$^{-1}$, so that conductivity
is practically zero below 150 cm$^{-1}$, which gives an evidence of 
superconducting $s$ -- wave gap opening. Gap value determined by absorption
edge is 2$\Delta\simeq$150 cm$^{-1}$, which correlates well with ARPES data to
be discussed below.

Thus, at temperatures significantly lower T$_c$ a considerable suppression of
low frequency conductivity is observed, which is connected with the formation of
condensate of Cooper pairs. According to the well known 
Ferrell--Glover--Tinkham sum rule \cite{Ferrell,Tinkham2} the difference
of conductivities at T$\simeq$$T_c$ and T$\ll$T$_c$ (i.e. the so called
``missing area'' between appropriate curves, shown at the insert in Fig.  
\ref{op_cond}) directly determines the value of condensate density via:
\begin{equation} 
\omega_{ps}^2=8\int_{0^+}^{\omega_c}[\sigma_1(\omega, T\simeq
T_c)-\sigma_1(\omega, T\ll T_c)]d\omega 
\label{chik}
\end{equation}
where $\omega_{ps}^2=4\pi n_se^2/m^*$ -- is the square of plasma frequency of
superconducting carriers, $n_s$ is the their density, and $\omega_c$ -- is
a cut-off frequency, which is chosen to guarantee the convergence of
$\omega_{ps}^2$. Then we can determine the penetration depth as
$\lambda=c/\omega_{ps}$. Eq. (\ref{chik}) defines (via the general optical sum 
rule) the fraction of electron (carrier) density, which is transferred to
$\delta(\omega)$ singularity of $\sigma_1(\omega)$, corresponding to
superconducting response of the condensate. Direct estimate of the value of
``missing area'' gives for penetration depth the value $\lambda$=2080 \AA, 
which agrees well with other data \cite{Li1094}.

\subsection{Phonons and spin excitations: neutron spectroscopy}

Up to now a number of experiments have already been done to study collective
excitations, i.e. phonons and spin waves, in new superconductors, which
is of principal importance for the clarification of the nature of Cooper
pairing in these systems. Below we shall mainly deal with experiments on
inelastic neutron scattering.

In Ref. \cite{Qiu1062} inelastic neutron scattering was studied on
$LaO_{0.87}F_{0.13}FeAs$ with $T_c\approx 26$K. Characteristic maxima of
phonon density of states were observed at 12 and 17 meV.

\begin{figure}[!h]
\includegraphics[clip=true,width=0.4\columnwidth]{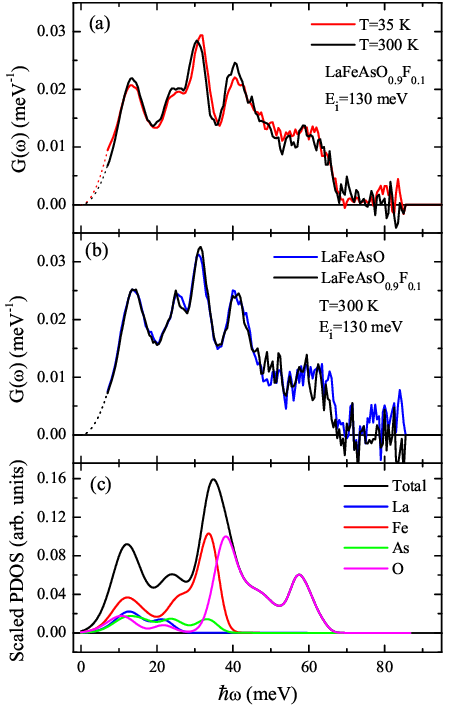}
\includegraphics[clip=true,width=0.4\columnwidth]{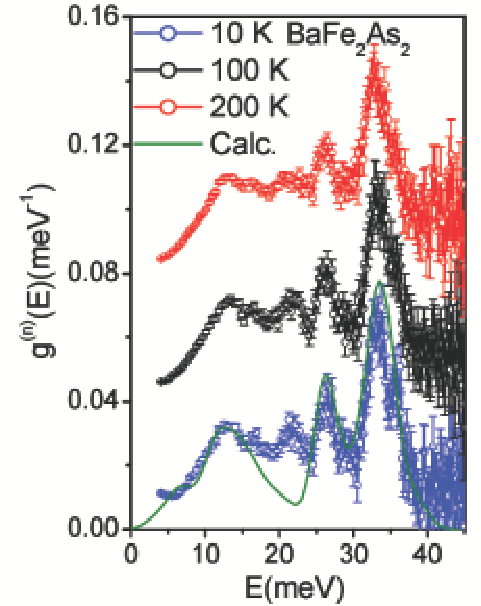}
\caption{(a)-(b) phonon density of states in $LaO_{1-x}F_xFeAs$, obtained
from inelastic neutron scattering \cite{Christ3370} and compared with
calculated (á) in Ref. \cite{Singh08}, (d) comparison of experimental and
calculated phonon density of states in $BaFe_2As_2$ \cite{Mittal3172}.} 
\label{ph_neutr} 
\end{figure}

In more details phonon density of states in $LaO_{1-x}F_xFeAs$ (for $x=0$ and
$x\sim 0.1$) was studied in Ref. \cite{Christ3370}, where it was also compared
with results of phonon spectrum calculations done in Ref. \cite{Singh08}. 
Main results of this work are shown in Fig. \ref{ph_neutr} (a-c), where we can
see both the general structure of phonon density of states and a satisfactory
agreement with calculations of Ref. \cite{Singh08}. Also we can observe that
the phonon spectrum of prototype 1111 system is differs very little from that
observed in doped (superconducting) sample. The origin of peaks in phonon
density of states is well explained on the basis of theoretical calculations.

Phonon density of states in 1111 systems was also studied by nuclear resonance
inelastic synchrotron radiation scattering \cite{Higa3968} 
(in systems based on $La$) as well as by inelastic X-ray scattering 
\cite{Fukuda0838} (systems based on $La$ and $Pr$). In all cases, results 
found are quite similar to that obtained by inelastic neutron scattering and
are in satisfactory agreement with theoretical calculations.

As to 122 systems, up to now there are works on inelastic neutron scattering
on the prototype system $BaFe_2As_2$ \cite{Mittal3172, Zbiri4429}, where
consistent results were obtained, which are also in agreement with results of
theoretical calculations of phonon spectrum made in these works. As an example,
in Fig. \ref{ph_neutr} (d) we show a comparison of calculated and experimental
phonon densities of states for this system made in Ref. \cite{Mittal3172}. 
We see quite satisfactory agreement, except an additional peak at frequency
of the order of 21.5 meV, observed in the experiments.

Dynamics of spin excitations in new superconductors was studied by
neutronography in $SrFe_2As_2$ \cite{Zhao2455} and $BaFe_2As_2$
\cite{Ewi2836}, i.e. on undoped samples, where antiferromagnetic ordering
takes place at low temperatures.

In particular, in Ref. \cite{Zhao2455} it was shown that the spectrum of
magnetic excitations is characterized by a gap $\Delta\leq 6.5$ meV, 
while above this gap the well defined spin waves are observed, and the
measurement of their velocity allowed to estimate exchange integrals
(in localized spins model). In the vicinity of the temperature of 
antiferromagnetic transition no signs of critical scattering was observed, 
which in the opinion of authors show that magnetic transition here is of the
first order \cite{Zhao2455}.

Qualitatively similar results (though without detailed measurements of
spin wave dispersion) were obtained in Ref. \cite{Ewi2836}, where it was shown
that magnon spectrum continues up to energies, which are significantly higher
than typical phonon frequencies ($\sim$ 40 meV) and ends at energies about
170 meV.
                                   
\subsection{Other experiments}

In our rather short review of experiments on new superconductors we could not
pay attention to a number of important studies. Outside our presentation
remained more detailed discussion of experiments on critical magnetic fields
(in particular measurements of $H_{c1}$), direct measurements of penetration
depth, experiments on X-ray photoemission. Practically we have not paid any
attention to experiments on traditional transport properties in the normal
state (such as Hall effect, thermoelectricity etc.). It is connected mainly
with the limited size of this review, as well as by personal preferences of the
author. Below, in theoretical part of this review we shall return to
discussion of a number of extra experiments. In particular, we shall pay
much attention to experiments of angle resolved photoemission (ARPES), which
are more appropriately discussed in parallel with discussion of electronic
spectrum of these systems. The same applies to some other experiments on
determination of Fermi surfaces (quantum oscillation effects in strong magnetic
fields).


\section{Electronic spectrum and magnetism}

\subsection{Band structure (LDA)}

Clarification of the structure of electronic spectrum of new superconductors
is crucial for explanation of their physical properties. Accordingly, since
the first days, different groups have started the detailed band -- structure
calculations for all classes of these compounds, based primarily on different
realizations of general LDA approach.

First calculations of electronic spectrum of iron oxypnictide $LaOFeP$ were
performed in Ref. \cite{lebegue}, before the discovery of high temperature
superconductivity in FeAs based systems. For $LaOFeAs$ such calculations were,
almost simultaneously, were done in Refs. \cite{Singh08,dolg,mazin,Xu1282}. 
In the following, similar calculations were performed also for the other
1111 systems, as well as for 122 \cite{Shein,Singh2643},
111 \cite{Singh2643,SheinIv,Ma3526} and $\alpha$-FeSe \cite{Singh4312}.
As results obtained in all these references were more or less similar, in
the following we shall concentrate in more details on our group works
\cite{Nek1239,Nek2630,Nek1010,Nek_Sr}, referring to other authors where
necessary. We shall also limit ourselves mainly to the results obtained for
nonmagnetic tetragonal phase of 1111 and 122 systems (as well as 111), as 
superconductivity is realized in this phase.

In Ref. \cite{Nek1239} we have performed {\it ab initio} calculations of
electronic structure for a number of oxypnictides from the series
REO$_{1-x}$F$_x$FeAs (where RE=La, Ce, Nd, Pr, Sm, and also for the
hypothetical at a time case of RE=Y) in the framework of the standard
LDA-LMTO approach \cite{LMTO}.

\begin{figure}
\includegraphics[clip=true,width=0.36\columnwidth,angle=270]{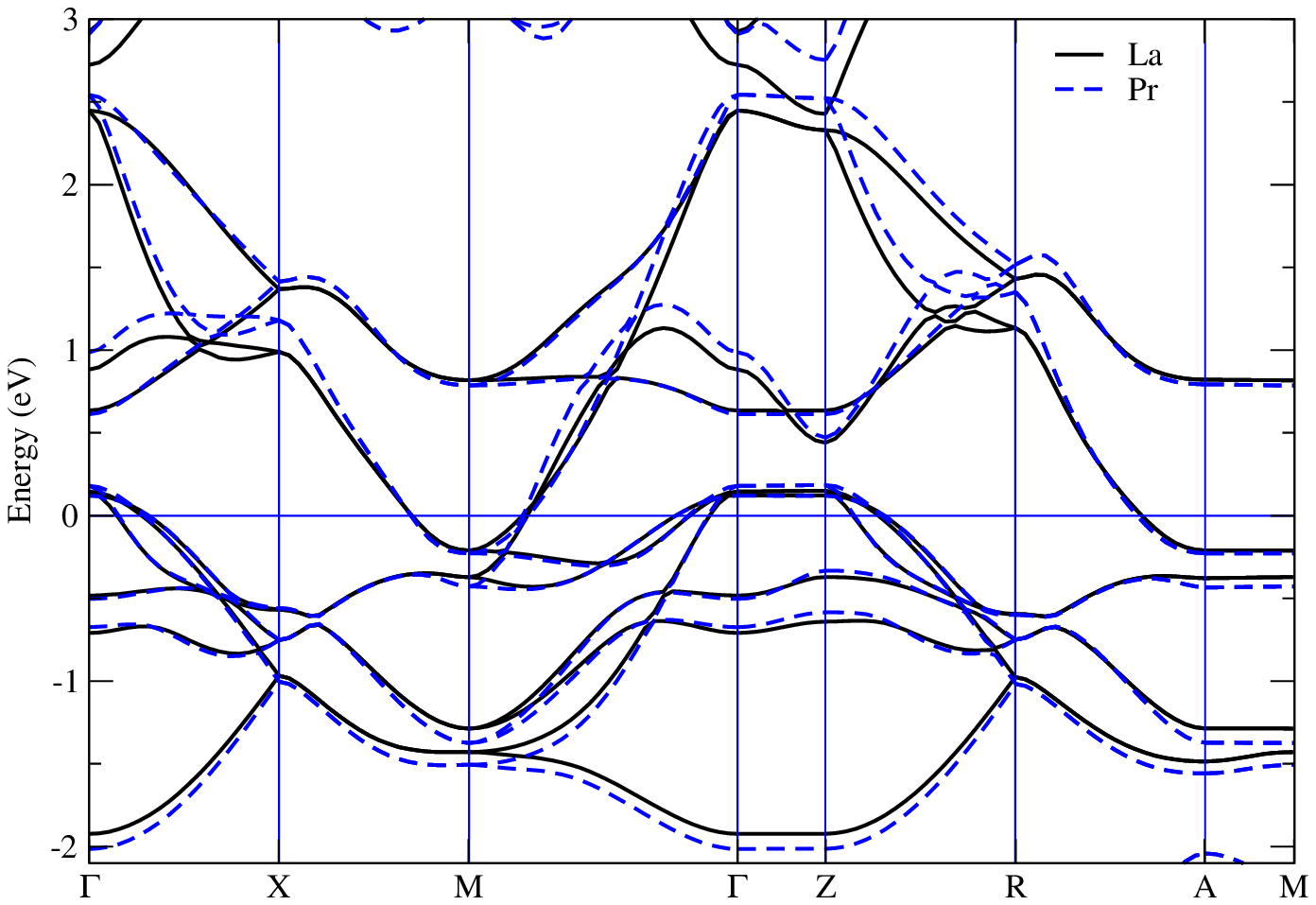}
\includegraphics[clip=true,width=0.55\columnwidth,angle=270]{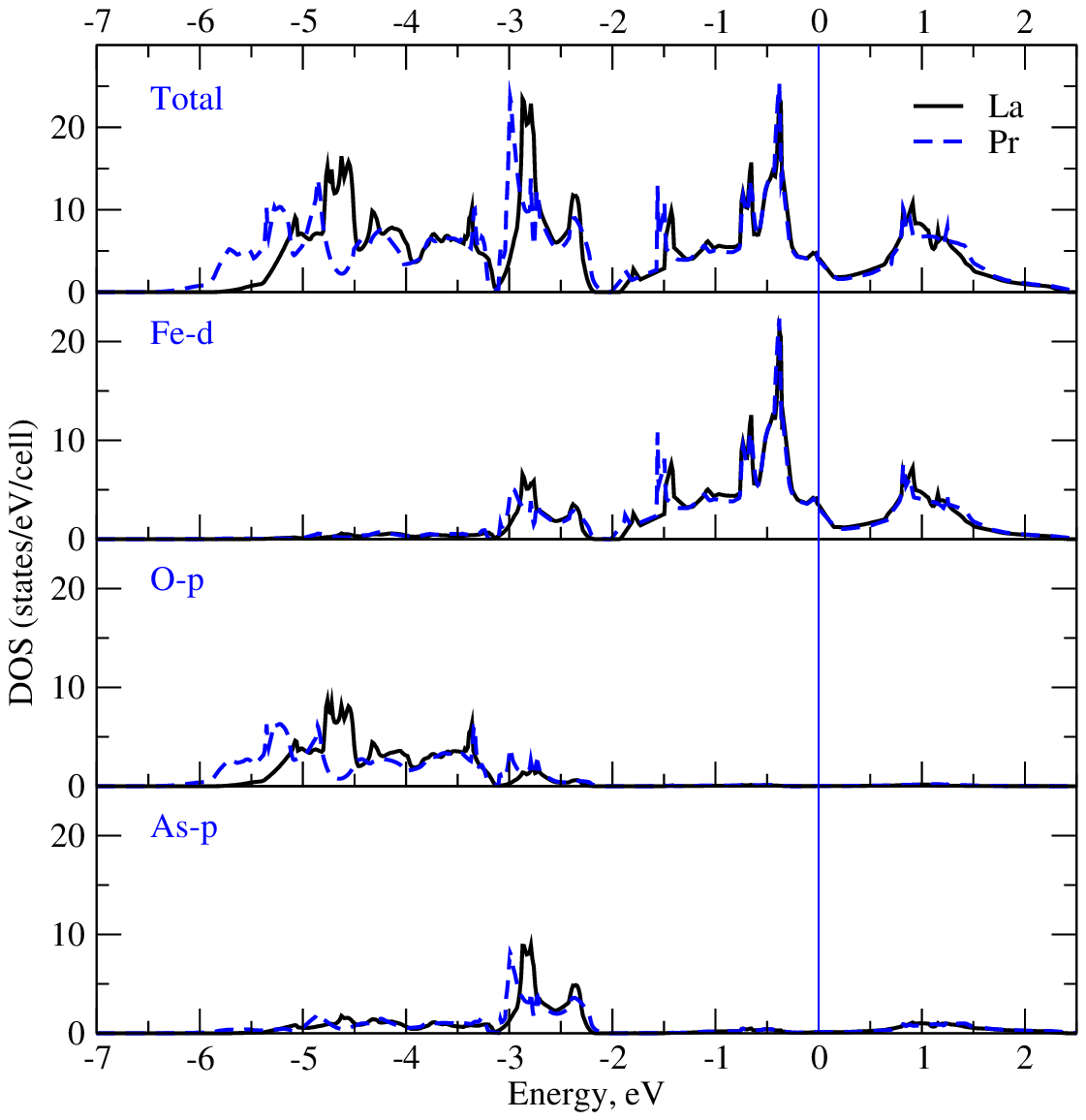}
\caption{(a) electronic spectra of $LaOFeAs$ and $PrOFeAs$ in high - symmetry
directions in the Brillouin zone of tetragonal lattice, obtained in LDA 
approximation cite{Nek1239},  (b) comparison of the total and partial
densities of states in $LaOFeAs$ and $PrOFeAs$ \cite{Nek1239}.} 
\label{la_pr_comp} 
\end{figure}

In Fig. \ref{la_pr_comp} (a) we show the comparison of electronic spectra of 
$LaOFeAs$ and $PrOFeAs$ \cite{Nek1239} in main symmetry directions in 
Brillouin zone. It can be seen that differences in spectra due the change of
rare -- earth ion (as well as a small change in lattice constants) are rather
small. In a narrow enough energy interval close to the Fermi energy, which
is relevant to superconductivity (of the order of $\pm$ 0.2 eV), these spectra
practically coincide.

This is also clearly seen from the comparison of the densities of states shown
in Fig. \ref{la_pr_comp} (b). In fact, densities of states of both compounds
close to the Fermi level are just the same (up to few percents). This is typical
also for the other compounds from the rare -- earth series
REO$_{1-x}$F$_x$FeAs \cite{Nek1239}.

The only noticeable difference in spectra of these systems with different
rare -- earth ions manifests itself in the growth of tetrahedral splitting
due to lattice compression, which appears at energies of the order of -1.5 eV 
for $d$-states of Fe and -3 eV for $p$-states of As.

From comparison of partial densities of states we can also see that the value 
of the density of states close to the Fermi level is determined almost entirely
by $d$-states of Fe (with very insignificant contribution from $p$-states of As).
In this sense we can say that all phenomena related to superconductivity in
these compounds take place in the square lattice of Fe within FeAs layer.

Naturally, most of these peculiarities of electronic spectra can be attributed
to the quasi -- two -- dimensional character of compounds under study. For
example, the insensitivity of electronic spectra to the type of rare -- earth
ion is simply due to the fact that electronic states of REO layers are far
from the Fermi level and $p$-states of O only weakly overlap with
$d$-states of Fe and $p$-states of As in FeAs layers. Accordingly, hybridization
of $d$-states of Fe and $p$-states of As is more significant, but still is not
very strong, as demonstrated by band structure calculations.

Thus, situation with rare -- earth substitutions in REOFeAs series seems to be
largely analogous to the similar one in cuprates like REBa$_2$Cu$_3$O$_{7-x}$, 
which were studied in the early days of HTSC research \cite{MDM,trst}. 
In these compounds electronic states of rare -- earth ions also do not overlap
with electronic states in conducting CuO$_2$ planes, which leads to the well
known fact of almost complete independence of superconducting $T_c\sim 92K$ 
on the type of rare -- earth RE=Y,Nd,Sm,Eu,Gd,Ho,Er,Tm,Yb,Lu,Dy \cite{trst},
with only two exceptions -- that of much lower $T_c\sim 60K$ in case of La and
complete absence of superconductivity in case of Pr based compound \cite{MDM}.

Similarly, almost identical electronic structure of iron oxypnictides like
REOFeAs with different RE in a wide enough energy interval around the Fermi
level seems to lead inevitably to approximately the same values of
superconducting transition temperature T$_c$ (in any BCS -- like microscopic
mechanism of pairing). Different rare -- earth ions just do not influence
electronic structure, at least in this energy interval around the Fermi level, 
and, accordingly, do not change the value of the pairing coupling constant.
Also, there is no special reasons to believe that the change of rare -- earth
ion will change much the phonon spectrum of these systems, as well as the
spectrum of magnetic excitations in FeAs layer.
 
Thus, we have a kind of rare -- earth puzzle --- in contrast to cuprate
series REBa$_2$Cu$_3$O$_{7-x}$ different rare -- earth substitutions in
REOFeAs series lead to rather wide distribution of $T_c$ values, from
$\sim$ 26K in case of La system to $\sim$ 55K in case of Nd and Sm.
At the moment we can propose two possible explanations of this puzzle:
 
\begin{enumerate}

\item{Different quality of samples (disorder effects) can lead to rather
wide distribution of $T_c$ values, as internal disorder can strongly influence
on the value of critical temperature, especially in case of anomalous
pairings (anisotropic $s$ - wave, even more so $p$ or $d$-wave pairing, 
triplet pairing etc.), which are widely discussed at present for FeAs 
superconductors \cite{mazin,aoki}. This possibility is qualitatively analogous
to the case of copper oxides, where $d$-wave pairing is realized, which is 
strongly suppressed by disorder. This argumentation was used e.g. to explain
lower values of $T_c$ in LaBa$_2$Cu$_3$O$_{7-x}$, which were attributed to
disorder in positions of La and Ba ions, as well as to oxygen vacancies
\cite{MDM}. In fact, this point of view is confirmed by the reports on the
synthesis of $LaO_{1-x}FeAs$ with $T_c\sim 41$K \cite{Lu3725}, as well as
by the synthesis of initially hypothetical \cite{Nek1239} $YO_{1-x}FeAs$ 
system, first with $T_c\sim 10$K \cite{Chong0288}, and later (synthesis under
high pressure) with $T_c\sim 46$K \cite{Yang3582}. In this last paper 1111
compounds based on Ho, Dy and Tb were also synthesized, leading to $T_c$  
values of the order of 50, 52 and 48K correspondingly. It seems quite
probable that the best prepared samples of 1111 series may achieve the values
of $T_s\sim 55$K, as obtained in most ``favorable'' cases of Sm and Nd. 
This stresses the necessity of systematic studies of disorder effects in
new superconductors.
}
 
\item{However, we cannot exclude also the other physical reasons of 
$T_c$ distribution in REOFeAs series, connected e.g. with spin ordering on
rare -- earth ions (like Ce, Pr, Nd, Sm, Gd, which possess magnetic moments).
Above we have already mentioned the unusually high coupling of these moments
with moments on Fe. Besides, the temperature of rare -- earth moments ordering
is known experimentally to be an order of magnitude higher than in
REBa$_2$Cu$_3$O$_{7-x}$ systems \cite{MDM}, which also indicates to rather
strong magnetic couplings, which may significantly influence e.g. on spin
fluctuation spectrum in FeAs layers (and $T_c$ values in case of magnetic
mechanisms of pairing \cite{mazin,aoki}).}

\end{enumerate}

\begin{figure}
\includegraphics[clip=true,width=0.62\columnwidth,angle=270]{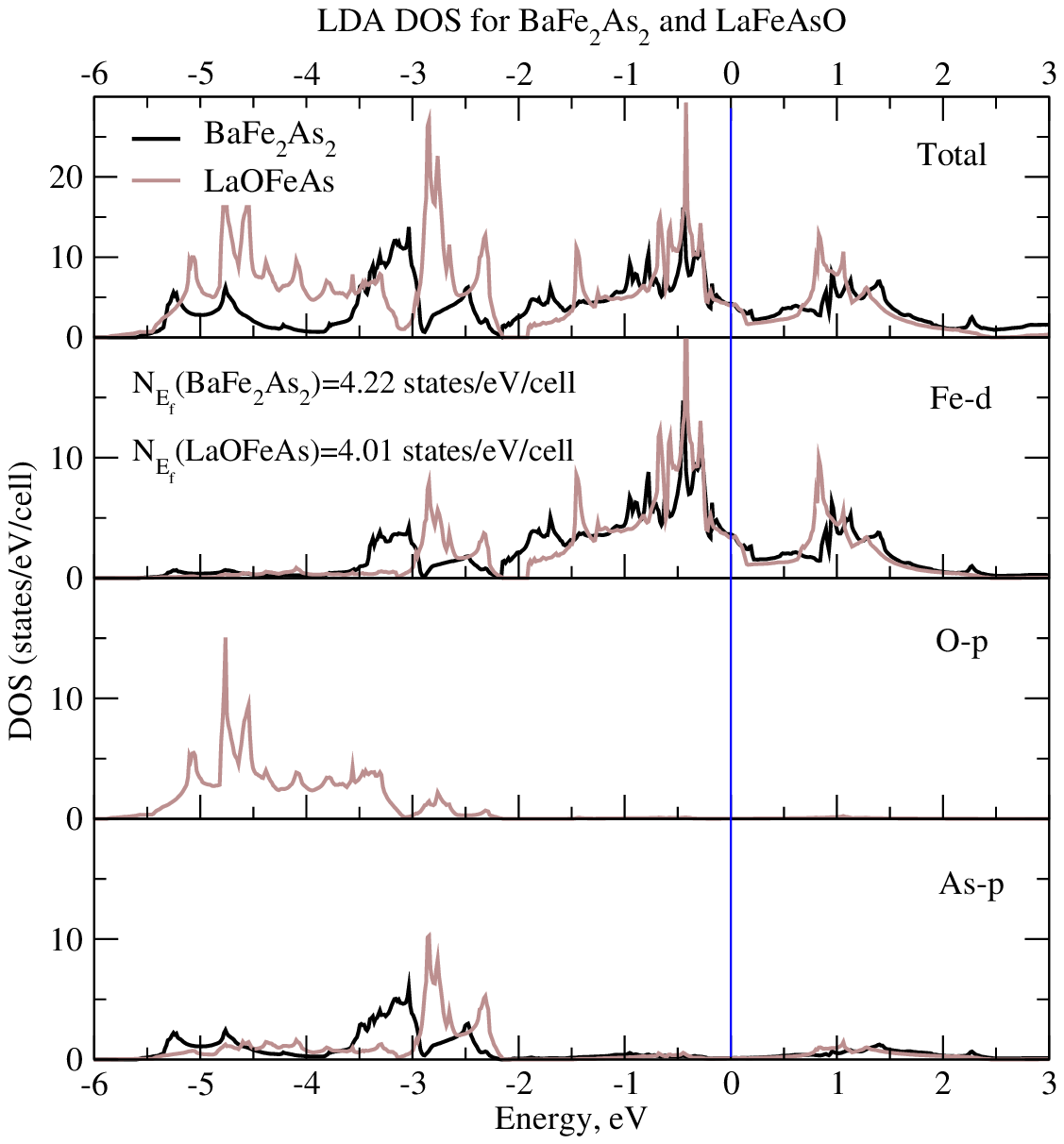}
\includegraphics[clip=true,width=0.55\columnwidth]{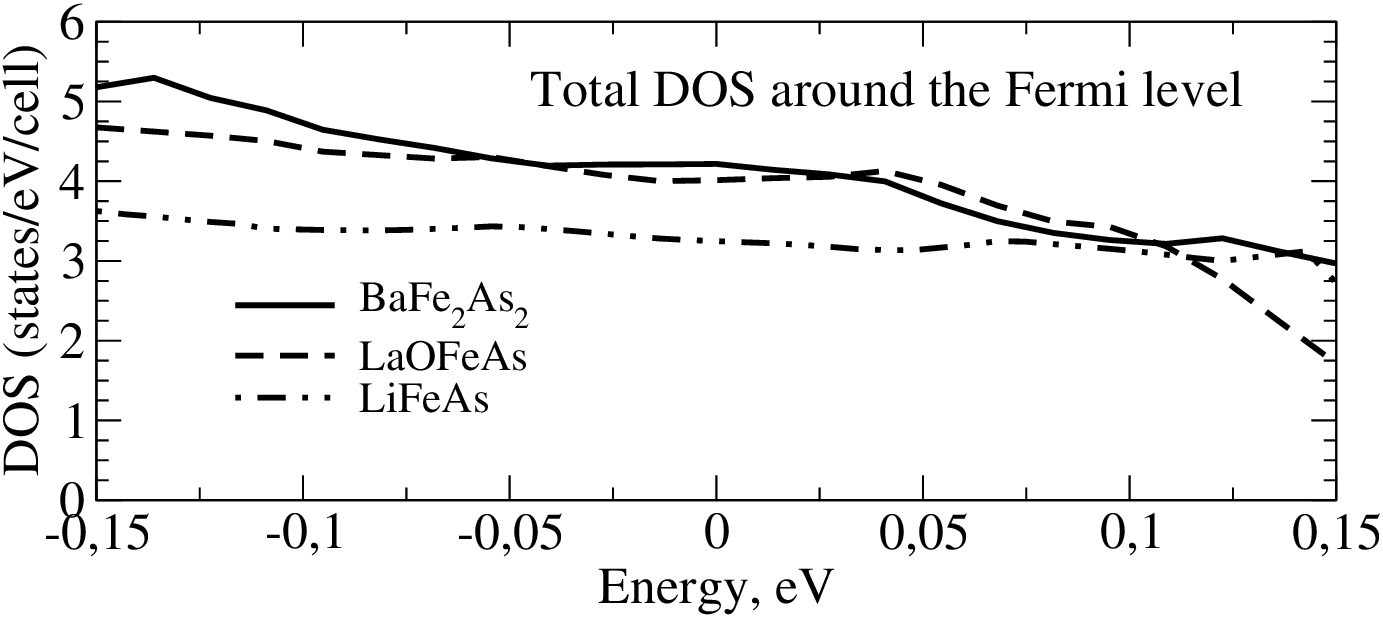}
\caption{(a) comparison of the total electronic density of states and partial
densities of states in $LaOFeAs$ and $BaFe_2As_2$  \cite{Nek2630}, 
(b) comparison of densities of states in a narrow energy interval around the 
Fermi level in $LaOFeAs$, $BaFe_2As_2$ and $LiFeAs$
\cite{Nek1010}.} 
\label{la_ba_comp} 
\end{figure}

\begin{figure}[!h]
\includegraphics[clip=true,width=0.60\columnwidth]{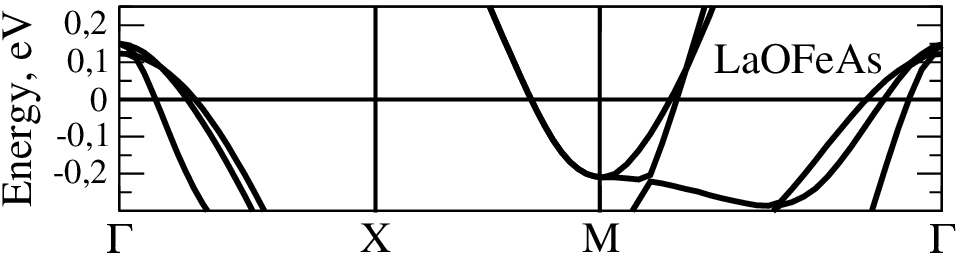}
\includegraphics[clip=true,width=0.25\columnwidth]{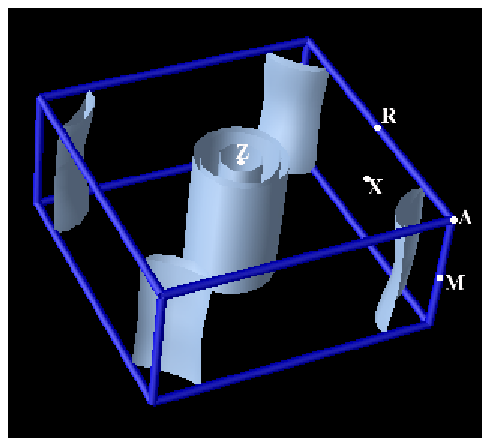}
\includegraphics[clip=true,width=0.60\columnwidth]{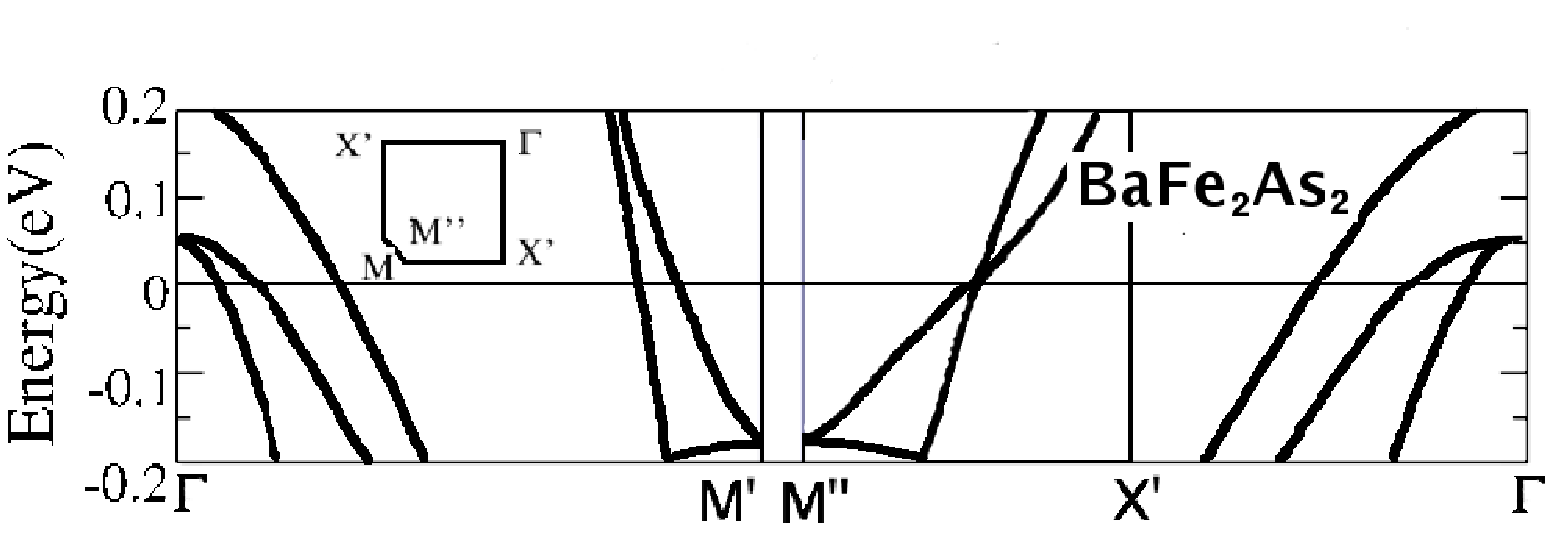}
\includegraphics[clip=true,width=0.25\columnwidth]{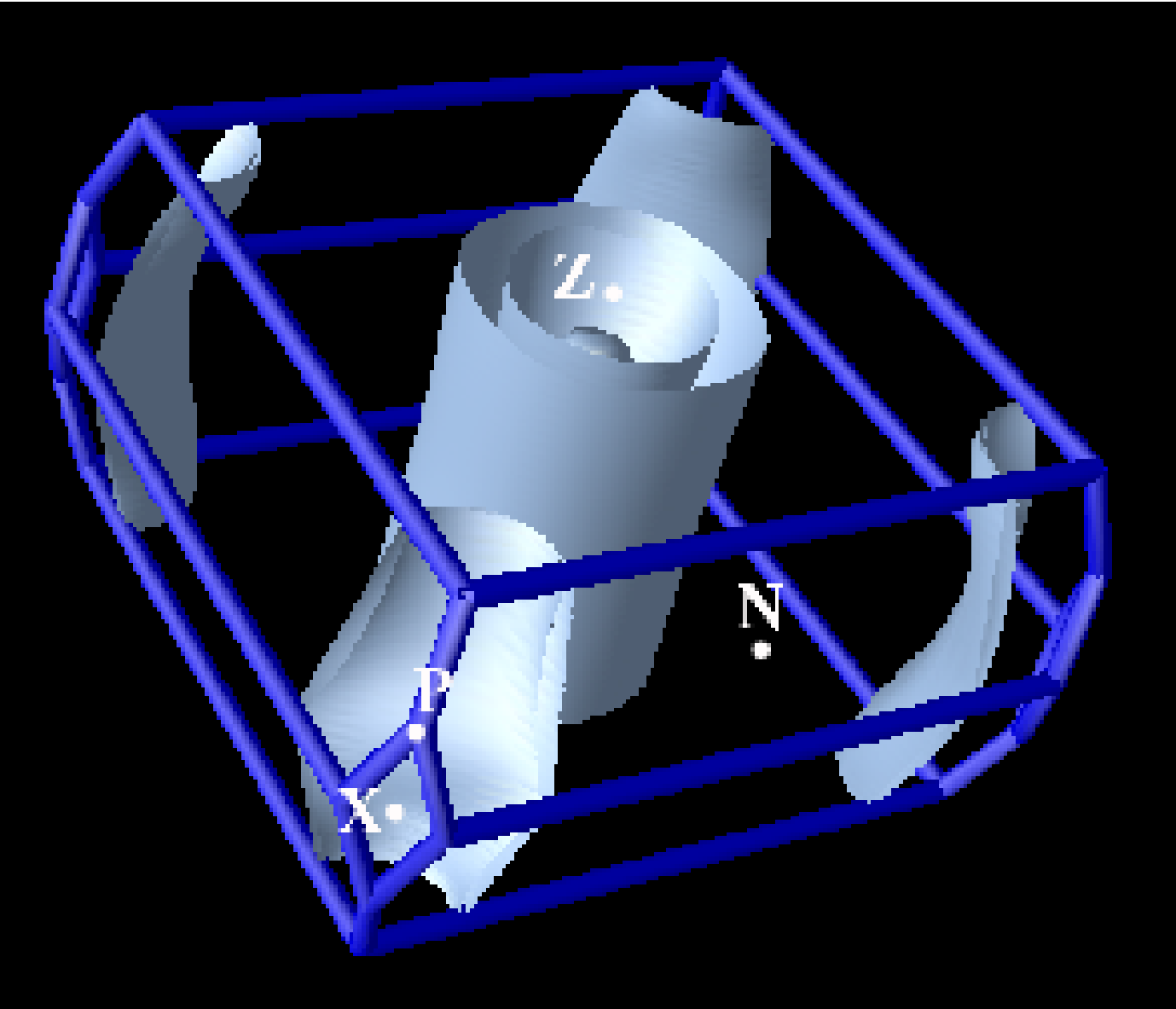}
\includegraphics[clip=true,width=0.60\columnwidth]{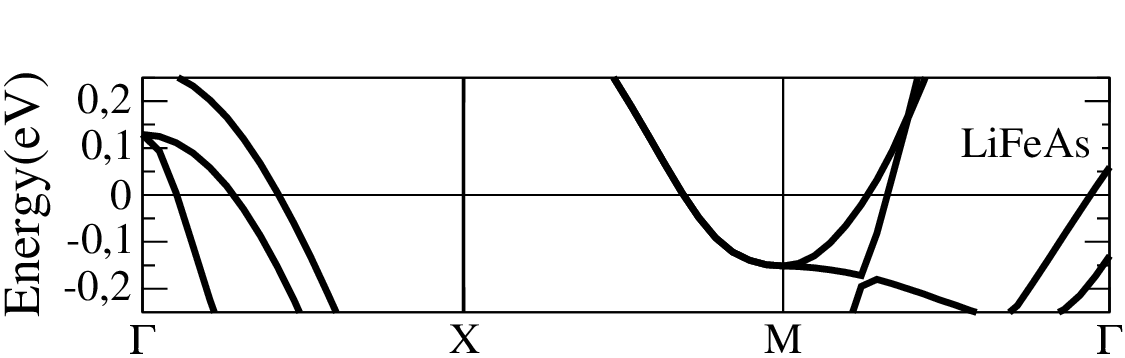}
\includegraphics[clip=true,width=0.25\columnwidth]{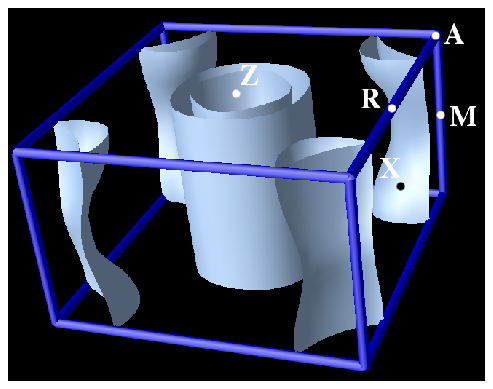}
\caption{Left -- electronic spectrum of $LaO_{1-x}F_xFeAs$, $BaFe_2As_2$ and
$LiFeAs$ in a narrow interval of energies close to the Fermi level, relevant
to the formation of superconducting state. Right -- Fermi surfaces of these
compounds \cite{Nek1239,Nek1010}.} 
\label{F_level_comp} 
\end{figure}

In Fig. \ref{la_ba_comp} (a) we show the comparison of total electronic 
density of states and partial densities of states in $LaOFeAs$ and $BaFe_2As_2$
\cite{Nek2630}. It is seen that again we have almost the same values of DOS'es
in an energy interval around the Fermi level, relevant to
superconductivity. In more details we can see it in Fig. \ref{la_ba_comp} (b),
where densities of states are compared in a narrow energy interval
($\pm$ 0.15 eV) close to the Fermi level in $LaOFeAs$, $BaFe_2As_2$ and 
$LiFeAs$ \cite{Nek1010}. The densities of states in this energy interval are
almost energy independent (quasi two -- dimensionality!) and only slightly 
different (though, in principle, we can notice some correlation of these
values of DOS at the Fermi level and the values of $T_c$ in these compounds).

The bandwidth of $d$-states of Fe in BaFe$_2$As$_2$ is approximately
0.3 eV larger than in LaOFeAs, which may be connected with shorter
Fe-As bonds, i.e. with larger Fe-$d$-As-$p$ hybridization. In both compounds 
bands crossing the Fermi level are formed mainly from three
$d$-orbitals of Fe with $t_{2g}$ symmetry -- $xz$, $yz$, $xy$. 
Similar situation is realized also in case of $LiFeAs$.

In Fig. \ref{F_level_comp} (left) we show electronic dispersions in high
symmetry directions in all three main classes of new superconductors
(1111, 122 ¨ 111) in a narrow ($\pm$ 0.2 eV) energy interval around the Fermi
level, where superconducting state is formed \cite{Nek2630,Nek1010}. 
It can be seen that electronic spectra of all systems in this energy interval 
are very close to each other. In general case, the Fermi level is crossed
by five bands, formed by $d$-states of Fe. Of these, three form hole -- like 
Fermi surface pockets close to $\Gamma$ -- point, and the other two --
electron -- like pockets at the corners of Brillouin zone (note that Brillouin
zones of 1111, 111 and 122 systems are slightly different due to differences
in lattice symmetry).

It is clear that this kind of a band structure leads to similar Fermi 
surfaces of these compounds --- appropriate calculation results are shown
in the right side of Fig. \ref{F_level_comp}: there are three hole--like
cylinders at the center of Brillouin zone and two electron -- like at the
corners. Almost cylindrical form of the Fermi surface reflects quasi
two -- dimensional nature of electronic spectrum in new superconductors.
The smallest of hole -- like cylinders is usually neglected in the analysis of
superconducting pairings, as its contribution to electronic properties is rather
small (smallness of its phase space volume). At the same time, from the general
picture of electronic spectrum it is clear that superconductivity is formed in
multiple band system with several Fermi surfaces of different (electron or 
hole -- like) nature, which is drastically different from the simple
one -- band situation in cuprates. Below we shall see that results of LDA
calculations of electronic structure correlate rather well with experiments on
angle resolved photoemission (ARPES).
 
LDA calculations of band structure of $\alpha$-FeSe were performed in a recent
paper \cite{Singh4312}. Dropping the details we note that the results are 
qualitatively quite similar to those described above for 1111, 122 and 111
systems. In particular, the form of Fermi surfaces is qualitatively the same,
while conduction bands near the Fermi level are formed from $d$-states of Fe.

First calculations of band structure of Sa(Ca)FFeAs compounds were done in
Refs. \cite{Nek_Sr,Iv_Sr}. Naturally enough, the band structure and Fermi
surfaces in these compounds are also very similar to those obtained before
for REOFeAs systems. The only difference is slightly more pronounced quasi
two -- dimensional nature of the spectra in these compounds.

\subsection{``Minimal'' model}

Relative simplicity of electronic spectrum of FeAs superconductors close to
the Fermi level (Fig. \ref{F_level_comp}) suggests a possibility to formulate
a kind of ``minimal''  analytic or semi -- analytic model of the spectrum
(e.g. in the tight -- binding approximation), which will provide 
semi -- quantitative description of electrons in the vicinity of the Fermi
level, sufficient for theoretical description of superconducting state and
magnetic properties of FeAs planes. Up to now several variants of such model
were already proposed \cite{Scal,Cao3236,Aoki3325,Korsh1793}. Below we shall
limit ourselves to brief description of the simplest (and most crude) variant
of such model proposed by Scalapino et al. \cite{Scal}.

\begin{figure}[!h]
\includegraphics[clip=true,width=0.6\columnwidth]{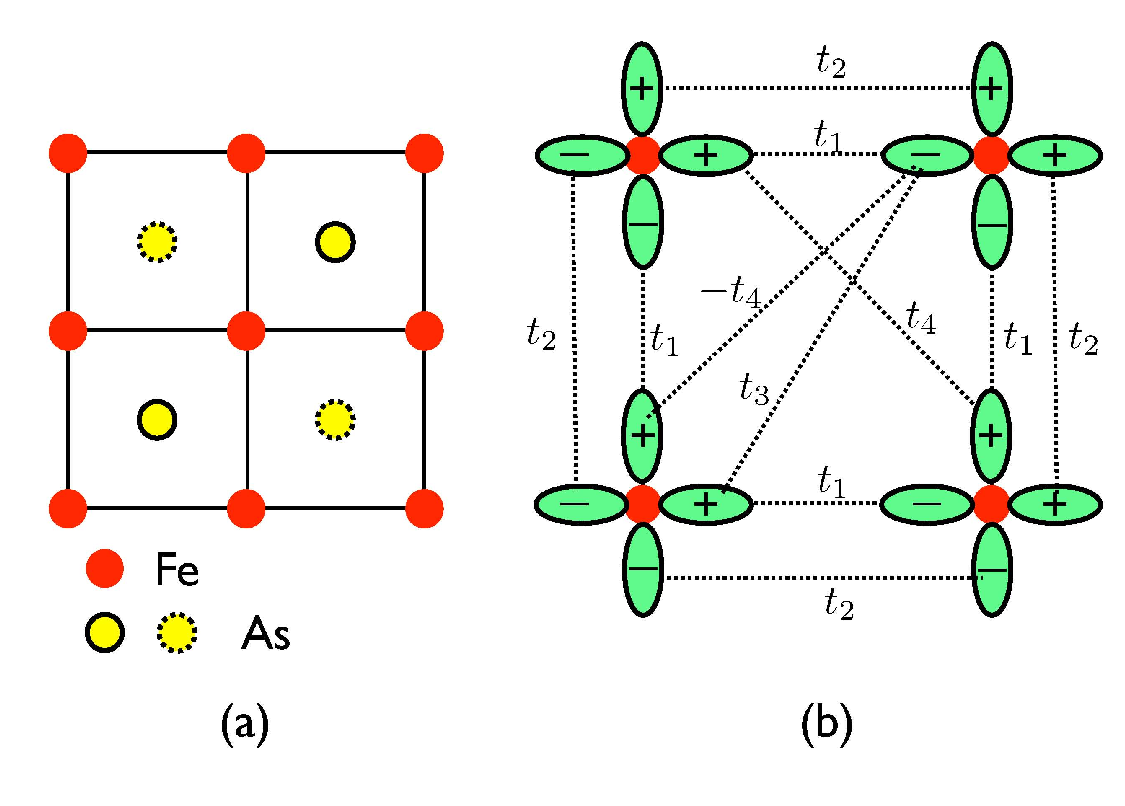}
\caption{(a) Fe ions in FeAs layer form quadratic lattice, containing two 
Fe ions and two As ions in elementary cell. As ions are placed above
(filled circles) or under (dashed circles) centers of the squares formed by Fe,
(b) transfer integrals taken into account in two -- orbital ($d_{xz},d_{yz}$)
model on the square lattice of Fe. Here $t_1$  -- transfer integral between
nearest $\sigma$-orbitals, and  $t_2$ -- transfer integral between nearest
$\pi$-orbitals. Also taken into account are transfer integrals $t_4$ between
different orbitals and $t_3$ between identical orbitals on the second
nearest neighbors. Shown are projections of $d_{xz},d_{yz}$ orbitals on $xy$ 
plane \cite{Scal}.} 
\label{min_str} 
\end{figure}

Schematic structure of a single FeAs layer is shown in Fig. \ref{min_str} (a).
Fe ions form the square lattice, surrounded by As layers, which also form
the square lattice and are placed in the centers of squares of Fe ions and
are displaced upwards or downwards with respect to Fe lattice in a checkerboard
order, as shown in Fig. \ref{min_str} (a). This leads to two inequivalent
positions of Fe, so that there are two ions of Fe and two As ions in an 
elementary cell. LDA calculations described in the previous section show that
the main contribution to electronic density of states in a wide enough energy
interval around the Fermi level as due to $d$ - states of Fe. Thus, we can
consider the simplified model which primarily takes into account three orbitals
of Fe, i.e. $d_{xz}$, $d_{yz}$ and $d_{xy}$ (or $d_{x^2-y^2}$, which is just
the same). As a further simplification the role of $d_{xy}$ (or $d_{x^2-y^2}$) 
orbitals can be effectively taken into account introducing transfer integrals
between $d_{xz},d_{zy}$ orbitals on the second nearest neighbors. Accordingly,
we may consider the square lattice with two degenerate ``$d_{xz},d_{yz}$" 
orbitals at each site, with transfer integrals shown in Fig. \ref{min_str} (b).
As we shall see below, such model produces the picture of two -- dimensional
Fermi surfaces of FeAs layer, which is in qualitative agreement with LDA
results.

\begin{figure}[!h]
\includegraphics[clip=true,width=0.8\columnwidth]{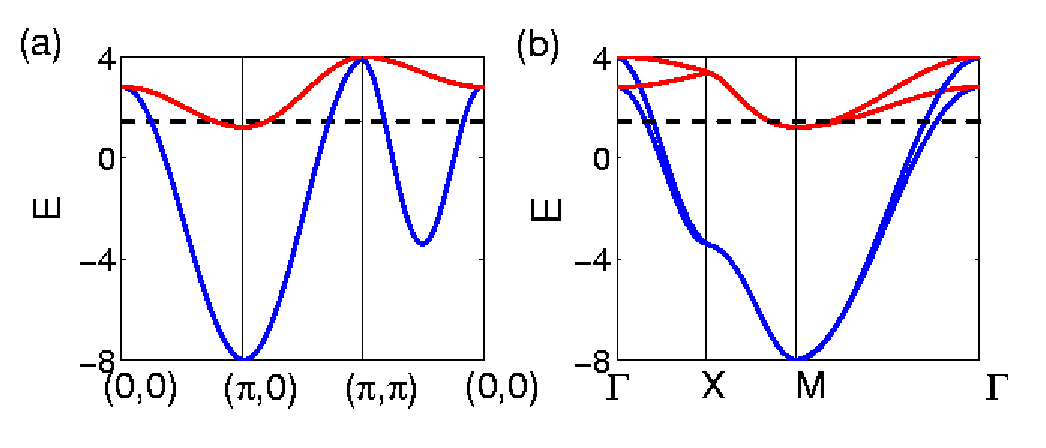}
\includegraphics[clip=true,width=0.4\columnwidth]{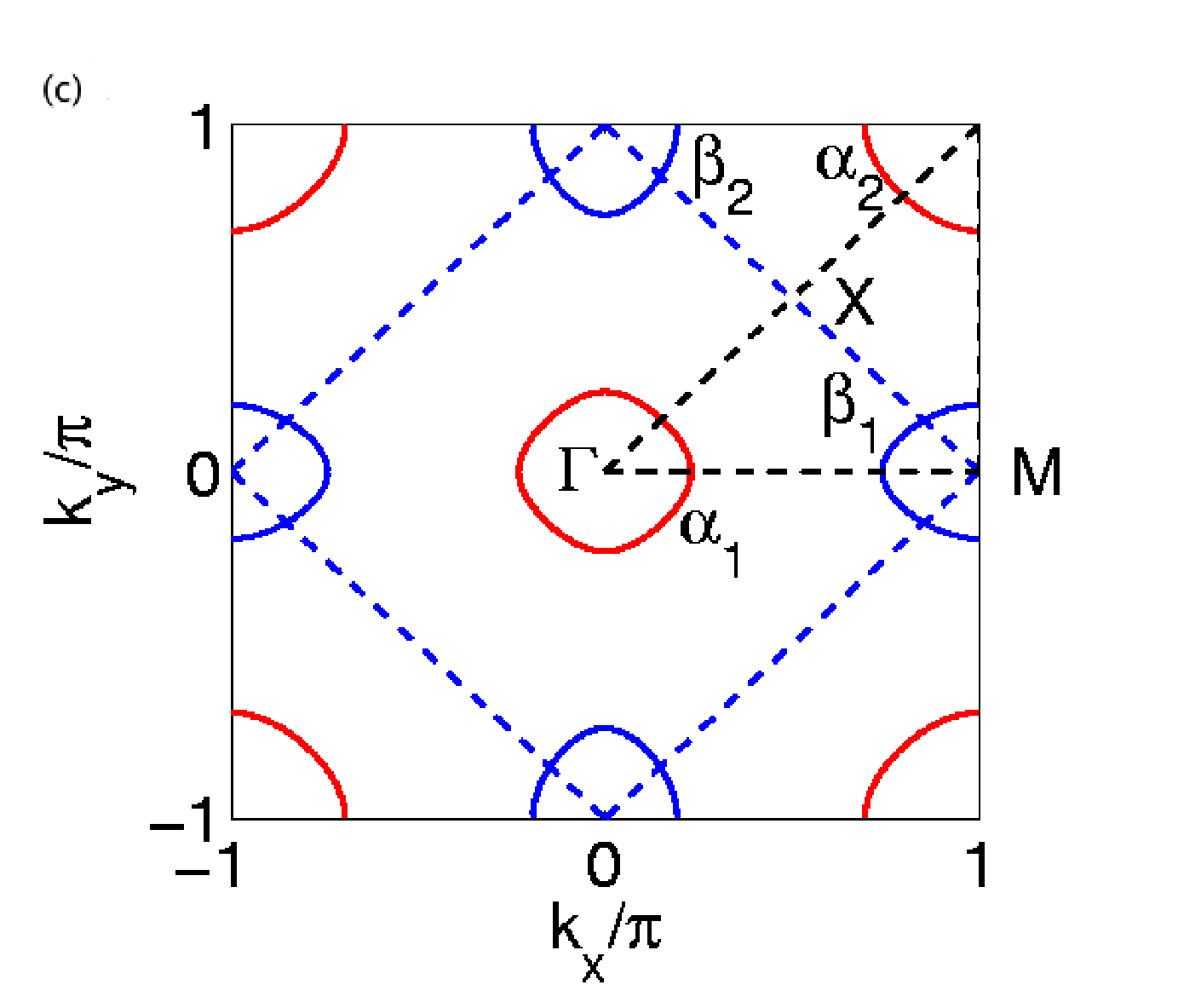}
\includegraphics[clip=true,height=0.25\textheight,width=0.4\columnwidth]{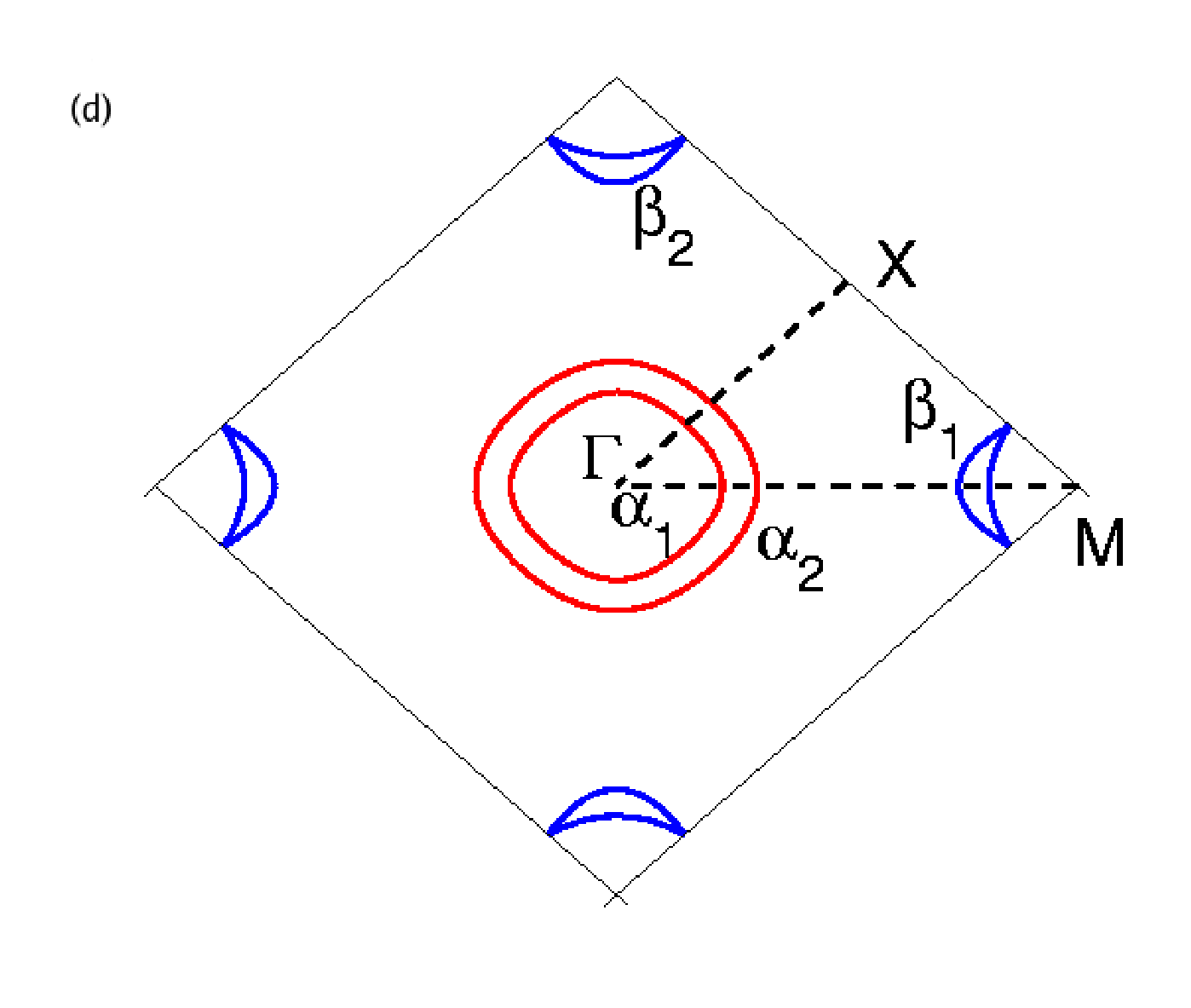}
\caption{(a) electronic spectrum in two -- orbital model with transfer integrals 
$t_1=-1,t_2 = 1.3, t_3=t_4=-0.85$ (in units of $\vert t_1 \vert$) and chemical
potential $\mu = 1.45$, along directions 
$(0,0)\rightarrow (\pi/a,0)\rightarrow (\pi/a,\pi/a)\rightarrow (0,0)$,
(b) the same spectrum in downfolded Brillouin zone, with appropriate
redefinition of $\Gamma, X, M$ points,
(c) Fermi surface in two -- orbital model in the Brillouin zone, corresponding 
to one Fe ion per elementary cell.
$\alpha_{1,2}$ -- hole -- like Fermi surfaces, defined by
$E_-(\bm k_F) = 0$, $\beta_{1,2}$ -- electron -- like Fermi surfaces, defined
by $E_+(\bm k_F) = 0$. Dashed lines show the Brillouin zone for the case of two
Fe ions in elementary cell, 
(d) Fermi surfaces in downfolded Brillouin zone, corresponding to two
Fe ions in elementary cell \cite{Scal}.} 
\label{min_spectr} 
\end{figure}

For analytical description of this model it is convenient to introduce a
two -- component spinor
\begin{equation}
\psi_{{\bf k}s}={d_{xs}({\bf k})\choose d_{ys}({\bf k})} \label{eq:1}
\end{equation}
where $d_{xs}({\bf k})$ ($d_{ys}({\bf k})$) annihilates $d_{xz}$ ($d_{yz}$) 
electron with spin $s$ and wave vector ${\bf k}$. Tight -- binding Hamiltonian
can be written as
\begin{equation}
H_0=\sum_{{\bf k}s}\psi^+_{{\bf k}s}\left[\left(\varepsilon_+({\bf k})
-\mu\right)1+\varepsilon_-({\bf k})\tau_3+\varepsilon_{xy}({\bf k})\tau_1\right]
\psi_{{\bf k}s}, \label{Htb}
\end{equation}
where $\tau_i$ are Pauli matrices,
\begin{eqnarray}
  \varepsilon_\pm({\bf k}) & = & \frac{\varepsilon_x({\bf k})\pm
\varepsilon_y({\bf k})}{2}, \nonumber\\
    \varepsilon_x({\bf k}) & = & -2t_1\cos k_xa-2t_2\cos k_ya-4t_3\cos k_xa\cos k_ya,\nonumber\\
    \varepsilon_y({\bf k}) & = & -2t_2\cos k_xa-2t_1\cos k_ya-4t_3\cos k_xa\cos k_ya,\nonumber\\
    \varepsilon_{xy}({\bf k}) & = & -4t_4\sin k_xa\sin k_ya.\nonumber
\end{eqnarray}
\label{eq:3}
Finally, the single -- particle Green's function in Matsubara representation
takes the following form
\begin{equation}
\hat{G}_s ({\bf k}, i \omega_n ) = \frac{\left( i \omega_n - \epsilon_+({\bf k}) 
\right) \hat{1} - \epsilon_-({\bf k}) \hat{\tau}_3 - \epsilon_{xy} ({\bf k}) 
\hat{\tau}_1}{\left(i \omega_n - E_+({\bf k}) \right) \left(i \omega_n - 
E_-({\bf k}) \right)}
\end{equation}
where
\begin{equation}
E_{\pm}({\bf k}) = \epsilon_+({\bf k}) \pm \sqrt{\epsilon_-^2({\bf k}) +
\epsilon_{xy}^2({\bf k})} - \mu\label{Ek}
\end{equation}
In Fig. \ref{min_spectr} (a) we show the appropriate electronic spectrum for 
the values of transfer integrals $t_1=-1,t_2 = 1.3, t_3=t_4=-0.85$ 
(in units of $\vert t_1 \vert$).

Now take into account the fact, mentioned above, that in real FeAs layer there
are two Fe ions per elementary cell. Accordingly, the Brillouin zone is
twice smaller and the spectrum must be folded down into this new zone 
as shown in Fig. \ref{min_spectr} (b). In Fig. \ref{min_spectr} (c,d) 
we show Fermi surfaces, which are obtained in this simplified model of
electronic spectrum. In large Brillouin zone, corresponding to the lattice with
one Fe ion per elementary cell, there are two hole -- like pockets, denoted as
$\alpha_1$ and $\alpha_2$, which are defined by the equation $E_-(\bm k) = 0$, 
and two electron -- like pockets $\beta_1$ and, defined by $E_+(\bm k) = 0$. 
To compare with the results of band structure calculations (LDA) these Fermi
surfaces should be folded down to twice smaller Brillouin zone, corresponding
to two Fe ions in elementary cell of the crystal, which is shown by dashed
lines in Fig. \ref{min_spectr} (c). The result of such downfolding is shown in
Fig. \ref{min_spectr} (d). We can see that the Fermi surfaces obtained in this
way are in qualitative agreement with the results of LDA calculations
(only the third, less relevant, small hole -- like pocket in the zone center
is absent). Despite its crudeness this model of the spectrum, proposed in
Ref. \cite{Scal}, is quite appropriate for qualitative analysis of electronic
properties of FeAs superconductors.

\subsection{Angle resolved photoemission spectroscopy (ARPES)}

At the moment there are already a number of papers, where electronic spectrum
and Fermi surfaces in new superconductors were studied using angle resolved
photoemission spectroscopy (ARPES) 
\cite{Liu2147,Yang2627,Liu3453,Liu4806,Zhao0398,Ding0419,Wray2185,Zhang2738,
Zab2454,Evt4455,Sato3047}, reliable method, which proved its effectiveness
in HTSC -- cuprates \cite{Schr,Shen}, due by the way to quasi
two -- dimensional nature of electronic spectrum in these systems.
In fact, for FeAs superconductors ARPES studies immediately provided valuable
information clarifying the general form of the spectrum, Fermi surfaces and
the values and peculiarities of superconducting gaps. Note that for 1111
systems up to now there is only one ARPES work \cite{Liu2147}, which is due to
the absence of good single crystals, so that all the remaining studies were
performed on single crystals of 122 systems. Below we shall discuss the
results of some of these papers in more details.

\begin{figure}[!h]
\includegraphics[clip=true,width=0.8\columnwidth]{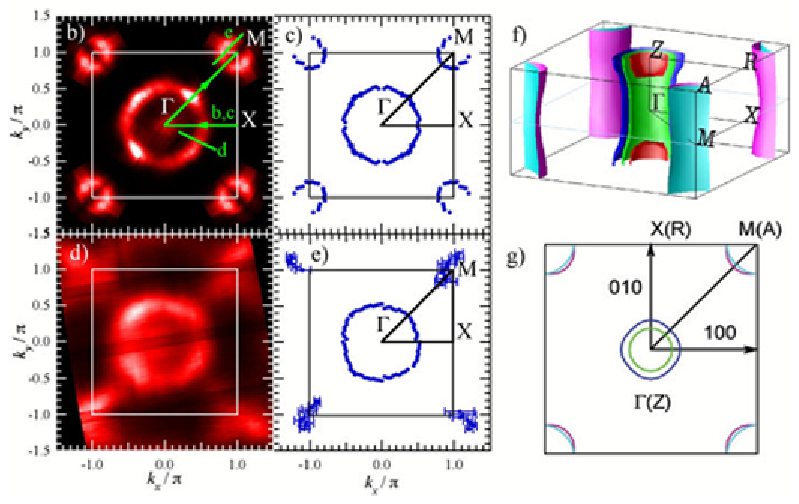}
\caption{(b) ARPES intensity map and (c) Fermi surfaces determined for
NdO$_{0.9}$F$_{0.1}$FeAs, at photon energy 22 eV and $T=70$K, (d,e) -- 
the same, but for photon energy 77 eV \cite{Liu2147}, (f,g) three -- dimensional
two -- dimensional picture of Fermi surfaces in NdOFeAs, obtained from LDA
calculations \cite{Liu2147}.} 
\label{ARPES_1111} 
\end{figure}

ARPES measurements in Ref. \cite{Liu2147} were performed on a single crystal
of micron sizes (of the order of 200$\times$200$\times$30 $\mu$m, with 
$T_c\sim 53$K). In Fig. \ref{ARPES_1111} we show ARPES intensity maps
(which is proportional to spectral density) in NdO$_{0.9}$F$_{0.1}$FeAs
system, allowing to determine the form of Fermi surfaces in two -- dimensional
Brillouin zone, corresponding to FeAs planes. It can be seen that the general
qualitative picture is in reasonable agreement with the results of LDA
calculations of the band structure, though around the $\Gamma$ point only one
hole -- like cylinder is resolved, and electron -- like cylinders in the
corners (point M) are resolved rather poorly. The value of superconducting gap
on the hole -- like cylinder estimated from ARPES spectra was determined to be
of the order of 20 meV \cite{Liu2147}, corresponding to $2\Delta/T_c\sim 8$.

\begin{figure}[!h]
\includegraphics[clip=true,width=0.4\columnwidth]{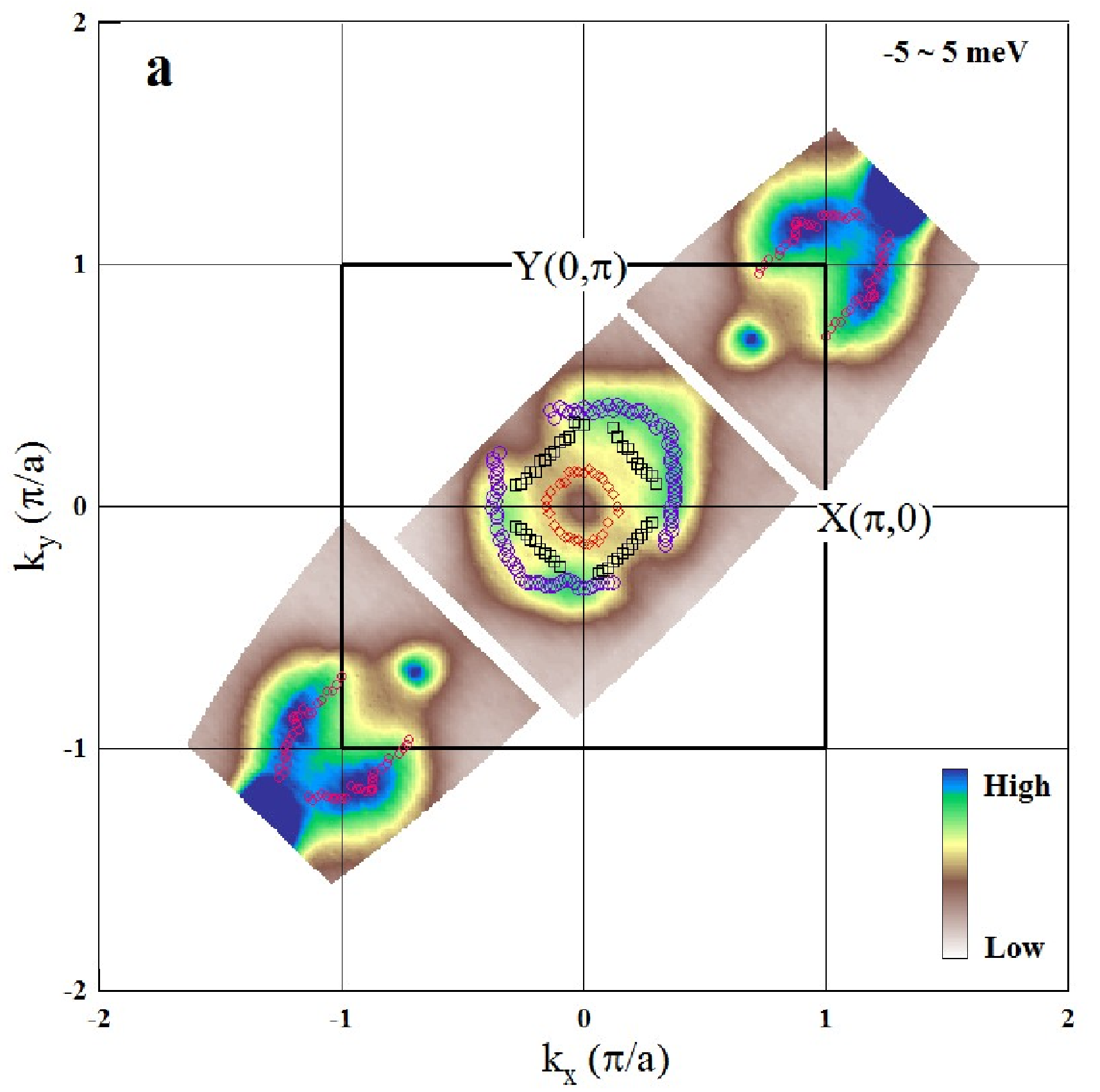}
\caption{ARPES map of Fermi surfaces in $(Sr,K)Fe_2As_2$ \cite{Liu4806}.} 
\label{ARPES_122a} 
\end{figure}

In Ref. \cite{Liu4806} ARPES measurements were performed on a single crystal of
superconducting (Sr,K)Fe$_2$As$_2$ with $T_c=21$K. ARPES map of the Fermi
surfaces obtained is shown in Fig. \ref{ARPES_122a}. In contrast to other
works, here the authors succeeded in resolving all three hole -- like
cylinders around the point $\Gamma$, in complete agreement with majority of
LDA calculations of the spectrum. Resolution in the corners of Brillouin zone
(point M) was much poorer, so that the topology of electronic sheets of the
Fermi surface remained unclear.

\begin{figure}[!h]
\includegraphics[clip=true,width=0.6\columnwidth]{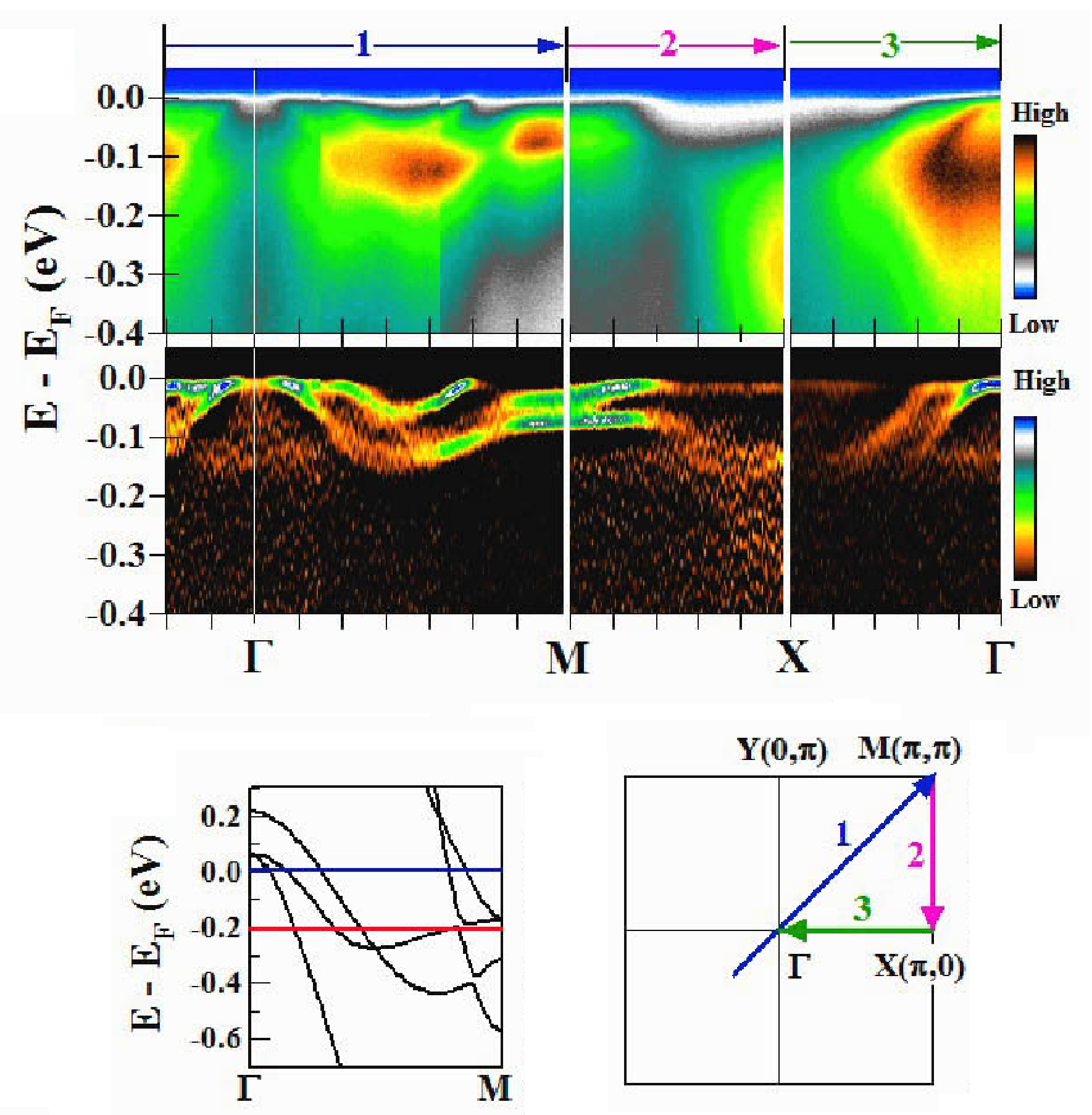}
\caption{Energy bands determined from ARPES data in 
$(Sr,K)Fe_2As_2$\cite{Liu4806}, in upper part of the figure -- raw ARPES data,
below -- their second derivative, allowing to follow dispersion curves.
Below to the left -- band spectrum obtained in Ref. \cite{Nek2630}, and to the
right -- directions in the Brillouin zone, along which measurements have been
made.}  
\label{ARPES_122b} 
\end{figure}

In Fig. \ref{ARPES_122b} we show energy bands in high symmetry directions of
Brillouin zone obtained from ARPES measurements \cite{Liu4806}. First of all,
these data correlate well with the results of band structure calculations of
Ref. \cite{Nek2630}, with the account of Fermi level shift downwards in 
energy by $\sim$ 0.2 eV (in complete accordance with hole doping of
superconducting sample). On the other hand, rather significant band narrowing
in comparison with LDA results is also observed, which can be attributed to 
strong electronic correlations (cf. below).

\begin{figure}[!h]
\includegraphics[clip=true,width=0.7\columnwidth]{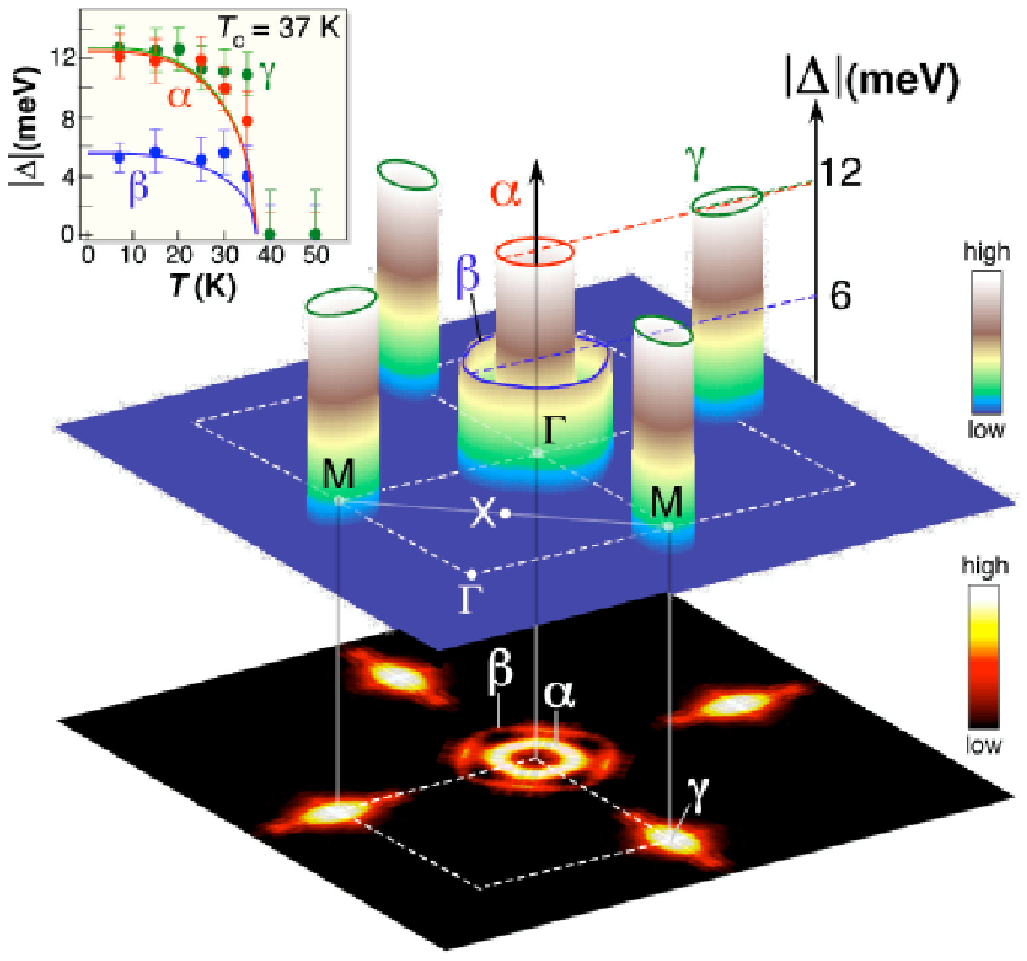}
\caption{Schematic three -- dimensional picture of superconducting gap in 
$Ba_{0.6}K_{0.4}Fe_2As_2$ according to ARPES measurements \cite{Ding0419}. 
Below -- Fermi surfaces (ARPES intensity), at the insert above -- temperature
dependences of gaps on different sheets of the Fermi surface.} 
\label{ARPES_122c} 
\end{figure}

In Ref. \cite{Ding0419} for the first time were studied in detail not only Fermi
surfaces of Ba$_{0.6}$K$_{0.4}$Fe$_2$As$_2$ ($T_c=37$K) \footnote{The third
small hole -- like cylinder around point $\Gamma$ was not observed, probably due
to insufficient resolution of ARPES spectra.}, but also ARPES measurements were 
done of superconducting gaps (and their temperature dependence) on different
sheets of the Fermi surface. Schematically, results of these measurements
are shown in Fig. \ref{ARPES_122c}. Two superconducting gaps were discovered --
a large one ($\Delta\sim$ 12 meV) on small hole -- like cylinder around point
$\Gamma$ and also on electron -- like cylinders around point M, and a small one
($\Delta\sim$ 6 meV) on big hole -- like cylinder around $\Gamma$ point. 
Both gaps close at the same temperature coinciding with $T_c$, have no zeroes
and are practically isotropic on appropriate sheets of the Fermi surface. 
Accordingly, $2\Delta/T_c$ ratio is different on different sheets (cylinders)
and formally are consistent with both strong (large gap, the ratio is 7.5) and
weak (small gap, the ratio is 3.7) coupling. These results correspond to the
picture of generalized $s$ - wave pairing, to be discussed below.                                                

Quite similar results on gap values on different sheets of the Fermi surface
were obtained also in Ref. \cite{Wray2185} via ARPES measurements on single
crystals of (Sr/Ba)$_{1-x}$K$_x$Fe$_2$As$_2$. Also in this work the electron
dispersion was measured by ARPES in rather wide energy interval, which
demonstrated characteristic ``kinks'', attributed to conduction electrons
interaction with collective oscillations (phonons or spin excitations),
allowing determination of electron velocity close to the Fermi level:
$v_F\sim 0.7\pm 0.1$ eV \AA, so that using the value of the ``large'' gap
$\Delta\sim 12\pm 2$ meV gives an estimate of coherence length
(the size of Cooper pairs) $\xi_0=\frac{\hbar v_F}{\Delta} < 20$ \AA, i.e. 
relatively small value (compact pairs).
 
Rather unexpected results for the topology of Fermi surfaces were obtained in
Refs. \cite{Zab2454,Evt4455} for Ba$_{1-x}$K$_x$Fe$_2$As$_2$, where Fermi
surface sheets close to the M point were discovered to have a characteristic
``propeller'' -- like form, which does not agree with any LDA calculations. 
Measurements of superconducting gap in Ref. \cite{Evt4455} has given for the 
``large'' gap on an ``internal'' hole -- like cylinder around $\Gamma$ point
the value of $\Delta\sim$ 9 meV and the same value on ``propellers'' close to
M points, which corresponds to the value of $2\Delta/T_c=6.8$. On an ``external''
hole -- like cylinder around point $\Gamma$ the value $\Delta<$ 4 meV was
obtained, corresponding to $2\Delta/T_c<$ 3.

Note also Ref. \cite{Sato3047}, where ARPES measurements were performed on
``maximally doped'' variant of Ba$_{1-x}$K$_x$Fe$_2$As$_2$ system with
$x=1$, i.e. on superconducting KFe$_2$As$_2$ ($T_c=3$K). It was discovered that
the form of hole -- like Fermi surfaces (sheets) surrounding $\Gamma$ point
is qualitatively the same as in Ba$_{1-x}$K$_x$Fe$_2$As$_2$ with $x=0.4$ 
($T_c$=37K), while electron -- like cylinders surrounding M points are just
absent. This is a natural consequence of the downward shift of the Fermi level
due to hole doping (cf. Fig. \ref{F_level_comp}, where it is clearly seen that
electronic branches of the spectrum are above the Fermi level in case of its
big enough downward energy shift). Besides that, similarly to Ref. 
\cite{Liu4806}, the observed bands are significantly (2-4 times) narrower, than
obtained in all LDA -- type calculations. This fact, as already noted above, is
most probably due to strong enough electronic correlations (see next section).
The absence of electronic pockets of the Fermi surface leads to the
disappearance of interband mechanisms of pairing (see below) and corresponding
significant lowering of $T_c$.

Summing up, it can be noted that the results of ARPES studies of Fermi
surfaces and electronic spectrum of FeAs superconductors are in rather 
satisfactory with LDA calculations of band structure. Remaining inconsistencies
are most probably due to unaccounted role of electronic correlations and,
sometimes, due to insufficient resolution in ARPES experiments.

Unfortunately, up to now there are almost no experiments on determination of
Fermi surfaces from low temperature quantum oscillations (like de Haas --
van Alfen). We can only note Ref. \cite{Lon4726}, where non superconducting
phase of SrFe$_2$As$_2$ was studied and oscillation periods discovered
corresponded to small Fermi surface pockets, which apparently can be  
attributed to hole -- like cylinders close to $\Gamma$ point, almost completely
``closed'' by antiferromagnetic gap. In Ref. \cite{Cold4890} quantum
oscillations were studied in LaOFeP and found to be in agreement with Fermi 
surface predicted by LDA calculations, with additional two times mass 
enhancement.

\subsection{Correlations (LDA+DMFT)}

LDA calculations of electronic spectrum neglect potentially strong effects
of local electron correlations (Hubbard -- like repulsion of electrons), which
can be naturally expected in bands, formed mainly from $d$-states of Fe in
FeAs layers. Most consistent approach to the analysis of such correlations
at present is considered to be the dynamical mean field theory (DMFT)
\cite{MetzVoll89,vollha93,pruschke,Iz95,georges96}, including its variant taking
into account the LDA band structure of real systems (LDA+DMFT) 
\cite{PT,KotSav,Held}.

LDA+DMFT approach was used to calculate electronic structure of FeAs compounds 
in Refs. \cite{Haule1279,Anis3283,Anis0547,Miy2442,Anis2629}. In all of these,
except Ref. \cite{Miy2442}, only LaOFeAs system was studied, while in Ref.
\cite{Miy2442} the authors analyzed the series of REOFeAs (RE=La,Ce,Pr,Nd).

\begin{figure}[!h]
\includegraphics[clip=true,width=0.6\columnwidth]{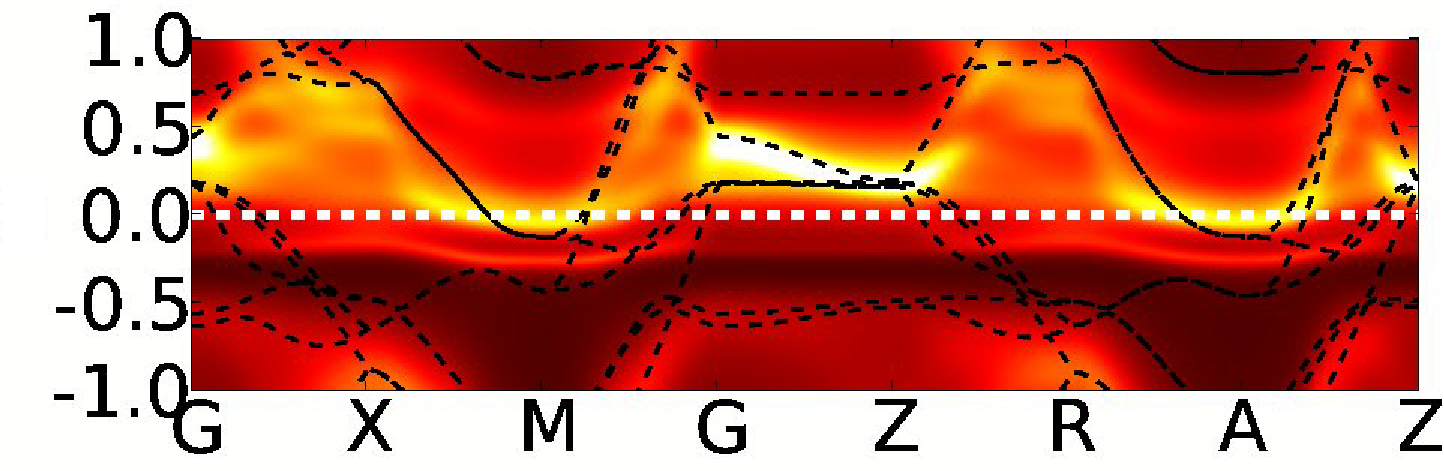}
\caption{Momentum dependence of spectral density (positions of the maxima of
define effective dispersion of quasiparticles) in LDA+DMFT approximation
for LaOFeAs with 10\% doping. Dashed lines -- results of LDA approximation 
\cite{Haule1279}.} 
\label{DMFT_spectr} 
\end{figure}

In Ref. \cite{Haule1279} LDA+DMFT was used to calculate spectral density and
optical conductivity of LaO$_{1-x}$F$_x$FeAs. The value of Hubbard repulsion
was taken to be $U=$4 eV, while Hund (exchange) coupling was assumed to be
equal to $J=$0.7 eV
\footnote{DMFT impurity problem was solved by quantum Monte Carlo (QMC)
with temperature taken to be 116 K.}. 
In Fig.  \ref{DMFT_spectr} we show the results for momentum dependence of the
spectral density in high symmetry directions in the Brillouin zone. Position
of the maxima of spectral density describe the effective dispersion of
(damped) quasiparticles, which may be compared with results of LDA 
calculations (infinite lifetime quasiparticles), which are also show.
Doping was described in virtual crystal approximation. It can be seen that
electron correlations lead to a significant (3-5 times) enhancement of effective
masses and strong damping of quasiparticles. System remains metallic, though a
kind of a ``bad'' metal with strongly renormalized quasiparticle amplitude
(residue at the Green's function pole) $Z\sim 0.2-0.3$. 

The general picture of the spectrum shown in Fig. \ref{DMFT_spectr} can be
qualitatively compared with ARPES data, shown in the lower part of Fig.
\ref{ARPES_122b}. Though the experiment was done on another system certain
similarity is obvious --- conduction bands are significantly narrowed
(in comparison with LDA results), while in the energy region around -0.5 eV 
the bands are just ``destroyed'' and there a kind of energy gap there.

According to Ref. \cite{Haule1279} even rather small enhancement of Hubbard
repulsion up to $U=$4.5 eV makes transforms this system into a kind of
Mott insulator with an energy gap at the Fermi level. Thus it was concluded
that systems under consideration are characterized by intermediate correlations
and are close to Mott insulator, making them partly similar to HTSC cuprates.   
At the same time, contrary to cuprates, prototype (undoped) compounds are
metals, not Mott insulators.

Quite different conclusion was reached by the authors of Ref. \cite{Anis3283},
which performed formally the same type of calculations (by the same methods and
with the same choice of parameters). In this work it was claimed that even
the use of $U$ values up to $U$ ¤® 5 eV does not transform a system into Mott
insulator and changes in spectral density (electron dispersion) and density of
states due to correlations are rather insignificant.

At present the reasons for such drastic differences of results obtained by two
leading groups making LDA+DMFT calculations are unclear.

In general it should be noted that ``ab initio'' nature of LDA+DMFT approach
is rather relative, as e.g. the value of Hubbard interaction $U$ is, in fact,
a kind of (semi)phenomenological parameter. There exists a number of 
``ab initio'' approaches to calculate its value, of which most consistent is
assumed to be the so called method of constrained random phase approximation
(constrained RPA) \cite{CRPA,RPAC}. Within this approach the value of local
Hubbard repulsion is calculated from the ``bare'' Coulomb repulsion $V$
using the RPA expression:
\begin{equation}
U=\frac{V}{1-V\Pi_r}
\label{CRPA}
\end{equation}
where $\Pi_r$ is polarization operator, taking into account screening by
electrons from the outside of (correlated) bands of interest. For example, if
we limit ourselves only to the analysis of bands formed by $d$-states of Fe, 
we have to use $\Pi_r=\Pi-\Pi_d$, where $\Pi_d$ is polarization operator
calculated only on $d$-states. It is clear that the value of $U$ defined in this
way depends on the assumed scheme of calculations (the account of different
groups of screening electrons).

For FeAs compounds this problem was studied in Ref. \cite{Miy2442}.
The values of different interaction parameters entering to LDA+DMFT calculation
scheme, obtained by exclusion of different groups of states from screening of
the ``bare'' Coulomb interaction, are given in Table IV.

\vskip 0.5cm
Table IV. ``Bare'' and partly screened Coulomb interactions V and U, and
Hund coupling J for LaOFeAs, obtained in Ref. \cite{Miy2442} by exclusion 
different groups of states (listed in the first column) from screening. 
\begin{center}
\begin{tabular}{|c|c|c|c|c|}
\hline
 & V (eV) &  U (eV) & J (eV) \tabularnewline
\hline
\hline
\emph{d} & 15.99 & 2.92 &  0.43 \tabularnewline
\hline
\emph{dpp} & 20.31 & 4.83 &  0.61 \tabularnewline
\hline
\emph{d-dpp} & 20.31 & 3.69  & 0.58\tabularnewline
\hline
\end{tabular}
\end{center}
\vskip 0.5cm

We can see that the values of interaction parameters, entering LDA+DMFT
scheme, can change in rather wide intervals.

The same questions were discussed in Refs. \cite{Anis0547,Anis2629}
(within constrained DFT approach \cite{CDFT}), with authors coming to the
conclusion that effective parameters of Coulomb (Hubbard) repulsion are
relatively small, if we limit ourselves within the basis of $d$-states of Fe.
Accordingly, LDA+DMFT calculations, performed in these works, produced
electronic structure, which is only slightly different from LDA results.

In the opinion of authors of Refs. \cite{Anis3283,Anis0547,Anis2629}, 
their conclusion about the smallness of electronic correlations in FeAs
systems is confirmed by experiments on X-Ray absorption qnd emission
spectroscopy of Ref. \cite{Kur0668}, however it is in sharp contrast with
ARPES data \cite{Liu4806,Sato3047}, which definitely indicate rather strong
renormalization of electronic spectrum due to correlations. It should be
stressed that data on the topology of Fermi surfaces are, in fact, insufficient
to make any judgements on the role of correlations --- Fermi surfaces in 
LDA+DMFT approach are just the same as in LDA approximation. It is important to
study in detail electronic dispersion, bandwidths and quasiparticle damping
far enough from the Fermi level, as well as quasiparticle residue Z.
In our opinion, ARPES measurements have many advantages here and can help to
solve this important problem.

Summarizing, the question of the role of electronic correlations in new
superconductors is still under discussion. It is most probable that correlations
in these systems are of intermediate strength between typical metals and systems
like Mott insulators, so that we are dealing here with the state of
{\em correlated} metal \cite{pruschke,georges96}.

\subsection{Spin ordering: localized or itinerant spins?}

In this section we shall briefly consider theoretical ideas on the nature
of antiferromagnetic ordering in undoped FeAs compounds and closely related
problem of structural transition from tetragonal to orthorhombic phase.
Strictly speaking, these questions are slightly outside the scope of our
review (superconductivity in FeAs compounds), so that our presentation will be
very short.

Possibility of antiferromagnetic ordering in systems under consideration was
noted even before direct neutronographic observations discussed above.
Already in one of the earliest papers on electronic structure of iron
oxypnictides \cite{mazin}, as well as during the analysis of ``minimal''
two -- band model \cite{Scal}, it was stressed that there is an approximate 
``nesting'' of hole -- like Fermi surfaces around $\Gamma$ point and 
electron -- like Fermi surfaces around M point. It is most easily seen e.g.
in Fig. \ref{min_spectr} (c) --- the shift of hole -- like cylinder by vector
$(\pi,0)$ or $(0,\pi)$ leads to its approximate coincidence with electron --
like cylinder. Direct calculations show \cite{mazin,Scal} that this fact leads
to formation of rather wide, but still quite noticeable, peak in static
magnetic susceptibility $\chi_0({\bf q})$ (determined by appropriate loop
diagram) at ${\bf q}=(\pi, 0)$ and ${\bf q}=(0,\pi)$. In its turn, this may
lead to antiferromagnetic instability towards formation of spin density wave
(SDW) with appropriate wave vector, at least in case of strong enough exchange
interaction. Above we have seen that precisely this type of spin ordering is
observed experimentally in FeAs layers
\cite{Cruz0795,Zhao2528,Chen0662,Huang2776,Zhao1077,Chen3950}. Also observed
rather small values of magnetic moments on Fe also give an evidence for the
itinerant nature of magnetism (SDW). It is also natural to assume that in
doped (superconducting) compounds well developed spin fluctuation of SDW type
persist, giving a way to pairing interaction of electrons.

\begin{figure}[!h]
\includegraphics[clip=true,width=0.25\columnwidth]{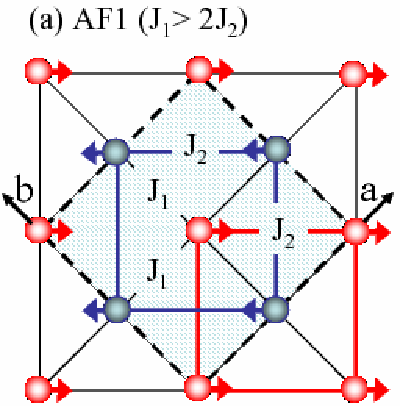}
\includegraphics[clip=true,width=0.25\columnwidth]{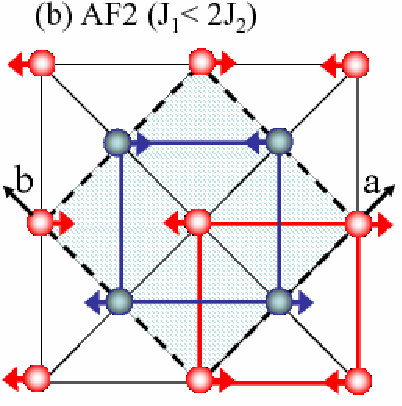}
\caption{Two alternative spin configurations in antiferromagnetic FeAs layer: 
(a) -- antiparallel spins on nearest neighbors, (b) -- antiparallel spins on
on second nearest neighbors \cite{Yil2252}.} 
\label{exch_int} 
\end{figure}

At the same time, magnetic ordering in FeAs layers can be analyzed also within
more traditional approach, based on the qualitative picture of localized spins
on Fe ions, interacting via the usual Heisenberg exchange between nearest and
second nearest neighbors. Such analysis was performed e.g. in Ref. 
\cite{Yil2252} for LaOFeAs, as well as ``ab initio'' calculations of appropriate
exchange integrals and values of magnetic moments. For us more important are
qualitative aspects of this analysis, which are illustrated by Fig. 
\ref{exch_int} \cite{Yil2252}.

In this figure we show two possible antiferromagnetic configurations of spins
in FeAs layer. In the experiments \cite{Cruz0795} the AF2 (see Fig. \ref{exch_int}) 
type spin structure is observed, which is realized when inequality
$J_1<2J_2$ holds, where $J_1$ is an exchange integral between nearest, 
and $J_2$ -- between second nearest neighbors, with both integrals assumed
positive (antiferromagnetism). Direct (FP-LAPW) calculation of ground state
energy of LaOFeAs, performed in Ref. \cite{Yil2252}, confirmed the greater
stability of this state. AF2 configuration can be considered as two 
interpenetrating antiferromagnetic square sublattices, shown by different
colors in Fig. \ref{exch_int}. In this case each Fe ion is placed in the
center of antiferromagnetically ordered call, so that the mean molecular field 
on its spin is just zero. In this case each sublattice may freely 
(with no cost in energy) rotate with respect to the other. This is a situation
of complete frustration. It is well known that such a state is usually unstable
with respect to structural distortions. Thus, it can be expected that a
structural distortion will appear in our system, making spins at a pair
of Fe ions closer to each other, while the pair on the other side of a square
slightly farther apart. This is just the type of distortion (tetra -- ortho
transition) observed in the experiment \cite{Cruz0795}. Direct calculations of
total energy confirm this guess \cite{Yil2252}.

However, the general situation with the nature of magnetic ordering and
structural transition in FeAs layers is rather far from being completely
clear. ``Ab initio'' calculations as a rule produce strongly overestimated
values of magnetic moments of Fe, while the relative stability of magnetic
structures strongly depends on details of methods used. Apparently, this is due
to the itinerant nature of magnetism in these systems. Detailed discussion
of these problems can be found in interesting papers \cite{maz,maz_jh}, 
where an original qualitative picture of strong magnetic fluctuations is 
proposed, allowing, in the authors opinion, to explain all the anomalies of
magnetic properties.


\section{Mechanisms and types of pairing}

After the discovery of high -- temperature superconductivity in iron based
layered compounds a dozens of theoretical papers appeared with different
proposals on possible microscopic mechanisms and types of Cooper pairing in
these systems. The review of all these papers here seems impossible and below
we shall deal only with very few, in our opinion, most important works.

\subsection{Multi -- band superconductivity}

Main peculiarity of new superconductors is their multiple -- band nature.
Electronic structure in a narrow enough energy interval around the Fermi level
is formed practically only from  $d$-states of Fe. Fermi surface consists of
several hole -- like and electron -- like cylinders and on each its ``own''
energy gap can be formed. From the general point of view this situation is not
new and was already analyzed in the literature \cite{Mosk}. However, for the 
case of specific band structure typical for FeAs layers we need an additional
analysis. 

\begin{figure}[!h]
\includegraphics[clip=true,width=0.4\columnwidth]{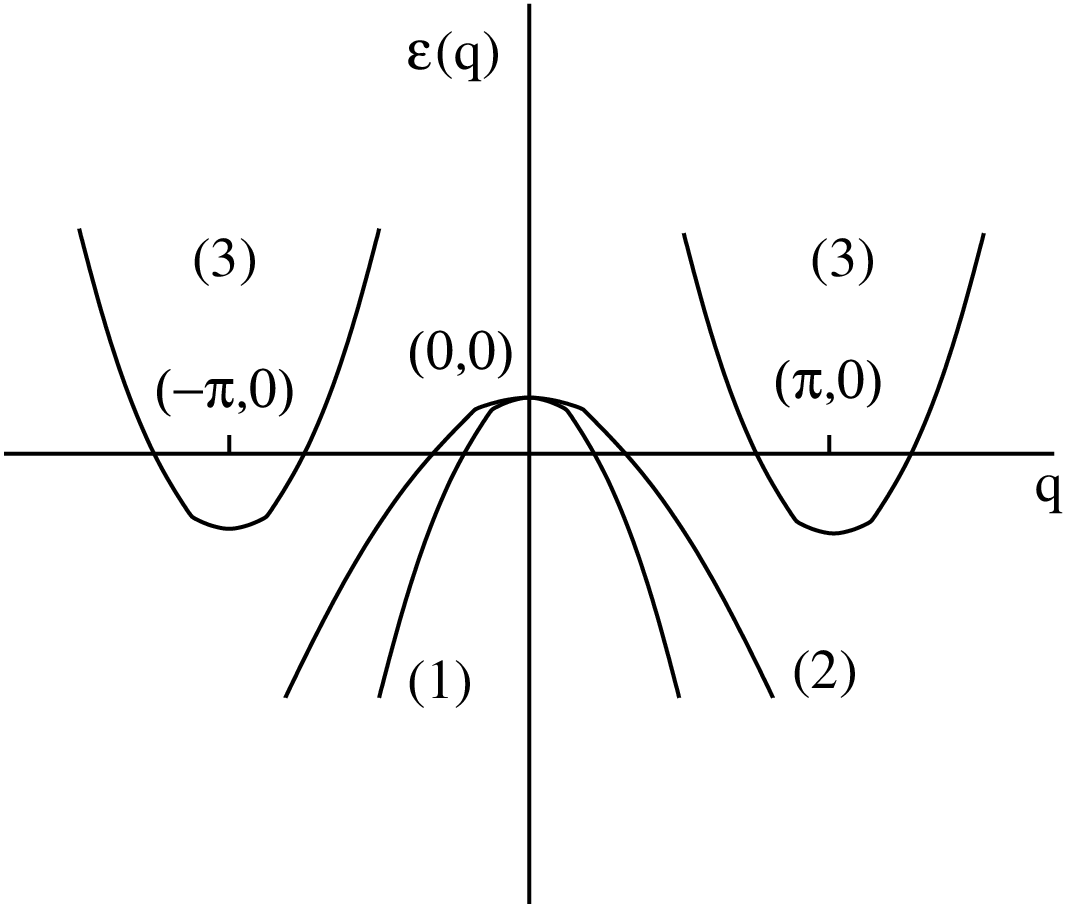}
\includegraphics[clip=true,width=0.4\columnwidth]{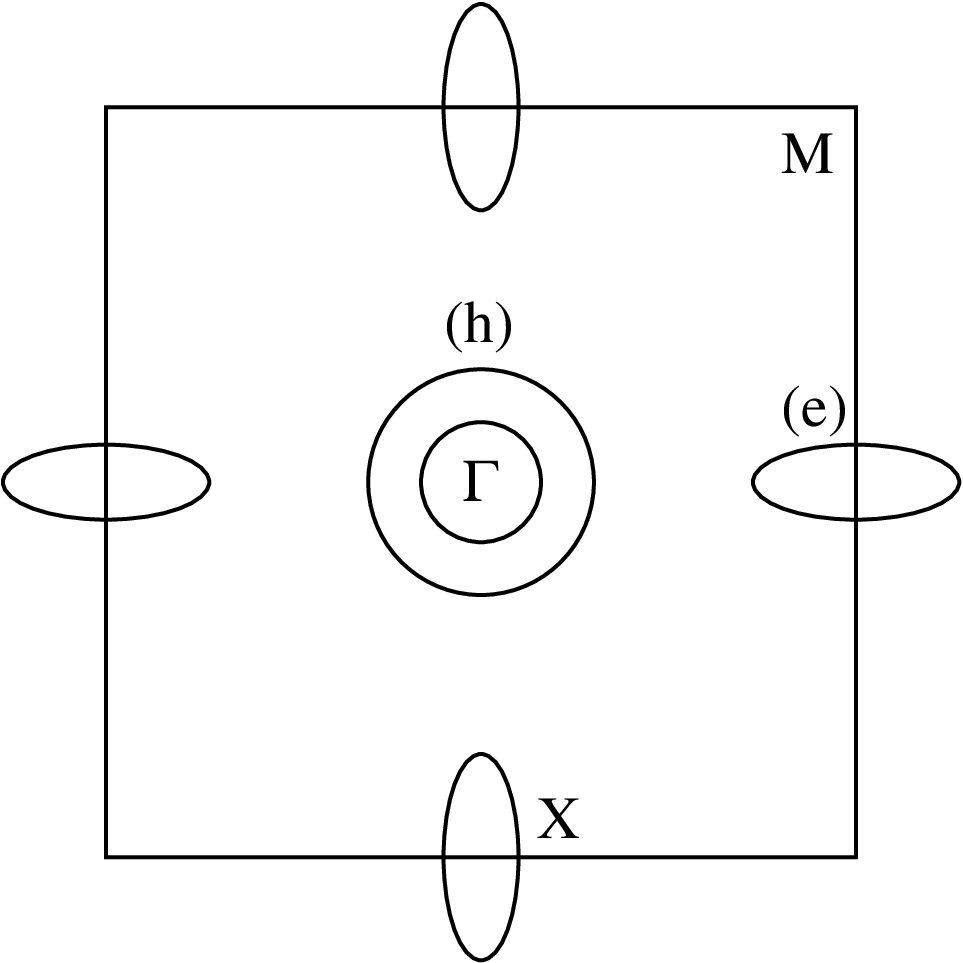}
\caption{Schematic view of electronic spectrum (a) and Fermi surfaces for 
LaOFeAs in extended band picture.  Around point $\Gamma$ there are two
hole -- like sheets, while electron -- like sheets are around
$X$ points \cite{Gork}.} 
\label{barz_gork} 
\end{figure}

In general enough formulation this problem was considered in a paper by
Barzykin and Gor'kov \cite{Gork} and some of the results will be presented below.
Typical electronic spectrum of FeAs layered systems in the relevant (for
superconductivity) energy interval was shown in Fig. \ref{F_level_comp}. 
Similar, in principle, spectrum was obtained also in the 
``minimal'' model of Ref. \cite{Scal}. An oversimplified view of this spectrum
is shown in Fig.  \ref{barz_gork} \cite{Gork}.

This form of Fermi surfaces corresponds to stoichiometric (undoped) composition
of these compounds, and electrons and holes occupy the same volumes
(compensated semi -- metal)\footnote{In fact, just this form of the spectrum
was assumed in numerous papers on excitonic instability and excitonic insulator
\cite{KK,KM,Halp,Kop,Volk}. This instability may be considered as an alternative
explanation of antiferromagnetic ordering and structural transition in these
systems \cite{Gork}.} Electronic doping shrinks hole -- like pockets, while
hole doping --- electron -- like pockets.

In Ref. \cite{Gork} a symmetry analysis of possible types of superconducting 
order parameter was performed, along the lines of Refs. \cite{VG,GV}, and also
the explicit solutions of BCS equations were found for the system with 
electronic spectrum shown in Fig. \ref{barz_gork} . 

Let $\Delta_i (\bm{p})$ be a superconducting order parameter (gap) on the
$i$-th sheet of the Fermi surface. The value of $\Delta_i ({\bf p})$ is 
determined by self -- consistency equation for the anomalous Gor'kov's
function $F_i(\omega_n, {\bf p})$:  
\begin{equation} 
\Delta_i ({\bf p})=T \sum_{j; \omega_n} \int 
V^{i,j}({\bf p}-{\b p'}) d {\bf p'} F_j(\omega_n, {\bf p'}) \label{gapeq} 
\end{equation}
where $V^{i,j}({\bf p}-{\bf p'})$ is the pairing interaction. If Fermi
surface pockets are sufficiently small (as it seems to be in FeAs systems),
transferred momentum ${\bf p}-{\bf p'}$ within each pockets is also small and
we can replace $V^{i,j}({\bf p}-{\bf p'})$ by $V^{i,j}(0)$, which is favorable
for the formation of momentum independent gaps.

Pairing BCS interaction in this model can be represented by the following
matrix:
\begin{equation}
V=\left(\begin{array}{cccc} 
u & u & t & t \\
u & u & t & t \\ 
t & t & \lambda & \mu \\
t & t & \mu & \lambda 
\end{array}\right).
\label{mmatr}
\end{equation}
where $\lambda=V^{eX,eX}=V^{eY,eY}$ defines interaction on an electron --
like pockets at point $X$, $\mu=V^{eX,eY}$ connects electrons from different
pockets at points $(\pi, 0)$ and $(0, \pi)$,  $u=V^{h1,h1}=V^{h2,h2}=V^{h1,h2}$ 
characterizes BCS interaction on two hole -- like pockets\footnote{An assumption
that these couplings are the same seems to us to be too rigid.}, surrounding the
point $\Gamma$, while $t=V^{h,eX}=V^{h,eY}$ connects electrons from points
$X$ and $\Gamma$. 

Critical temperature of superconducting transition $T_c$ is determined by the
solution of the system of linearized gap equations:
\begin{equation} 
\Delta_{i}=\sum_{j}\bar{V}^{i,j} 
\Delta_{j}\ln{\frac{2 \gamma \bar{\omega}}{\pi T_c}\,}\,, \label{gapeqlin} 
\end{equation}
where $\bar{\omega}$ is the usual cut -- off frequency of logarithmic
divergence in Cooper channel, and 
\begin{equation}
\bar{V}^{i,j} \equiv - \frac{1}{2}\, V^{i,j} \nu_{j},
\label{intmatr}
\end{equation}
where $\nu_j$ is the density of states on the $j$-th sheet (pocket) of the 
Fermi surface.

Introducing an effective coupling constant $g$ and writing down $T_c$ as:
\begin{equation}
T_c = \frac{2 \gamma \bar{\omega}}{\pi}\,e^{- 2/g}, 
\label{TC1} 
\end{equation}
we obtain solutions of three types:

1) solution corresponding to $d_{x^2 - y^2}$ symmetry, when gap on different
pockets at points $X$ change signs, while gaps on hole -- like pockets are zero:

\begin{equation}
\Delta_1=\Delta_2 = 0, \ \ \Delta_3=-\Delta_4 =\Delta,
\end{equation}

\begin{equation}
g=(\mu - \lambda)\nu_3. 
\label{dwve}
\end{equation}

Possibility of such solution follows also from the general symmetry analysis
\cite{Gork}.

2) two solutions, corresponding to the so called $s^{\pm}$ -- pairing, 
when gaps at points $X$ have the same sign, while gaps on Fermi surfaces,
surrounding point $\Gamma$ may have opposite sign, and 

\begin{eqnarray}
\label{swaveg}
2 g_{+,-} &=& - u (\nu_1 + \nu_2) - (\lambda + \mu) \nu_3 \pm  \\
& & \sqrt{(u (\nu_1 + \nu_2) - (\lambda + \mu) \nu_3)^2 + 8 t^2 \nu_3 
(\nu_1 + \nu_2)} \nonumber  
\end{eqnarray}
and\footnote{Here we have corrected small misprints of Ref. \cite{Gork}.}
\begin{equation}
\Delta_1 = \Delta_2 = \kappa \Delta , \ \ \Delta_3 = \Delta_4 = \Delta,
\label{swaveord}
\end{equation}
where $\kappa^{-1} = - (g_{+,-} + u (\nu_1 + \nu_2))/(t \nu_3)$. 

For the first time possibility of  $s^{\pm}$ -- pairing in FeAs compounds was
noted in Ref. \cite{mazin}. This solution is in qualitative agreement with
ARPES data \cite{Ding0419,Wray2185,Evt4455}, except the results
$\Delta_1 = \Delta_2$ from (\ref{swaveord}), which contradicts the established
fact -- the gap on a small hole -- like cylinder is approximately twice as
large as on a big cylinder. In our opinion, this deficiency is due to 
unnecessary limitation of $u = V^{h1,h1}=V^{h2,h2}=V^{h1,h2}$, used in Ref.
\cite{Gork}. In case of different BCS constants on hole -- like cylinders we
can easily obtain different values of gaps in agreement with experiments.

The presence of practically isotropic gaps (possibly of different signs)
on hole -- like and electron -- like pockets of the Fermi surface,
corresponding to $s^{\pm}$ -- pairing, seems at first to contradict NMR (NQR)
data discussed above, which indicate possible gap $d$-wave gap symmetry. 
This contradiction was studied in Ref. \cite{Dolg3729}, where it was shown
rather convincingly that the absence of Gebel -- Slicter peak and power like
temperature dependence  of NMR relaxation may be easily explained also in
the case of $s^{\pm}$ -- pairing, with an account of impurity scattering.

\subsection{Electron -- phonon mechanism}

General discussion of the previous section tells us nothing about the origin
of pairing coupling constants entering the matrix (\ref{mmatr}), i.e. about
microscopic mechanism of Cooper pairing in new superconductors.

As a first candidate of such mechanism we have to consider the usual
electron -- phonon interaction. Such analysis for LaOFeAs was performed in
Ref. \cite{dolg}, using ab initio calculations of electron and phonon spectra
and Eliashberg function $\alpha^2 F(\omega)$, determining electron -- phonon
pairing constant as:
\begin{equation}
\lambda(\omega)=2\int_{0}^{\omega } d\Omega \alpha ^{2}F(\Omega )/\Omega
\label{eq:eq2}
\end{equation}
The total electron -- phonon pairing constant $\lambda$ was obtained by 
numerical integration in (\ref{eq:eq2}) up to $\omega=\infty$ and was found to
be 0.21. To estimate superconducting critical temperature $T_c$ we can use a
popular Allen -- Dynes interpolation formula \cite{AD}:
\begin{equation}
T_c=\frac{f_1f_2\omega_{ln}}{1.20}\exp\left(-\frac{1.04(1+\lambda)}
{\lambda-\mu^*-0.62\lambda\mu^*}\right),
\label{AllenDynes}
\end{equation}
where
\begin{eqnarray}
f_1=[1+(\lambda/\Lambda_1)^{3/2}]^{1/3},\qquad
\Lambda_1=2.46(1+3.8\mu^*),\nonumber\\
f_2=1+\frac{(\bar\omega_2/\omega_{ln}-1)\lambda^2}{\lambda^2+\Lambda^2_{2}},
\qquad
\Lambda_2=1.82(1+6.3\mu^*)(\bar\omega_2/\omega_{ln}),\qquad
\bar\omega_2=<\omega^2>^{1/2}\nonumber
\label{AllDyn}
\end{eqnarray}
and $<\omega^2>$ is an average square of phonon frequency. Taking into account
the value of average logarithmic frequency of phonons found in Ref.
\cite{dolg} $\omega_{ln}=205 K$ (and assuming $\omega_{ln}\approx\bar\omega_2$), 
with optimistic choice of Coulomb pseudopotential $\mu^*=0$, (\ref{AllenDynes}) 
gives the value of $T_c=0.5$K. Numerical solution of Eliashberg equations with
calculated $\alpha^2 F(\omega)$ gave the value of $T_c=0.8$K \cite{dolg}. 
Actually, to reproduce the experimental value of $T_c=26$K coupling constant
$\lambda$ should be approximately five times larger, even if we use $\mu^*=0$. 
In opinion of the authors of Ref. \cite{dolg} such a strong discrepancy
clearly shows that the usual picture of Cooper pairing, based on electron --
phonon coupling, is invalid.

Despite quite convincing estimates of Ref. \cite{dolg} we should note,
that these are based upon the standard approach of Eliashberg theory, which
does not take into account, in particular, an important role of multi -- band
nature of superconductivity in these compounds, e.g. the importance of
interband pairing interactions. Besides that, the estimates of the value of
electron -- phonon coupling constant has also met some objections. Thus, in
Ref. \cite{Eschr0186} it was argued that strong enough coupling exists between
electrons and a certain mode of Fe oscillations in FeAs plane.

However, probably most convincing objections to the claims of irrelevance of
electron -- phonon coupling can be based on experiments. For example, we can
estimate the electron -- phonon coupling constant from the temperature dependence
of resistivity. In Ref. \cite{Bhoi2695} the measurements of resistivity of
PrFeAsO$_{1-x}$F$_x$ were done in a wide enough temperature interval. Linear
growth of resistivity with temperature for $T>$170K saturated, which by itself
suggests strong enough electron -- phonon coupling. As a simple estimate we
can write resistivity as:
\begin{equation} 
\rho(T)=\frac{4\pi}{\omega_p^2\tau} 
\label{rhoT} 
\end{equation} 
where $\omega_p$ is plasma frequency, $\tau$ is carriers relaxation time. 
In the region of high enough temperatures, where resistivity grows linearly
with temperature, relaxation frequency of carriers on phonons is given by
\cite{AlKrk}:
\begin{equation}
\frac{\hbar}{\tau_{ep}}=2\pi\lambda_{tr}k_BT
\label{tauph}
\end{equation}
where ``transport'' constant of electron -- phonon coupling $\lambda_{tr}$, 
is naturally of the order of $\lambda$ of interest to us, differing from it
usually not more than by 10\%. From (\ref{rhoT}) and (\ref{tauph}) we have:
\begin{equation}
\lambda_{tr}=\frac{\hbar\omega_p^2}{8\pi^2k_B}\frac{d\rho}{dT}
\label{lamtr}
\end{equation}
Using the slope of the temperature dependence of resistivity determined in
Ref. \cite{Bhoi2695} $d\rho/dT\sim$ 8.6 ($\mu\Omega$/K) and the estimate of
plasma frequency $\omega_p\sim$ 0.8 eV (which follows from measurements of
penetration depth), we get $\lambda\sim$ 1.3, which according to 
(\ref{AllenDynes}) is quite sufficient to get the observable values of
transition temperatures in new superconductors.

Decisive evidence for electron -- phonon mechanism of Cooper pairing was
always considered to be the observation of isotope effect. Appropriate 
measurements were performed recently in Ref. \cite{Liu2694} on 
SmFeAsO$_{1-x}$F$_{x}$, where $^{16}$O was substituted by $^{18}$O, 
and on Ba$_{1-x}$K$_x$Fe$_2$As$_2$, where $^{56}$Fe was substituted by $^{54}$Fe. 
A finite shift of superconducting transition temperature was observed, which
can be characterized in a standard way by isotope effect exponent
$\alpha=-\frac{d\ln T_c}{d\ln M}$. For SmFeAsO$_{1-x}$F$_{x}$ isotope effect
was small enough, with $\alpha\sim 0.08$, which is quite natural as O ions
are outside the conducting FeAs layer. At the same time, the change of Fe ions
in FeAs layers in Ba$_{1-x}$K$_x$Fe$_2$As$_2$ has lead to a large isotope effect
with $\alpha\approx 0.4$, which is close to the ``ideal'' value of $\alpha=0.5$. 

Thus, rather wide pessimism in the literature on the role of electron --
phonon interaction in new superconductors seems to be rather premature.

\subsection{Magnetic fluctuations}

The pessimism with respect to the role of electron -- phonon interaction
mentioned above, as well as the closeness of superconducting phase to
antiferromagnetic on the phase diagram of new superconductors, has lead to the
growth of popularity of pairing models based upon the decisive role of
magnetic (spin) fluctuations, in many respects similar to those already
considered for HTSC cuprates \cite{I99,I91}.

Apparently, one of the first papers where the possible role of magnetic 
fluctuations in formation of Cooper pairs of $s^{\pm}$ -- type was stressed on
qualitative level was Ref. \cite{mazin}. Similar conclusions were reached by the
authors of Ref. \cite{aoki}, where a possibility of $d_{x^2 - y^2}$ -- pairing
was also noticed.

Rather detailed analysis of possible electronic mechanisms of pairing within
the generalized Hubbard model applied to the ``minimal'' model of Ref.
\cite{Scal} was performed in Ref. \cite{Scal2}. In fact, this analysis was done
in the framework of generalized RPA approximation, which takes into account
an exchange by spin (and orbital) fluctuations in particle -- hole channel
(cf. the review of similar one -- band models used for cuprates \cite{I91}).  
It was shown that the pairing interaction due to these fluctuations leads to
effective attraction in case of singlet $d$-wave pairing and triplet
$p$-wave pairing, with tendency to $d$-wave pairing instability becoming 
stronger as system moves towards magnetic (SDW) instability. In rather general
formulation, using the general renormalization group approach, the similar
model was analyzed in Ref. \cite{Chub3735}.

Unfortunately, in most of the papers devoted to pairing mechanism due to
exchange of magnetic fluctuations there are no direct calculations of $T_c$ 
allowing comparison with experiments. Thus we shall limit ourselves to these
short comments.


\section{Conclusion: end of cuprate monopoly} 

Let us briefly summarize. During the first half a year of studies of new
superconductors we have observed rather impressive progress in learning on
their basic physical properties. Very few problems remain uninvestigated,
though many results obtained require further clarification\footnote{
In this respect we want to stress the absence of any systematic studies of
effects of disordering, which may be unusual due to anomalous nature of
$s^{\pm}$ -- pairing}. The main result of all these studies is certainly
the end of cuprate monopoly in physics of high -- temperature superconductors.
A new wide class of iron based systems was discovered with high enough values 
of $T_c$ and variety of physical properties which reminds copper oxides,
which for more than 20 years were in the center of interests of superconductor
community. There are rather well founded expectations that in some near future
new systems will be discovered, though there is now a certain impression that
in the subclass of layered FeAs systems we have already reached the maximum
values of $T_c\sim 50$K, so that for further enhancement of $T_c$ we need some
kind of a new approaches. It is doubtless that work already done significantly
deepened our understanding of the nature of high -- temperature 
superconductivity, though we are still far from formulation of definite
``recipies'' for the search of new superconductors with higher values of $T_c$.

Let us formulate

\subsubsection{What is in common between iron based and cuprate superconductors?}

\begin{itemize}

\item{Both classes are represented by quasi two -- dimensional (layered)
systems from the point of view of their electronic properties, which leads to
more or less strong anisotropy.}

\item{In both classes superconducting region on the phase diagram is close to
the region of antiferromagnetic ordering. Prototype phase for both classes of
superconductors is antiferromagnetic.}

\item{Cooper pairing in both classes is singlet type, but ``anomalous'', i.e.
different from simple $s$-wave pairing, characteristic for traditional
(low -- temperature) superconductors.}

\item{Basic properties of superconducting state is more or less the same as in 
typical type II superconductors.}

\end{itemize}

From the point of view of our understanding of basic nature of high --
temperature superconductivity the meaning of these common properties remains
rather unclear. Why do we need (and do we need?) two -- dimensionality? 
Historically, the importance of two -- dimensionality was first stressed
in connection with the proposed high -- temperature superconductivity based
on excitonic mechanism of Ginzburg and Little \cite{GK}, but does it play any
significant role for superconductivity in cuprates and new iron based
superconductors? Do we need ``closeness'' to antiferromagnetism? 
Is antiferromagnetism just a competing phase, or it is helpful for HTSC, e.g.
via the replacement of electron -- phonon pairing by mechanism based upon
spin (antiferromagnetic) fluctuations? The answers to these questions are still
unclear, though the presence of such coincidences in rather different classes of
physical systems seem rather significant.

Let us look now

\subsubsection{What is different between iron based and cuprate superconductors?}

\begin{itemize}

\item{Prototype phases of HTSC cuprates are antiferromagnetic (strongly
correlated, Mott type) insulators, while for new superconductors these are
antiferromagnetic (intermediately correlated?) metals.}

\item{Cuprates in superconducting state are one band metals with a single
Fermi surface (hole -- like or electron -- like), while new superconductors
are multiple band metals with several hole -- like and electron -- like 
Fermi surfaces.}

\item{In cuprates we have anisotropic $d$-wave pairing, while in new
superconductors, almost surely, we have (almost?) isotropic
$s^{\pm}$ -- pairing.}

\item{It is quite possible that the microscopic mechanism of pairing in both
classes of superconductors is different --- in cuprates it is almost certainly
electronic mechanism (spin fluctuations), while in iron based superconductors
the role of electron -- phonon coupling can be quite important 
(isotope effect!).}

\end{itemize}

We see that differences between cuprates and new superconductors are probably
more pronounced than common properties. In this sense, one of the main
conclusions which can already be made is that HTSC is not a unique property of
cuprates, i.e. strongly correlated systems close to insulating state.
In some respects new superconductors are simpler and easier to understand ---
their normal state is not so mysterious as in the case of cuprates\footnote{
Some evidence for the pseudogap state is observed only in NMR data and it is
unclear, whether in new superconductors we have an additional pseudogap region
on the phase diagram as in cuprates, with appropriate renormalization of
electronic spectrum, like formation of ``Fermi arcs'' etc.}, though multiple
band structure complicates situation.

To conclude, we can expect that high -- temperature superconductivity is much 
more common, than it was assumed during the last 20 years, so that 
superconducting community may look into the future with certain optimism. 

The author is grateful to L.P. Gor'kov and I.I. Mazin for numerous discussions
on physics of new superconductors. He is also grateful to I.A. Nekrasov and 
Z.V. Pchelkina who co-authored papers on calculations of electronic spectra of
FeAs based systems. This work was partly supported by Russian Foundation of
Basic Research grant 08-02-00021 and by the programs of fundamental research of
the Russian Academy of Sciences ``Quantum macrophysics'' and ``Strongly
correlated electrons in semiconductors, metals, superconductors and magnetic
materials''. 



\end{document}